\documentclass[fleqn,usenatbib]{mnras}

\usepackage{newtxtext,newtxmath}

\usepackage[T1]{fontenc}

\DeclareRobustCommand{\VAN}[3]{#2}
\let\VANthebibliography\thebibliography
\def\thebibliography{\DeclareRobustCommand{\VAN}[3]{##3}\VANthebibliography}

\usepackage{graphicx}	
\usepackage{amsmath}	
\usepackage{lscape}     
\usepackage{longtable} 
\usepackage{subfigure}  
\usepackage{ulem}    
\setcitestyle{notesep={ }} 
\newcommand{\totalinspectedlightcurves}{20\,427 }

\newcommand{\totalsignals}{42 }
\newcommand{\totaltargets}{37 }

\newcommand{\newsignals}{18 }
\newcommand{\knownsignals}{24 }

\newcommand{\newVP}{four }
\newcommand{\newPC}{14 }

\newcommand{\medianfactorP}{30}
\newcommand{\medianfactorTo}{1.6 }
\newcommand{\medianfactorrprs}{1.4 }

\newcommand{\targetstoimprove}{22}

\newcommand{\starsCfive}{25\,137 }
\newcommand{\starsCsixteen}{29\,888 }
\newcommand{\starsCeigtheen}{20\,427 }

\newcommand{\starsCfiveeighteen}{11\,444 }
\newcommand{\starsCfivesixteeneighteen}{3261 }

\title[The K2-OjOS Project]{The K2-OjOS Project: New and revisited planets and candidates in \textit{K2} campaigns 5, 16, \& 18}

\author[A. Castro-Gonz{\'a}lez et al.]{A. Castro-Gonz{\'a}lez$^{1}$\thanks{E-mail: \url{amadeo@eimem.es}},
E. D{\'i}ez Alonso$^{1}$,
J. Men{\'e}ndez Blanco$^{1}$,
J.~Livingston$^{2}$,
J.~P.~de~Leon$^{2}$,
\newauthor
J. Lillo-Box$^{3}$,
J.~Korth$^{4}$,
S. Fern{\'a}ndez Men{\'e}ndez$^{1}$,
J. M. Recio$^{5}$,
F. Izquierdo-Ruiz$^{6}$,
\newauthor
A. Coya Lozano$^{7}$,
F. Garc{\'i}a de la Cuesta$^{7}$,
N. G{\'o}mez Hern{\'a}ndez$^{7}$,
J. R. Vidal Blanco$^{7}$,
\newauthor
R. Hevia D{\'i}az$^{7}$,
R. Pardo Silva$^{7}$,
S. P{\'e}rez Acevedo$^{7}$, 
J. Polancos Ruiz$^{7}$,
P. Padilla Tijer{\'i}n$^{7}$,
\newauthor
D. V{\'a}zquez Garc{\'i}a$^{7}$,
S. L. Su{\'a}rez G{\'o}mez$^{1}$,
F. Garc{\'i}a Riesgo$^{1}$,
C. Gonz{\'a}lez Guti{\'e}rrez$^{1}$,
\newauthor
L. Bonavera$^{1}$,
J. Gonz{\'a}lez-Nuevo$^{1}$,
C. Rodr{\'i}guez Pereira$^{1}$,
F. S{\'a}nchez Lasheras$^{1}$,
\newauthor
M. L. S{\'a}nchez Rodr{\'i}guez$^{1}$,
R. Muñiz$^{1}$,
J. D. Santos Rodr{\'i}guez$^{1}$ and
F. J. de Cos Juez$^{1}$
\\
$^{1}$Instituto Universitario de Ciencias y Tecnologías Espaciales de Asturias (ICTEA), C. Independencia 13, E-33004 Oviedo, Spain\\
$^{2}$Department of Astronomy, University of Tokyo, 7-3-1 Hongo, Bunkyo-ku, Tokyo 113-0033, Japan\\
$^{3}$Centro de Astrobiología (CSIC-INTA), ESAC campus, Villanueva de la Cañada, E-28692 Madrid, Spain\\
$^{4}$Department of Space, Earth and Environment, Astronomy and Plasma Physics, Chalmers University of Technology, SE-412 96 Gothenburg, Sweden\\
$^{5}$Malta-Consolider Team and Departmento de Química Física y Analítica, Universidad de Oviedo, E-33006 Oviedo, Spain\\
$^{6}$Malta-Consolider Team and Department of Chemistry and Chemical Engineering, Chalmers University of Technology, SE-412 96 Gothenburg, Sweden\\
$^{7}$Sociedad Astronómica Asturiana Omega, C. Alegría 45, 33209 Gijón, Spain\\
}

\date{Accepted 2021 September 2. Received 2021 August 14; in original form 2021 June 23}

\pubyear{2021}

\begin{document}
\label{firstpage}
\pagerange{\pageref{firstpage}--\pageref{lastpage}}
\maketitle
\begin{abstract}
We present the first results of K2-OjOS, a collaborative project between professional and amateur astronomers primarily aimed to detect, characterize, and validate new extrasolar planets. For this work, 10 amateur astronomers looked for planetary signals by visually inspecting the \totalinspectedlightcurves light curves of \textit{K2} campaign 18 (C18). They found \totalsignals planet candidates, of which \newsignals are new detections and \knownsignals had been detected in the overlapping C5 by previous works. We used archival photometric and spectroscopic observations, as well as new high-spatial resolution images in order to carry out a complete analysis of the candidates found, including a homogeneous characterization of the host stars, transit modelling, search for transit timing variations and statistical validation. As a result, we report \newVP new planets (K2-355 b, K2-356 b, K2-357 b, and K2-358 b) and \newPC planet candidates. Besides, we refine the transit ephemeris of the previously published planets and candidates by modelling C5, C16 (when available) and C18 photometric data jointly, largely improving the period and mid-transit time precision. Regarding individual systems, we highlight the new planet K2-356 b and candidate EPIC 211537087.02 being near a 2:1 period commensurability, the detection of significant TTVs in the bright star K2-184 (V = 10.35), the location of K2-103 b inside the habitable zone according to optimistic models, the detection of a new single transit in the known system K2-274, and the disposition reassignment of K2-120 b, which we consider as a planet candidate as the origin of the signal cannot be ascertained.



\end{abstract}




\section{I\,n\,t\,r\,o\,d\,u\,c\,t\,i\,o\,n}

\begin{figure}
    \centering
    \includegraphics[scale = 0.32]{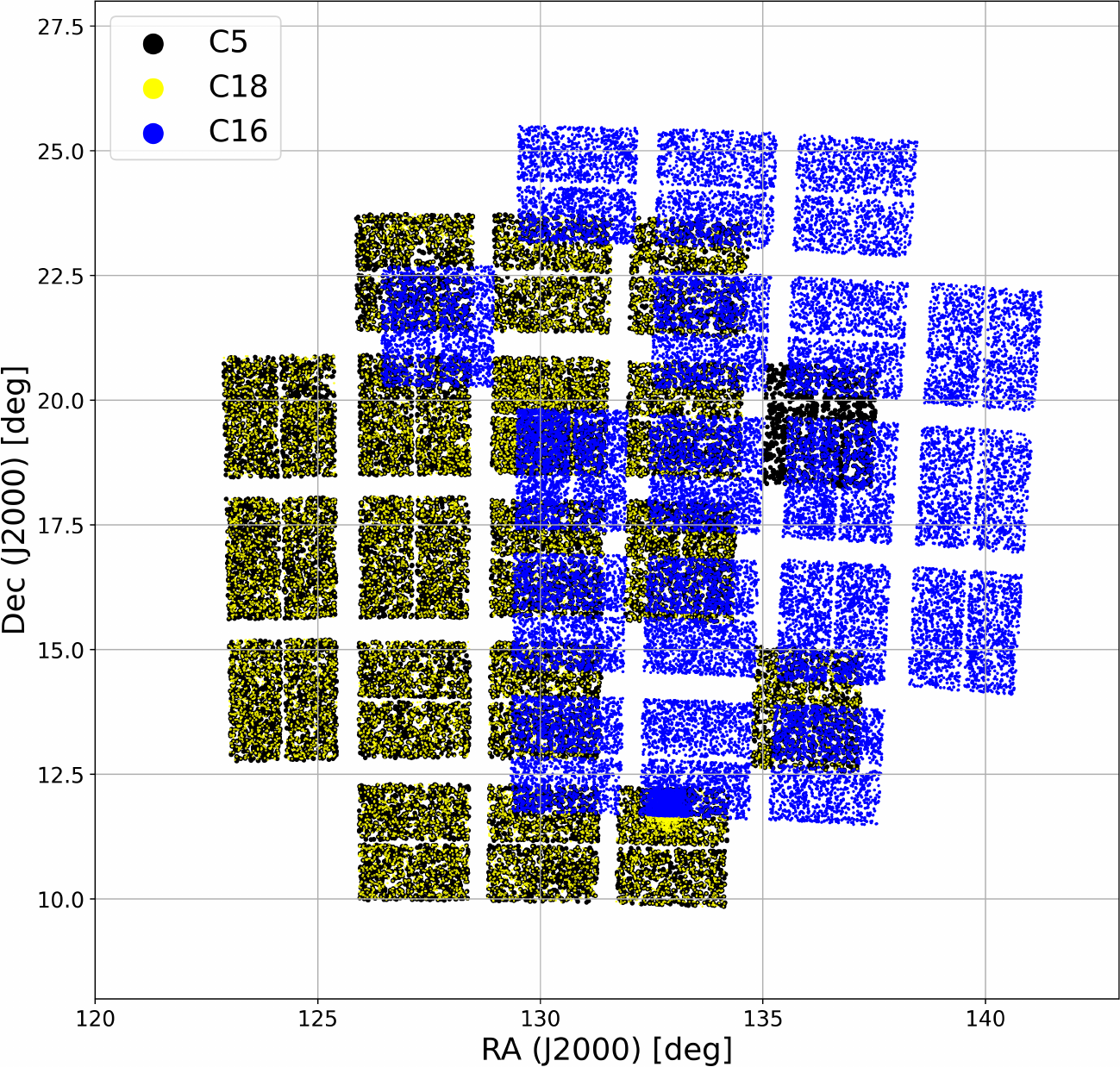}
    \caption{Sky position of the stellar targets observed in C5, C16, and C18. 95 per cent of C5 field is covered by C18, and 30 per cent is covered by C16.}
    \label{fig:C5_C16_C18_fields}
\end{figure}

The \textit{K2} mission \citep{2014PASP..126..398H} observed different ecliptic fields over 19 campaigns between 2014 and 2018 aiming to further enlarge the great success achieved by the primary \textit{Kepler} mission \citep{2010Sci...327..977B}. Both missions shared the main scientific goal of detecting and characterizing a large amount of transiting exoplanets in order to explore the structure and diversity of planetary systems. Their findings marked a real breakthrough in exoplanetary research, bringing a new perspective on the planet occurrence rates \citep[e.g.][]{2014PNAS..11112647B,2020AJ....159..248K}, multiplanetary systems \citep[e.g.][]{2015ApJ...799..170C,2018A&A...613A..68G}, and planet formation and evolution  \citep[e.g.][]{2014A&A...562A.109L,2014A&A...568L...1L}. 

Confirming the planetary nature of thousands of \textit{Kepler} and \textit{K2} transiting candidates by determining that the planet lies in the substellar regime via radial velocities measurements became impractical, especially for faint and magnetically active host stars. In this context arose the concept of statistical validation of transit signals, which consists of probing the planetary origin of the transit signal through discarding other possible non-planetary scenarios \citep[e.g.][]{2011ApJ...727...24T,2012ApJ...761....6M,2014MNRAS.441..983D}.

According to the NASA Exoplanet Archive \citep{2013PASP..125..989A}, 426 validated and confirmed planets have been detected using \textit{K2} data \citep[e.g.][]{2016ApJS..226....7C,2018A&A...620A..77L,2018MNRAS.476L..50D,2018MNRAS.480L...1D,2019MNRAS.489.5928D,2019MNRAS.482.1807K,2020MNRAS.499.5416C,2021MNRAS.508..195D}, representing 10 per cent of the total known planets. There are also nearly a thousand \textit{K2} candidates, which either do not meet the imposed validation criteria \citep[e.g. candidates reported in][]{2015ApJ...809...25M,2018AJ....156...78L,2018AJ....155..136M}, or are awaiting for a validation analysis \citep[e.g. many candidates reported in][]{2016A&A...594A.100B,2016MNRAS.461.3399P,2016ApJS..222...14V,2018AJ....156...22Y,2019ApJS..244...11K}. Even though a fairly extensive analysis of \textit{K2} data has been carried out so far, several campaigns still can be studied in greater detail.

In general, \textit{K2} fields are uniformly distributed along the ecliptic, so most of the targets were observed in just one campaign  (i.e. $\sim$80 d). Nevertheless, certain fields partially overlap, providing unique science opportunities due to the longer duty cycles and temporal baselines.
The most optimal campaigns  to take advantage of the existing overlaps in \textit{K2} fields are C5 (observations from 2015 April 27 to July 10), C16 (observations from 2017 December 7 to 2018 February 25), and C18  (observations from 2018 May 12 to July 2), which observed \starsCfive (C5), \starsCsixteen (C16) and \starsCeigtheen (C18) stellar targets in the  \textit{K2} long-cadence mode (30 min). C18 field covers 95 per cent of C5 field (see Fig. \ref{fig:C5_C16_C18_fields}) and both campaigns have \starsCfiveeighteen stellar targets in common, which were observed with a 3-yr temporal baseline and a 4-month duty cycle. The C16 field covers 30 per cent of C5 and C18, and the three campaigns have \starsCfivesixteeneighteen stellar targets in common, increasing their duty cycle up to nearly 7 months. Joining photometric data from these three campaigns allows us to  search for long-period planets not being identified in data from just one campaign, search for long-term transit timing variations (TTVs), and precisely measure the transit ephemeris, which is essential for future follow-up studies scheduled by the next ground- and space-based telescope generation. Although several works have been published reporting candidates, validated and confirmed transiting planets starting from C5 and C16 data \citep[e.g.][]{2017AJ....154..207D,2017AJ....153...64M,2018AJ....156..277L,2018AJ....155..136M,2018AJ....155...21P,2018AJ....156...22Y}, there are still no works aimed to an exhaustive analysis of C18.

\begin{figure*}
    \centering
    \includegraphics[scale=0.55]{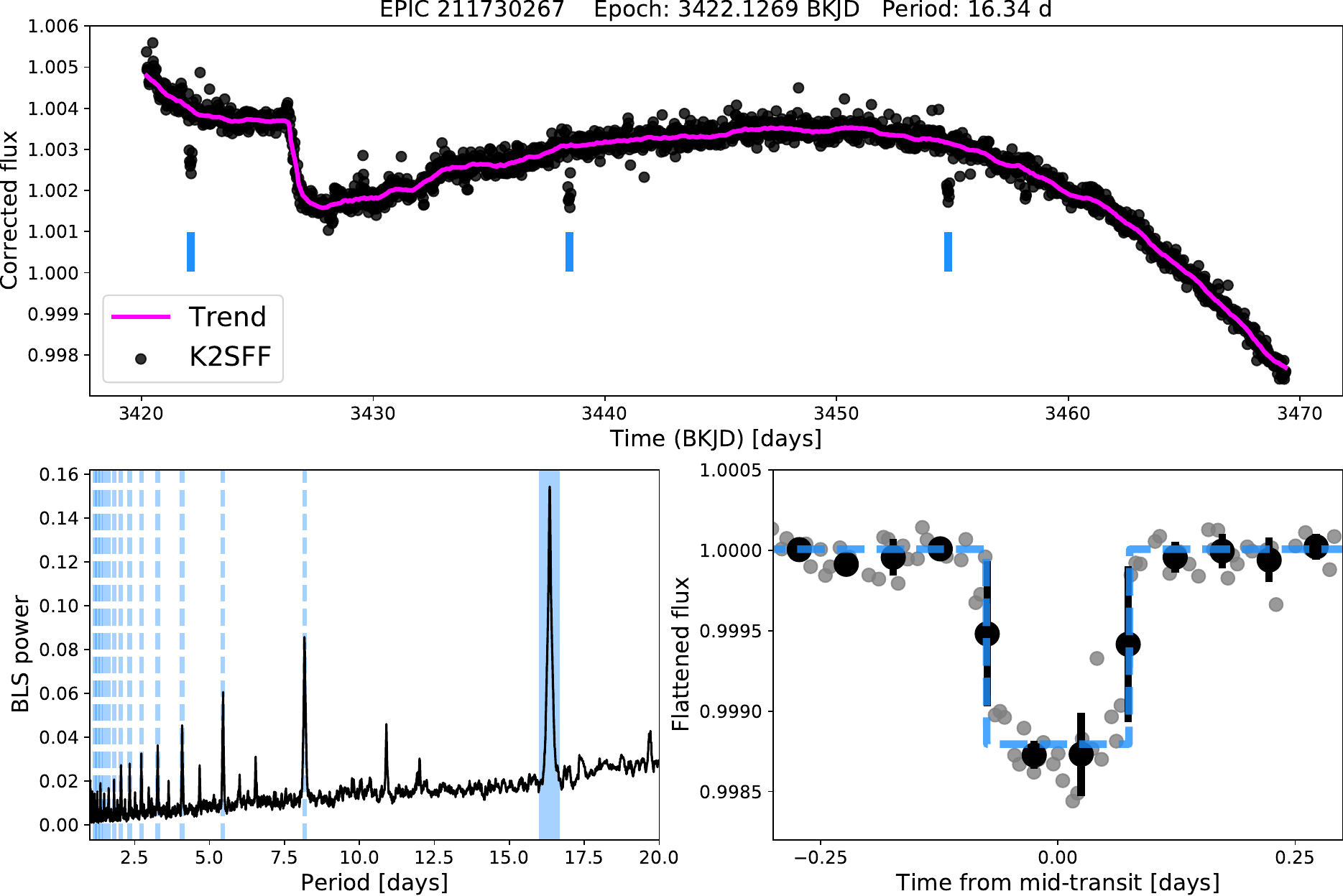}
    \caption{Summary plot of the preliminary vetting carried out by the K2-OjOS members for the newly detected planet K2-357 b (EPIC 211730267.01). Top: Long cadence K2SFF-corrected light curve along with the trend line. Blue vertical lines indicate the transits locations. Bottom left-hand panel: Box Least Squares Periodogram in which the main period ($P$ = 16.34 d, blue vertical line) and its harmonics (blue vertical dashed lines) are highlighted. Bottom right-hand panel: Phase-folded transit together with the BLS square model (0.12 per cent depth and 3.6 h duration). Black dots correspond to 1.2-h binned data.}
    \label{fig:K2-OjOS_summary_plot}
\end{figure*}


In this context, we started K2-OjOS\footnote{ \url{https://sites.google.com/view/k2-ojos/english}} \citep[][]{2020sea..confE..97C}, a Professional-Amateur (Pro-Am) project in which 10 amateur astronomers visually inspected the light curves of each C18 star in order to: (1) detect, characterize, and validate new extrasolar planets, (2) revisit the orbital and physical parameters of planets detected in C5 with new data from C16 and C18, (3) detect variable stars, and (4) compare visual and automated detection procedures.

Contributions of citizen scientists to the exoplanetary research has been remarkable along the last decade through projects like Planet Hunters \citep[e.g.][]{2012MNRAS.419.2900F,2012ApJ...754..129S,2014ApJ...795..167S,2020arXiv201113944E,2020MNRAS.494..750E} and Exoplanet Explorers \citep[e.g.][]{2018AJ....155...57C,2019AJ....157...40F,2019AAS...23316407H,2019RNAAS...3...43Z}. The K2-OjOS project was designed to have quite a few less members than the aforementioned projects (tens versus thousands). Besides searching for signals, the K2-OjOS members took part of the preliminary vetting process, carried out ground-based photometric follow-up of several variable stars, and participated in several scientific discussions with professional astronomers.


In Section \ref{project}, we describe the K2-OjOS project, including the signal detection methodology, preliminary vetting, independent search through BLS algorithm, and the transit injection and recovery. In Section \ref{sec:Observations}, we describe the observations and data reduction for both the \textit{K2} photometry and the high-resolution imaging. In Section \ref{analysis}, we present the analysis, in which we describe the stellar characterization of the host stars, transit modelling, search for transit timing variations and planet validation. In Section \ref{results_discussion}, we present the results, in which we quantify the transit ephemeris refinement of the already known planets, compare the recovery rates of both the K2-OjOS and BLS searches, contextualize and discuss the characteristics of the host stars, planets and candidates in our sample, and highlight interesting features of five individual systems. We conclude with a summary in Section \ref{summary}.

\section{T\,h\,e \, K\,2\,-\,O\,j\,O\,S \, p\,r\,o\,j\,e\,c\,t}
\label{project}

\subsection{Visual search for planetary signals}
\label{visual_search}


The K2-OjOS members had previous knowledge about the  typical photometric features of stars hosting
transiting exoplanets (i.e. periodicities, transit shapes, depths, and durations), as well as about the most recognizable types of variable stars. Furthermore, we developed tutorials with guidelines for proper identifications, which are available in the K2-OjOS website. We also established an online chat and scheduled frequent meetings to allow communication between K2-OjOS members and professional astronomers. 

We distributed the C18 20\,427 target stars in 21 batches (20 with 1000 light curves and 1 with 427). Then, we assigned one batch to each K2-OjOS member, and they systematically inspected the K2 Self Flat Fielding (K2SFF) corrected light curves \citep{2014PASP..126..948V} available on the Center for Astrophysics of the Harvard University and Smithsonian Institution website.\footnote{\url{https://www.cfa.harvard.edu/~avanderb/k2.html}} As they finished inspecting their batches, we assigned them more, until reaching all the 21 batches. As a result, six members analysed one batch and four members analysed more than one (2, 3, 4, and 6 batches).  The members carried out a preliminary classification of the stars with detected variability, discerning between the main following categories: \texttt{planets}, \texttt{eclipsing binaries}, \texttt{rotatings}, \texttt{pulsatings}, \texttt{irregulars}, and \texttt{artefacts}. Finally, they did a double check exchanging the targets found with a different member. The results were the following: 216 \texttt{planets}, 427 \texttt{eclipsing binaries}, 374 \texttt{rotatings}, 288 \texttt{pulsatings}, 195 \texttt{irregulars}, and 101 \texttt{artefacts}. The amateur astronomers jointly with professional astronomers subjected the 216 \texttt{planets} to a thorough vetting in order to avoid false detections or misclassifications and thus obtain the final sample of planet candidates (Section \ref{preliminary_vetting}). The remaining 1284 targets classified within the different categories of variable stars are being studied separately.

\subsection{Preliminary vetting and signal selection}
\label{preliminary_vetting}
The K2-OjOS members performed a preliminary vetting of the 216 targets classified as \texttt{planet} in the visual searching step, in order to create, together with professional astronomers, a high-quality planet candidate sample. In the following, we briefly summarize the procedure, which is fully available in the K2-OjOS website. (1) The members checked if the targets were observed in the overlapping C5 and C16, and if so, they checked that the signals found were also present in those campaigns. (2) They removed outliers and long-term trends from the light curves through the \textsc{wotan} package \citep[][]{2019AJ....158..143H}, being cautious of not overfit nor remove relevant data. (3) They assessed if the signals were periodic, and if so, computed relevant signal features as the orbital periods, mid-transit times, transit durations, and transit depths trough the BLS algorithm \citep[][]{2002A&A...391..369K}. (4) They produced phase-folded light curves to check their shapes. During this process,  we removed from the \texttt{planet} category 174 targets (80 per cent of the  \texttt{planet} targets), which were mostly misclassified eclipsing binaries or artefacts. As a result, we ended up with 42 planet candidates in 37 stars, being 18 of them new detections. In Fig. \ref{fig:K2-OjOS_summary_plot}, we show a summary plot of this process created by the K2-OjOS team for the newly detected planet K2-357 b.  

\subsection{Independent search through BLS algorithm}

We carried out an independent search by applying the BLS algorithm to the whole C18 sample, in order to cross-check the search. We used the BLS implementation of the \textsc{lctools} software \citep[][]{2019arXiv191008034S} considering low-restrictive signal properties: signal-to-noise ratio > 6, planet size > 0.5 $\rm R_{\oplus}$, orbital period > 0.25 d and transit duration > 1 h. As a result, the BLS search retrieved all the findings made by the K2-OjOS members except single transits, and no extra candidate was found.

\subsection{Transit injection and recovery}
\label{sec:injection_recovery}

To quantify and compare the K2-OjOS and BLS detection efficiencies, we simulated transit signals and injected them into real \textit{K2} light curves to be analysed by both search methods. In the following, we detail the transit injection procedure. 

First, we randomly selected 200 K2SFF corrected light curves of C18 stars that were analysed by \citet{2020ApJS..247...28H}, in order to acquire homogeneous estimates of their stellar parameters. Secondly, we exchanged the light curves with real planetary signals or eclipsing binaries for light curves without any hint of having eclipses. Thirdly, we randomly injected one simulated transit in 40 per cent of the 200 selected light curves generated through the \citet{2002ApJ...580L.171M} quadratic transit model as implemented in the \textsc{batman} package \citep{2015PASP..127.1161K}. We adopted stellar masses and radii from the \citet{2020ApJS..247...28H} catalogue and fixed the orbital inclination to $i=90^{\circ}$ and eccentricity to $e=0$. We randomly generated the orbital periods between 1 d and the 49-d temporal baseline of C18, the $R_{p} / R_{\star}$ ratios between 1 per cent and 10 per cent, and the mid-transit times between the C18 starting time and one orbital period later. We repeated the above-mentioned process generating different batches of 200 light curves with different injected transits. 

Each K2-OjOS member analysed the same number of batches as the number of original analysed batches with real C18 data.  Regarding the BLS search, the necessary condition for a signal to be considered as retrieved is that any of the 10 main BLS periods have to be within 1 per cent of the injected orbital period. The results of both independent searches are shown in Section \ref{sec:detection_efficiency}. 




     

\section{O\,b\,s\,e\,r\,v\,a\,t\,i\,o\,n\,s \, a\,n\,d \, d\,a\,t\,a \, r\,e\,d\,u\,c\,t\,i\,o\,n}
\label{sec:Observations}

\subsection{\textit{K2} photometry extraction and processing}
\label{K2_photometry}

\begin{figure}
    \centering
    \includegraphics[scale=0.40]{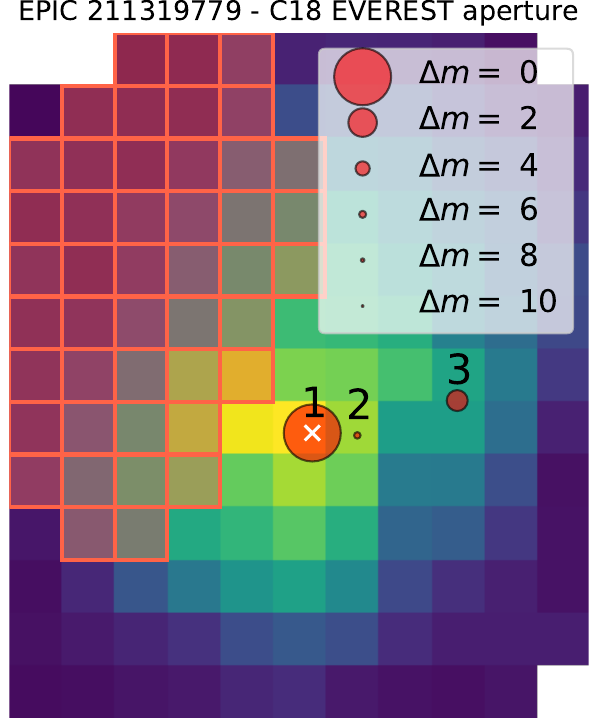}
    \includegraphics[scale=0.40]{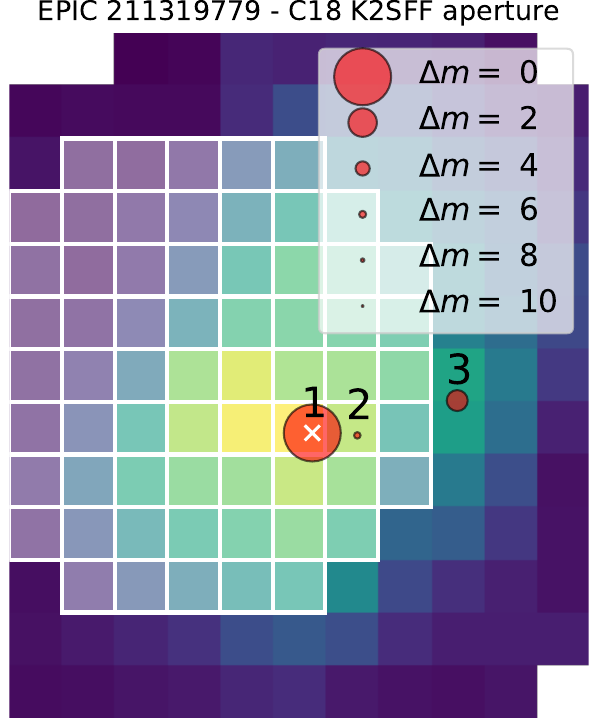}
    \centering
    \includegraphics[scale=0.17]{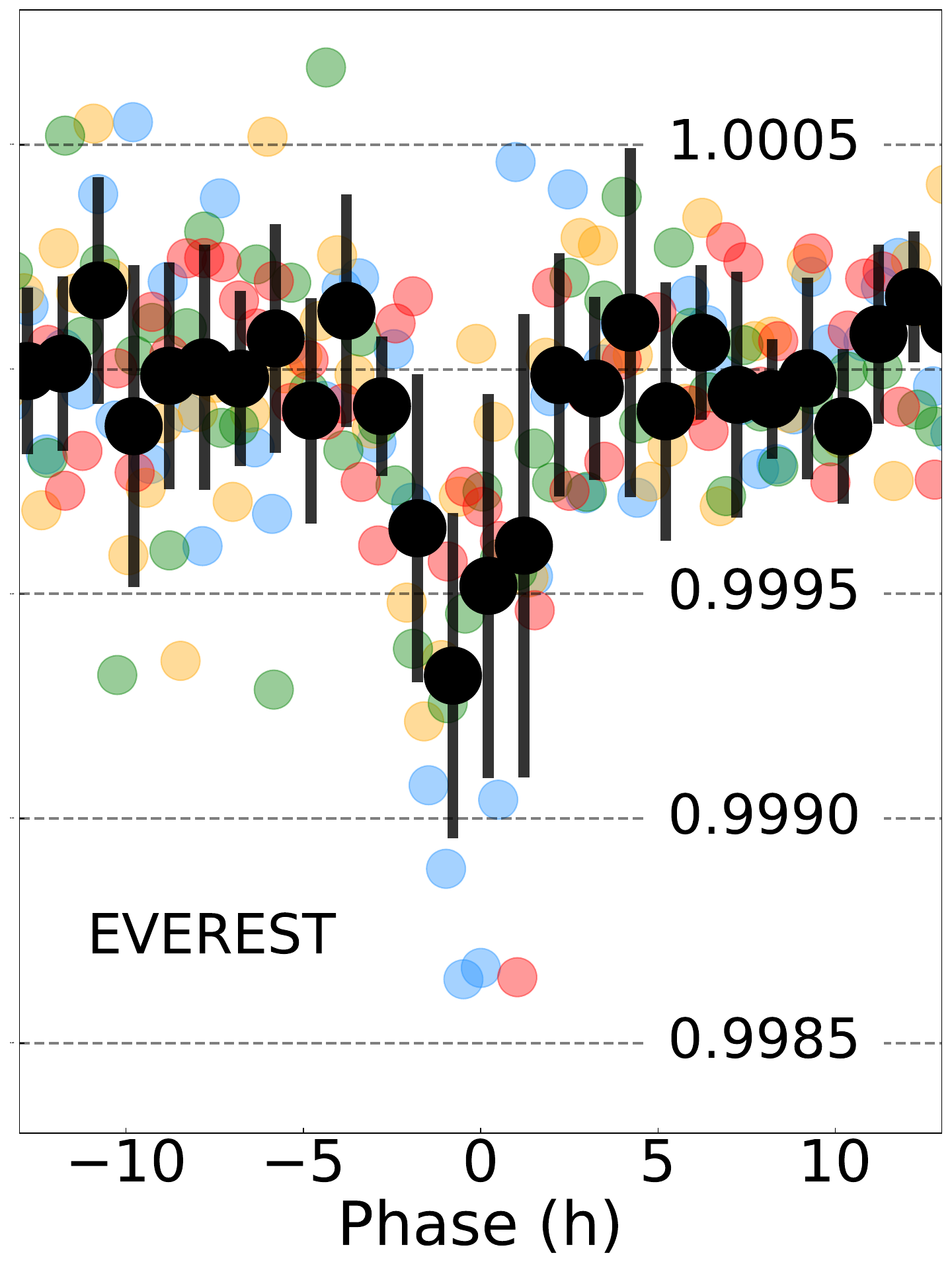}
    \includegraphics[scale=0.17]{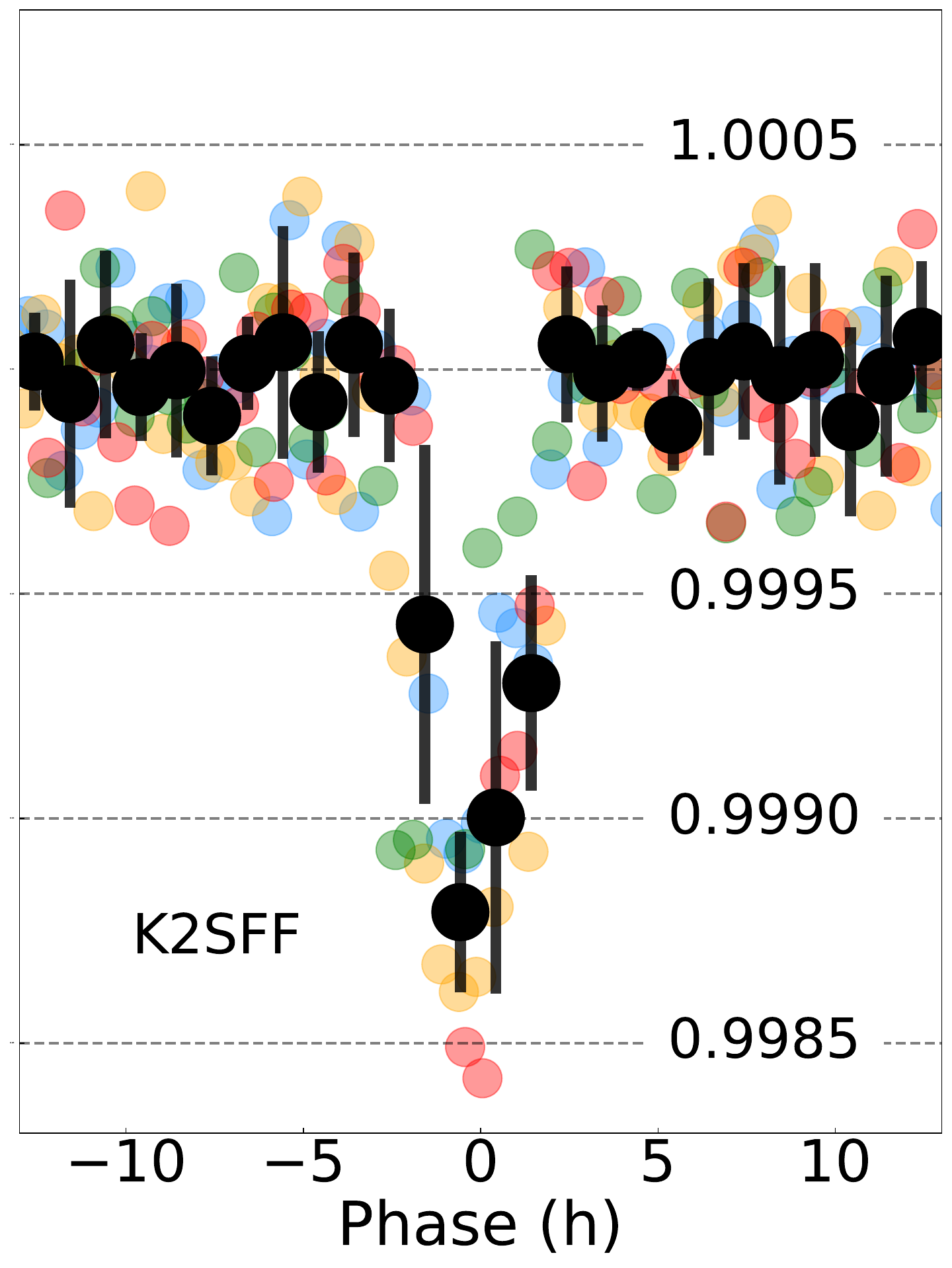}
    \caption{Top: Photometric apertures for EPIC 211319779 (star $\#$1) corresponding to the EVEREST and K2SFF pipelines. Both plots have been made through \textsc{tpfplotter}. Bottom: Phase-folded light curves obtained with each pipeline. For both plots, the first transit is plotted in blue, and the second, third, and fourth are in orange, green, and red, respectively. The black dots correspond to 1-h binned data.}
    \label{fig:EVEREST_vs_K2SFF}
\end{figure}

We downloaded from the Mikulsky Archive for Space Telescopes (MAST) the C18 (and C5 and C16 when available) light curves processed with the EVEREST pipeline \citep{2016AJ....152..100L,2018AJ....156...99L} for the \totaltargets targets found in the K2SFF light curves by the K2-OjOS team, with the objective of comparing both pipelines and deciding which one to use in the subsequent analysis. We checked that the C18 transit-like signals found starting from the K2SFF light curves were also detectable in the overlapping campaigns and in the EVEREST light curves. We found three cases (EPICs 211319779, 211606790, and 211407755) for which the EVEREST photometric apertures are not suitable to collect completely the stellar fluxes, as part of the edges of these apertures are located over the target PSFs. Fig. \ref{fig:EVEREST_vs_K2SFF} illustrates this phenomenon for EPIC 211319779, for which the unsuitable EVEREST aperture decreases the C18 phase-folded transit depth by a factor of two. In this particular case, the first EVEREST transit corresponds to a situation in which the target is located inside the aperture and skimming its edge. However, over the course of the campaign, the aperture is gradually separating from the target, causing a great flux dimming for the remaining three transits. Besides, we found one target (EPIC 212008766, $G_{\rm mag}$ = 12.87) whose C5 EVEREST photometry is contaminated, as its aperture, unlike that of K2SFF, encompasses the nearby star Gaia DR2 664406705976755840 ($G_{\rm mag}$ = 15.16). This shows the importance of a thorough cross-comparison between different photometric pipelines, in order to assess the quality of the signals. Due to the smaller out-of-transit scatter obtained with the EVEREST pipeline, we used its light curves for the subsequent analysis, except for the aforementioned four targets. For EPICs 211319779, 211606790, and 212008766, we used the K2SFF, and for EPIC 211407755, whose K2SFF aperture edge is also too close to the target, we applied the SFF corrector to the raw flux collected in a bigger aperture.

We removed long-term trends and normalized each light curve by using the \textsc{wotan} package \citep{2019AJ....158..143H}. Choosing a suitable de-trending procedure is highly important as it has a direct impact on the transit depth, and therefore on the derived planetary radius. First, we removed upper outliers that are more than 5-$\sigma$ above the running mean. Then, we used the robust time-windowed biweight method while cutting off the extremes of each time-series in order to avoid edge effects. We note that this method is not suitable for some of our targets, whose transit signals would disturb the trend (i.e. there is no window-length capable of reproducing the long-term trend while not being distorted by the transits). This phenomenon typically occurs in stars with strong stellar rotation modulations (e.g. EPICs 211335816, 211418290, and 211424769). In consequence, for these targets we used the cosine detrending method \citep{2019AJ....158..143H} by following a two-step process. First, we performed a preliminary detrending in which we used the Transit Least Squares algorithm \citep[TLS;][]{2019A&A...623A..39H} to identify transit-like signals that could disturb the trend. Secondly, we detrended the original time-series masking the transit data points found in the first step.

After the detrending process we joined the light curves from different campaigns generating a single light curve per target, and then used the TLS algorithm in order to obtain constraints on $P$ and $T_{0}$ to be used in the posterior transit modelling, as well as to search for additional shallow transit-like signals within the joined data.

\subsection{High-resolution imaging}
\label{sec:AstraLux}

We observed the four stars that meet all the validation criteria described in Section \ref{validation} (EPICs 211730267, 211914998, 211525753, and 211537087) with the high-spatial resolution camera AstraLux \citep{hormuth08}, located at the 2.2\,m telescope of the Calar Alto Observatory (Almería, Spain) on the nights of 22 and 23 of 2021 March. This camera uses the lucky-imaging approach to obtain diffraction-limited images based on the observation of a large number of frames with very short exposure below the coherence time. We used the  Sloan Digital Sky Survey z filter (SDSSz) and obtained 60\,000 frames with 20\,ms exposure times for EPIC\,211730267, 60\,000 $\times$ 30\,ms for EPIC\,211914998, 18\,400 $\times$ 30\,ms for EPIC\,211525753, and 23\,400 $\times$ 20\,ms for EPIC\,211537087. In order to focus on the closer regions, we restricted the field-of-view by windowing to $6\times6$ arcsec.

\begin{figure}
    \centering
    \includegraphics[scale=0.45]{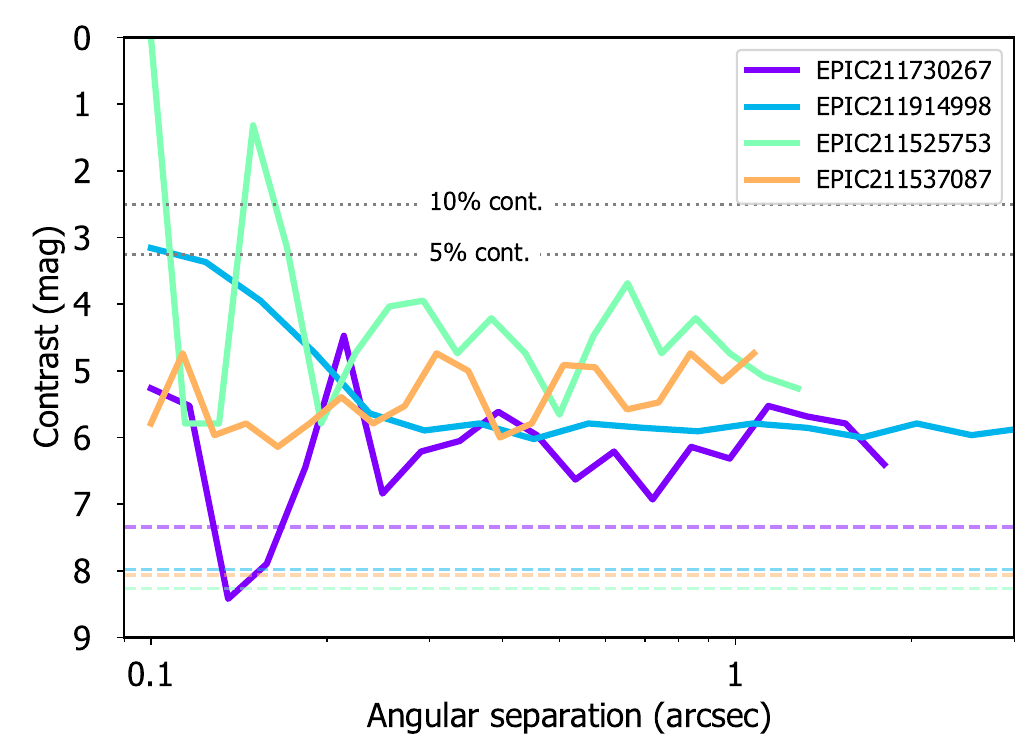}
    \caption{Contrast curves for EPICs 211730267, 211914998, 211525753, and 211537087 obtained from the AstraLux combined frames.}
    \label{fig:astralux}
\end{figure}

We used the instrument pipeline to select the 10 per cent frames with the highest Strehl ratio \citep{strehl1902} and combine them into a final high-spatial resolution image. Based on this final image, we computed the sensitivity curve by using our own developed \textsc{astrasens} package\footnote{\url{https://github.com/jlillo/astrasens}} with the procedure described in \cite{lillo-box12,lillo-box14b}. We find no evidence of additional sources within this field of view and within the computed sensitivity limits, shown in Fig. ~\ref{fig:astralux}. 

We use these contrast curves to estimate the probability of contamination from blended sources in the \textit{K2} aperture and undetectable from the public images. This probability is called the blended source confidence (BSC) and the steps for estimating it are fully described in \cite{lillo-box14b}. We use a \textsc{python} implementation of this approach (\textsc{bsc}) which uses the \textsc{trilegal}\footnote{\url{http://stev.oapd.inaf.it/cgi-bin/trilegal}} Galactic model \citep[v1.6;][]{girardi12} to retrieve a simulated source population of the region around the corresponding target.\footnote{This is done in \textsc{python} by using the \textsc{astrobase} implementation by \cite{astrobase}.} For instance, the transit signal in EPIC\,211730267 could be mimicked by a blended chance-aligned binary with a magnitude contrast up to $\Delta m=6.8$~mag. However, given the high-resolution image, we estimate using the \textsc{bsc} code that the probability of having an undetected source in the AstraLux image within this contrast range is 0.15 per cent. Similarly, we find such probability to be 2.83 per cent for EPIC\,211914998, 0.34 per cent for EPIC\,211525753, and 0.15 per cent for EPIC\,211537087.

\section{A\,n\,a\,l\,y\,s\,i\,s}
\label{analysis}

\subsection{Stellar characterization}
\label{stellar_charact}


\renewcommand{\arraystretch}{1.22}
\begin{table*}
\caption{\textsc{isochrones}-derived parameters.}
	\begin{center}
		\begin{tabular}{c|c|c|c|c|c|c|c|c}
		    \hline
			EPIC ID & $C$ & $K_p$ (mag) & $T_{\rm eff}$ (K) & [Fe/H] (dex) & log $g$ (cgs) & $R_{\star}(\rm R_{\odot})$ & $M_{\star}(\rm M_{\odot})$ & Spectroscopic parameters reference \\
			\hline
			211309648 & 18 & 13.848 & $6283^{+235}_{-232}$ & $0.05^{+0.15}_{-0.16}$ & $4.18^{+0.07}_{-0.07}$ & $1.50^{+0.15}_{-0.12}$ & $1.24^{+0.10}_{-0.08}$ & -- \\
			211317649 & 18 & 13.214 & $5693^{+62}_{-63}$ & $0.20^{+0.07}_{-0.07}$ & $4.38^{+0.05}_{-0.04}$ & $1.08^{+0.05}_{-0.05}$ & $1.03^{+0.04}_{-0.04}$ & \cite{Boisse_2013} \\
			211319617 & 5,18 & 12.393 & $5306^{+45}_{-40}$ & $-0.50^{+0.07}_{-0.06}$ & $4.57^{+0.03}_{-0.03}$ & $0.72^{+0.01}_{-0.01}$ & $0.72^{+0.04}_{-0.03}$ & \cite{2018AJ....155..136M} \\
			211319779 & 18 & 12.640 & $5163^{+161}_{-127}$ & $-0.20^{+0.14}_{-0.17}$ & $4.58^{+0.02}_{-0.03}$ & $0.75^{+0.02}_{-0.02}$ & $0.78^{+0.04}_{-0.05}$ & -- \\
			211335816 & 5,18 & 11.929 & $6223^{+105}_{-134}$ & $0.03^{+0.09}_{-0.12}$ & $4.07^{+0.04}_{-0.04}$ & $1.73^{+0.10}_{-0.10}$ & $1.27^{+0.07}_{-0.06}$ & \cite{2020ApJS..247...28H} \\
			211359660 & 5,18 & 11.742 & $5181^{+48}_{-42}$ & $-0.02^{+0.06}_{-0.07}$ & $4.56^{+0.02}_{-0.03}$ & $0.80^{+0.01}_{-0.01}$ & $0.84^{+0.03}_{-0.04}$ & \cite{2018AJ....155..136M} \\
			211393988 & 18 & 13.862 & $6972^{+456}_{-381}$ & $0.02^{+0.15}_{-0.16}$ & $4.10^{+0.10}_{-0.11}$ & $1.82^{+0.29}_{-0.22}$ & $1.53^{+0.15}_{-0.14}$ & -- \\
			211407755 & 5,16,18 & 14.573 & $4726^{+82}_{-74}$ & $-0.20^{+0.14}_{-0.18}$ & $4.63^{+0.03}_{-0.03}$ & $0.67^{+0.04}_{-0.04}$ & $0.68^{+0.04}_{-0.04}$ & \cite{2020ApJS..247...28H} \\
			211418290 & 5,18 & 11.504 & $5041^{+55}_{-46}$ & $-0.43^{+0.08}_{-0.09}$ & $3.64^{+0.06}_{-0.04}$ & $3.02^{+0.17}_{-0.16}$ & $1.47^{+0.12}_{-0.12}$ & \cite{2018AJ....155..136M} \\
			211418729 & 5,18 & 14.286 & $5039^{+52}_{-55}$ & $0.38^{+0.04}_{-0.04}$ & $4.51^{+0.03}_{-0.02}$ & $0.87^{+0.02}_{-0.02}$ & $0.89^{+0.03}_{-0.02}$ & \cite{2017AJ....154..188S} \\
			211424769 & 5,18 & 9.438 & $6217^{+46}_{-41}$ & $-0.06^{+0.06}_{-0.07}$ & $4.21^{+0.02}_{-0.02}$ & $1.38^{+0.03}_{-0.02}$ & $1.12^{+0.05}_{-0.04}$ & \cite{2018AJ....155..136M} \\
			211480861 & 18 & 9.961 & $7461^{+587}_{-401}$ & $0.03^{+0.14}_{-0.16}$ & $4.03^{+0.06}_{-0.05}$ & $2.11^{+0.12}_{-0.12}$ & $1.74^{+0.15}_{-0.11}$ & -- \\
			211525389 & 5,18 & 11.687 & $5636^{+49}_{-52}$ & $0.23^{+0.04}_{-0.03}$ & $4.49^{+0.01}_{-0.01}$ & $0.97^{+0.01}_{-0.01}$ & $1.05^{+0.02}_{-0.02}$ & \cite{2018AJ....155...21P} \\
			211525753 & 18 & 13.746 & $5790^{+177}_{-147}$ & $-0.12^{+0.15}_{-0.18}$ & $4.49^{+0.03}_{-0.04}$ & $0.92^{+0.04}_{-0.03}$ & $0.96^{+0.05}_{-0.06}$ & -- \\
			211537087 & 18 & 13.438 & $5568^{+178}_{-150}$ & $-0.11^{+0.15}_{-0.17}$ & $4.52^{+0.03}_{-0.04}$ & $0.86^{+0.03}_{-0.03}$ & $0.90^{+0.05}_{-0.06}$ & -- \\
			211590050 & 18 & 13.317 & $6414^{+310}_{-280}$ & $0.03^{+0.13}_{-0.16}$ & $4.16^{+0.07}_{-0.06}$ & $1.57^{+0.14}_{-0.12}$ & $1.29^{+0.10}_{-0.09}$ & -- \\
			211594205 & 5,18 & 10.680 & $5245^{+34}_{-35}$ & $-0.10^{+0.04}_{-0.04}$ & $4.62^{+0.00}_{-0.00}$ & $0.75^{+0.01}_{-0.01}$ & $0.84^{+0.01}_{-0.01}$ & \cite{2018AJ....155...21P} \\
			211606790 & 5,16,18 & 12.673 & $5456^{+49}_{-50}$ & $0.07^{+0.07}_{-0.07}$ & $4.12^{+0.03}_{-0.03}$ & $1.76^{+0.09}_{-0.09}$ & $1.47^{+0.07}_{-0.05}$ & \cite{2018AJ....155..136M} \\
			211644764 & 18 & 13.105 & $6688^{+377}_{-269}$ & $0.04^{+0.14}_{-0.14}$ & $4.05^{+0.08}_{-0.07}$ & $1.90^{+0.21}_{-0.18}$ & $1.48^{+0.12}_{-0.11}$ & -- \\
			211705502 & 18 & 13.216 & $6471^{+314}_{-269}$ & $0.01^{+0.15}_{-0.16}$ & $4.24^{+0.06}_{-0.06}$ & $1.39^{+0.12}_{-0.09}$ & $1.26^{+0.10}_{-0.10}$ & -- \\
			211724246 & 18 & 13.242 & $6486^{+290}_{-302}$ & $0.07^{+0.15}_{-0.14}$ & $3.97^{+0.07}_{-0.07}$ & $2.10^{+0.23}_{-0.19}$ & $1.50^{+0.13}_{-0.11}$ & -- \\
			211730267 & 18 & 13.459 & $5794^{+210}_{-169}$ & $-0.02^{+0.14}_{-0.16}$ & $4.43^{+0.04}_{-0.04}$ & $1.00^{+0.05}_{-0.05}$ & $0.98^{+0.07}_{-0.07}$ & -- \\
			211733267 & 5,16,18 & 12.150 & $5342^{+48}_{-47}$ & $0.02^{+0.06}_{-0.07}$ & $4.46^{+0.03}_{-0.02}$ & $0.90^{+0.02}_{-0.02}$ & $0.85^{+0.03}_{-0.02}$ & \cite{2018AJ....155..136M} \\
			211791178 & 5,18 & 13.648 & $4648^{+88}_{-76}$ & $-0.31^{+0.12}_{-0.10}$ & $4.50^{+0.03}_{-0.02}$ & $0.81^{+0.03}_{-0.03}$ & $0.74^{+0.02}_{-0.02}$ & \cite{2017ApJ...836..167D} \\
			211816003 & 5,16,18 & 13.654 & $5397^{+105}_{-96}$ & $-0.09^{+0.12}_{-0.12}$ & $4.53^{+0.03}_{-0.04}$ & $0.83^{+0.03}_{-0.03}$ & $0.86^{+0.05}_{-0.05}$ & \cite{2020ApJS..247...28H} \\
			211818569 & 5,18 & 12.935 & $4690^{+43}_{-41}$ & $-0.16^{+0.05}_{-0.06}$ & $4.63^{+0.02}_{-0.02}$ & $0.67^{+0.01}_{-0.01}$ & $0.69^{+0.03}_{-0.03}$ & \cite{2018AJ....155..136M} \\
			211822797 & 5,16,18 & 14.568 & $4057^{+37}_{-39}$ & $0.20^{+0.05}_{-0.07}$ & $4.70^{+0.01}_{-0.01}$ & $0.58^{+0.01}_{-0.01}$ & $0.62^{+0.02}_{-0.02}$ & \cite{2017ApJ...836..167D} \\
			211904310 & 18 & 13.636 & $6404^{+284}_{-261}$ & $0.05^{+0.15}_{-0.16}$ & $4.12^{+0.08}_{-0.07}$ & $1.65^{+0.19}_{-0.16}$ & $1.33^{+0.10}_{-0.11}$ & -- \\
			211913977 & 5,16,18 & 12.619 & $4927^{+47}_{-46}$ & $0.07^{+0.07}_{-0.07}$ & $4.59^{+0.01}_{-0.02}$ & $0.76^{+0.01}_{-0.01}$ & $0.82^{+0.02}_{-0.03}$ & \cite{2018AJ....155..136M} \\
			211914998 & 18 & 13.587 & $5650^{+193}_{-152}$ & $-0.07^{+0.15}_{-0.16}$ & $4.49^{+0.03}_{-0.04}$ & $0.91^{+0.04}_{-0.04}$ & $0.93^{+0.06}_{-0.06}$ & -- \\
			211916756 & 5,18 & 15.498 & $3566^{+34}_{-37}$ & $0.10^{+0.06}_{-0.06}$ & $4.85^{+0.01}_{-0.01}$ & $0.40^{+0.01}_{-0.01}$ & $0.41^{+0.01}_{-0.01}$ & \cite{2018AJ....156..277L} \\
			211919004 & 5,16,18 & 13.131 & $5161^{+53}_{-47}$ & $0.20^{+0.04}_{-0.04}$ & $4.53^{+0.03}_{-0.03}$ & $0.85^{+0.02}_{-0.02}$ & $0.89^{+0.03}_{-0.03}$ & \cite{2018AJ....155...21P} \\
			211969807 & 5,16,18 & 15.149 & $3712^{+54}_{-76}$ & $0.18^{+0.09}_{-0.10}$ & $4.78^{+0.02}_{-0.02}$ & $0.48^{+0.01}_{-0.01}$ & $0.51^{+0.02}_{-0.02}$ & \cite{2017ApJ...836..167D} \\
			212006344 & 5,16,18 & 12.466 & $4009^{+47}_{-44}$ & $0.32^{+0.06}_{-0.07}$ & $4.69^{+0.01}_{-0.01}$ & $0.59^{+0.01}_{-0.01}$ & $0.63^{+0.02}_{-0.02}$ & \cite{2017ApJ...836..167D} \\
			212008766 & 5,18 & 12.802 & $5044^{+43}_{-42}$ & $-0.16^{+0.03}_{-0.04}$ & $4.64^{+0.01}_{-0.01}$ & $0.70^{+0.01}_{-0.01}$ & $0.79^{+0.01}_{-0.01}$ & \cite{2018AJ....155...21P} \\
			212012119 & 5,18 & 11.753 & $4841^{+39}_{-37}$ & $-0.06^{+0.04}_{-0.03}$ & $4.65^{+0.00}_{-0.00}$ & $0.69^{+0.01}_{-0.01}$ & $0.77^{+0.01}_{-0.01}$ & \cite{2018AJ....155...21P} \\
			212110888 & 5,16,18 & 11.441 & $6168^{+46}_{-76}$ & $0.01^{+0.03}_{-0.04}$ & $4.20^{+0.03}_{-0.03}$ & $1.42^{+0.06}_{-0.05}$ & $1.16^{+0.04}_{-0.04}$ & \cite{2016PASP..128l4402B} \\
			\hline
		\end{tabular}
	\end{center}
	\label{tab:stellar_parameters}
\end{table*}

Obtaining reliable stellar parameters is highly important since planetary parameters and statistical validation analysis depends on them. The Ecliptic Plane Input Catalog \citep[EPIC; ][]{2016ApJS..224....2H}, whose stellar parameters were based on photometry, proper motions, and models of the distribution of stars in the Milky Way, misclassifies between 56 and 72 per cent of subgiants as dwarfs, and 9 per cent of dwarfs as subgiants \citep{2016ApJS..224....2H}. Besides, it underestimates the radii for low-mass stars, as a result of the choice of the isochrones from the Padova data base \citep{2008A&A...482..883M}, which tend to underpredict the radii of these stars \citep{2012ApJ...757..112B,2016ApJS..224....2H}. For M dwarf stars, this bias has been empirically estimated to be 39 per cent by \citet{2017ApJ...836..167D} and 43 per cent by \citet{2020MNRAS.499.5416C}. Given that $\sim$40 per cent of selected \textit{K2} targets are low-mass M and K dwarfs \citep{2016ApJS..224....2H}, improving the stellar radii estimates of these targets is crucial to accurately characterize the planets observed by \textit{K2}. In this section, we detail our procedure to infer the stellar radii ($R_{\rm \star}$), masses ($M_{\rm \star}$), effective temperatures ($T_{\rm eff}$), surface gravities (log $g$), and metalicities ([Fe/H]) for the stars in our sample.

Among the \totaltargets targets, 21 have published spectroscopic parameters derived from different spectrographs and pipelines, and 16 lack spectra. Due to this heterogeneity, we performed an independent and uniform stellar characterization utilizing the \textsc{isochrones} package \citep{2015ascl.soft03010M}. The package is an interpolation tool that fits photometric and/or spectroscopic parameters to the MIST (MESA Isochrones Stellar Tracks) stellar models \citep{2015ApJS..220...15P,2016ApJ...823..102C,2016ApJS..222....8D} by using \textsc{multinest} \citep{2008MNRAS.384..449F,2009MNRAS.398.1601F,2019OJAp....2E..10F}, and thus it is able to predict the value of any physical property derived by the models.
We ran \textsc{isochrones} by using the following data for all our \totaltargets target stars: 2MASS \textit{JHK} photometry \citep{2006AJ....131.1163S} and \textit{Gaia} DR2 parallaxes \citep{2018A&A...616A...1G}, accounting for the systematics reported in \citet{2018ApJ...862...61S} and \citet{2018A&A...616A...9L}. Including \textit{Gaia} parallaxes within the \textsc{isochrones} analysis has proven to be of crucial importance in order to remove most of the potential for misclassifying dwarfs and subgiants, as well as to correct the aforementioned radii underestimation, specially for stars lacking spectroscopy \citep{2018AJ....156..277L}. We included additional priors of inferred spectroscopic parameters $T_{\rm eff}$, [Fe/H], and log $g$ for the 21 targets with published spectra, and for three more targets (EPICs 211335816, 211407755, 211816003) without spectra but with $T_{\rm eff}$, [Fe/H], and log $g$ available from \citet{2020ApJS..247...28H}; the authors derived stellar parameters for 195\,250 \textit{K2} targets by using random forest regression on photometric colours, trained on a sample of 26\,838 \textit{K2} stars with spectroscopic measurements from the Large Sky Area Multi-Object Fibre Spectroscopic Telescope \citep[LAMOST;][]{2012RAA....12.1197C} DR5.  The resulting \textsc{isochrones}-derived stellar parameters, together with the references of the spectroscopic priors are shown in Table \ref{tab:stellar_parameters}. 
\vspace{-5pt}

\begin{figure}
\centering
    \includegraphics[scale = 0.49]{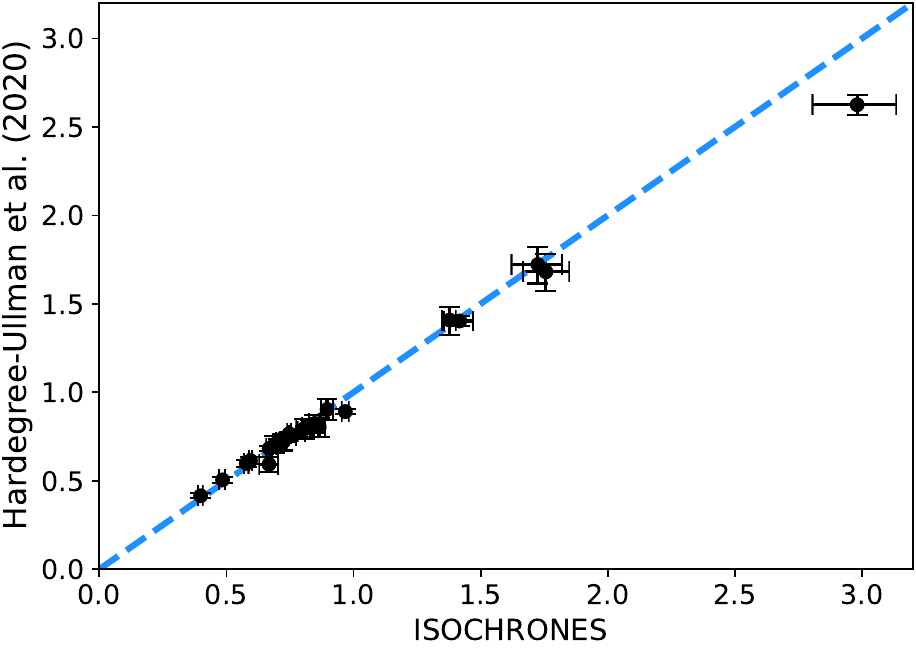}
    \caption{Comparison between \textsc{isochrones}-derived and \citet{2020ApJS..247...28H} reported stellar radii ($R_{\star}$). The dashed line represents the 1:1 relation.}
    \label{fig: R_Isoc_vs_HU}
\end{figure}

In order to check the consistency of our results and search for possible outliers, we compared the \textsc{isochrones}-derived stellar parameters with those obtained by \cite{2020ApJS..247...28H} for the 23 common targets in both samples.  In Fig. \ref{fig: R_Isoc_vs_HU}, we plot the comparison between stellar radii, which are typically consistent at the 1-$\sigma$ level, showing a strong agreement between both independent analysis. There is however an outlier (EPIC 211418290, the largest star of our sample), for which we derive $R_{\star}$ = $3.02^{+0.17}_{-0.16}$ $\rm R_{\odot}$ by using \citet{2018AJ....155..136M} spectroscopic values as priors, while \citet{2020ApJS..247...28H} derive $R_{\star}$ = $2.625^{+0.058}_{-0.055}$ $\rm R_{\odot}$, which are consistent at the 2-$\sigma$ level. 

\begin{landscape} \renewcommand{\arraystretch}{1.15}
\begin{table}
\caption{Planetary parameters and dispositions (CP = confirmed planet, VP = validated planet, PC = planet candidate) for the planets and candidates first detected in this work (Detection: New) and for the previously published planets and candidates (Detection: Known). }
	\begin{center}
		\begin{tabular}{c|c|c|c|c|c|c|c|c|c|c|c}
		    \hline
			ID & Name & $T_{0}$ (BKJD) & $P$ (d) & $R_{p}/R_{\star}$ (\%) & $a/R_{\star}$ & $b$ & $T_{eq}$ [A = 0] (K) & $a$ (AU) & $R_{\rm p}$ ($\rm R_{\oplus}$) & Detection & Disp \\
			\hline
			211309648.01 &   & $3421.4841636^{+0.0002984}_{-0.0002959}$ & $3.1393226^{+0.0000325}_{-0.0000316}$ & $9.42^{+0.14}_{-0.11}$ & $6.25^{+0.26}_{-0.33}$ & $0.31^{+0.13}_{-0.20}$ & $1780^{+80}_{-74}$ & \textbf{$0.0434^{+0.0047}_{-0.0047}$} & $15.43^{+1.56}_{-1.56}$ & New & PC \\
			211317649.01 & HAT-P-43 b & $3420.5965728^{+0.0001280}_{-0.0001230}$ & $3.3326414^{+0.0000151}_{-0.0000153}$ & $11.58^{+0.05}_{-0.05}$ & $8.86^{+0.05}_{-0.07}$ & $0.07^{+0.07}_{-0.05}$ & $1353^{+16}_{-16}$ & $0.0444^{+0.0021}_{-0.0021}$ & $13.63^{+0.64}_{-0.63}$ & Known & CP \\
			211319617.01 & K2-180 b & $2310.3959512^{+0.0012311}_{-0.0011967}$ & $8.8656511^{+0.0000151}_{-0.0000152}$ & $3.20^{+0.31}_{-0.11}$ & $21.93^{+2.36}_{-6.55}$ & $0.46^{+0.33}_{-0.31}$ & $802^{+55}_{-41}$ & $0.0734^{+0.0080}_{-0.0219}$ & $2.52^{+0.24}_{-0.10}$ & Known & VP \\
			211319779.01 &   & $3422.1814025^{+0.0039344}_{-0.0038465}$ & $13.8253157^{+0.0021214}_{-0.0020261}$ & -- & -- & -- & -- & -- & -- & New & PC \\
			211335816.01 &   & $2312.0099067^{+0.0001884}_{-0.0001853}$ & $4.9898611^{+0.0000014}_{-0.0000014}$ & $8.21^{+0.39}_{-0.47}$ & $11.30^{+0.22}_{-0.19}$ & $0.99^{+0.00}_{-0.01}$ & $1308^{+31}_{-31}$ & $0.0911^{+0.0055}_{-0.0055}$ & $15.44^{+1.22}_{-1.23}$ & Known & PC \\
			211359660.01 & K2-182 b & $2312.9420111^{+0.0002707}_{-0.0002661}$ & $4.7369729^{+0.0000020}_{-0.0000020}$ & $3.08^{+0.11}_{-0.04}$ & $13.98^{+0.55}_{-1.81}$ & $0.28^{+0.27}_{-0.20}$ & $982^{+9}_{-22}$ & $0.0519^{+0.0022}_{-0.0067}$ & $2.70^{+0.09}_{-0.05}$ & Known & VP \\
			211393988.01 &   & $3421.7347162^{+0.0006113}_{-0.0006035}$ & $4.8718955^{+0.0001146}_{-0.0001140}$ & $7.35^{+0.15}_{-0.09}$ & $8.94^{+0.43}_{-0.86}$ & $0.33^{+0.20}_{-0.21}$ & $1663^{+129}_{-119}$ & $0.0740^{+0.0133}_{-0.0130}$ & $14.64^{+2.33}_{-2.39}$ & New & PC \\
			211407755.01 &   & $2327.2333494^{+0.0017168}_{-0.0017028}$ & $36.0861607^{+0.0000913}_{-0.0000926}$ & -- & -- & -- & -- & -- & -- & New & PC \\
			211418290.01 &   & $2308.8080774^{+0.0002891}_{-0.0002927}$ & $5.0321716^{+0.0000020}_{-0.0000021}$ & $9.25^{+0.04}_{-0.03}$ & $4.88^{+0.03}_{-0.06}$ & $0.10^{+0.10}_{-0.07}$ & $1615^{+20}_{-19}$ & $0.0684^{+0.0039}_{-0.0039}$ & $30.48^{+1.71}_{-1.73}$ & Known & PC \\
			211418729.01 & K2-114 b & $2318.7149031^{+0.0001778}_{-0.0001781}$ & $11.3909344^{+0.0000039}_{-0.0000039}$ & $11.44^{+0.18}_{-0.12}$ & $24.38^{+0.69}_{-1.04}$ & $0.26^{+0.13}_{-0.16}$ & $722^{+7}_{-13}$ & $0.0983^{+0.0037}_{-0.0044}$ & $10.88^{+0.29}_{-0.28}$ & Known & CP \\
			211424769.01 &   & $2311.4983392^{+0.0002221}_{-0.0002172}$ & $5.1762331^{+0.0000015}_{-0.0000015}$ & $10.23^{+1.23}_{-1.27}$ & $12.94^{+0.55}_{-0.36}$ & $0.97^{+0.02}_{-0.02}$ & $1222^{+20}_{-26}$ & $0.0832^{+0.0039}_{-0.0030}$ & $15.39^{+1.88}_{-1.92}$ & Known & PC \\
			211480861.01 &   & $3421.5574015^{+0.0004884}_{-0.0004842}$ & $6.3508198^{+0.0001244}_{-0.0001214}$ & -- & -- & -- & -- & -- & -- & New & PC \\
			211525389.01 & K2-105 b & $2314.9895717^{+0.0004439}_{-0.0004516}$ & $8.2669928^{+0.0000070}_{-0.0000067}$ & $3.39^{+0.10}_{-0.05}$ & $18.48^{+0.85}_{-2.57}$ & $0.31^{+0.26}_{-0.21}$ & $928^{+1}_{-23}$ & $0.0833^{+0.0039}_{-0.0116}$ & $3.59^{+0.11}_{-0.07}$ & Known & CP \\
			211525753.01 & K2-355 b & $3419.1918298^{+0.0049327}_{-0.0055620}$ & $5.7385647^{+0.0013939}_{-0.0011048}$ & $2.23^{+0.19}_{-0.12}$ & $14.85^{+2.05}_{-4.15}$ & $0.43^{+0.35}_{-0.30}$ & $1066^{+190}_{-77}$ & $0.0632^{+0.0096}_{-0.0176}$ & $2.25^{+0.22}_{-0.16}$ & New & VP \\
			211537087.01 & K2-356 b & $3423.0408399^{+0.0045164}_{-0.0045890}$ & $21.0267366^{+0.0036912}_{-0.0038993}$ & $2.44^{+0.13}_{-0.10}$ & $40.15^{+4.07}_{-7.66}$ & $0.36^{+0.30}_{-0.25}$ & $624^{+0}_{-37}$ & $0.1599^{+0.0178}_{-0.0306}$ & $2.29^{+0.15}_{-0.12}$ & New & VP \\
			211537087.02 &   & $3420.0192397^{+0.0039220}_{-0.0038371}$ & $42.3822512^{+0.0059266}_{-0.0055157}$ & $2.74^{+0.18}_{-0.10}$ & $67.60^{+6.47}_{-15.11}$ & $0.40^{+0.31}_{-0.27}$ & $482^{+3}_{-29}$ & $0.2688^{+0.0288}_{-0.0596}$ & $2.58^{+0.19}_{-0.14}$ & New & PC \\
			211537087.03 &   & $3426.9974440^{+0.0055336}_{-0.0063669}$ & > 41.9 & $2.43^{+0.46}_{-0.26}$ & > 57.0 & $0.70^{+0.23}_{-0.44}$ & < 82 & > 0.228 & $2.28^{+0.43}_{-0.26}$ & New & PC \\
			211590050.01 &   & $3442.4009415^{+0.0021095}_{-0.0017790}$ & > 26.0 & $6.11^{+0.11}_{-0.77}$ & > 25.6 & $0.27^{+0.21}_{-0.18}$ & < 222 & > 0.187 & $10.48^{+0.96}_{-0.95}$ & New & PC \\
			211594205.01 & K2-184 b & $2315.5253369^{+0.0009324}_{-0.0009632}$ & $16.9780102^{+0.0000237}_{-0.0000242}$ & $1.80^{+0.13}_{-0.05}$ & $48.36^{+4.39}_{-11.75}$ & $0.42^{+0.31}_{-0.29}$ & $533^{+10}_{-23}$ & $0.1685^{+0.0155}_{-0.0409}$ & $1.47^{+0.11}_{-0.05}$ & Known & VP \\
			211606790.01 &   & $2317.0889820^{+0.0004416}_{-0.0004583}$ & $37.2470875^{+0.0000218}_{-0.0000211}$ & $12.56^{+1.16}_{-1.16}$ & $33.57^{+0.92}_{-0.64}$ & $0.97^{+0.02}_{-0.02}$ & $665^{+0}_{-10}$ & $0.2755^{+0.0156}_{-0.0148}$ & $24.07^{+2.52}_{-2.44}$ & Known & PC \\
			211644764.01 &   & $3435.7199786^{+0.0019066}_{-0.0018808}$ & $23.5306712^{+0.0020516}_{-0.0020952}$ & $4.71^{+0.21}_{-0.12}$ & $65.70^{+6.01}_{-13.98}$ & $0.42^{+0.29}_{-0.29}$ & $591^{+10}_{-47}$ & $0.5304^{+0.0901}_{-0.1078}$ & $9.25^{+1.03}_{-1.02}$ & New & PC \\
			211705502.01 &   & $3421.3479060^{+0.0008176}_{-0.0008147}$ & $2.5844972^{+0.0000782}_{-0.0000779}$ & $6.85^{+0.55}_{-0.72}$ & $4.03^{+0.16}_{-0.14}$ & $0.99^{+0.01}_{-0.01}$ & $2277^{+118}_{-116}$ & $0.0260^{+0.0025}_{-0.0024}$ & $10.29^{+1.31}_{-1.32}$ & New & PC \\
			211724246.01 &   & $3423.1623021^{+0.0006878}_{-0.0006778}$ & $4.7464962^{+0.0001276}_{-0.0001258}$ & $5.06^{+0.11}_{-0.10}$ & $6.64^{+0.40}_{-0.31}$ & $0.91^{+0.01}_{-0.01}$ & $1775^{+96}_{-97}$ & $0.0651^{+0.0082}_{-0.0077}$ & $11.60^{+1.30}_{-1.28}$ & New & PC \\
			211730267.01 & K2-357 b & $3422.1162073^{+0.0017974}_{-0.0017414}$ & $16.3488637^{+0.0013811}_{-0.0014022}$ & $3.40^{+0.13}_{-0.08}$ & $31.62^{+1.98}_{-5.24}$ & $0.35^{+0.28}_{-0.24}$ & $735^{+8}_{-39}$ & $0.1450^{+0.0136}_{-0.0235}$ & $3.72^{+0.23}_{-0.21}$ & New & VP \\
			211733267.01 &   & $2311.9322437^{+0.0002781}_{-0.0002844}$ & $8.6580805^{+0.0000030}_{-0.0000030}$ & $12.22^{+1.84}_{-1.66}$ & $24.56^{+1.23}_{-0.78}$ & $0.96^{+0.03}_{-0.03}$ & $762^{+4}_{-19}$ & $0.1030^{+0.0055}_{-0.0041}$ & $11.99^{+1.82}_{-1.64}$ & Known & PC \\
			211791178.01 &   & $2313.9228292^{+0.0015524}_{-0.0015507}$ & $9.5607076^{+0.0000408}_{-0.0000362}$ & -- & -- & -- & -- & -- & -- & Known & PC \\
			211816003.01 & K2-272 b & $2311.8541321^{+0.0014315}_{-0.0015034}$ & $14.4536756^{+0.0000243}_{-0.0000240}$ & $3.43^{+0.44}_{-0.18}$ & $26.16^{+5.13}_{-9.27}$ & $0.58^{+0.28}_{-0.39}$ & $747^{+82}_{-65}$ & $0.1007^{+0.0199}_{-0.0354}$ & $3.13^{+0.38}_{-0.21}$ & Known & VP \\
			211818569.01 & K2-121 b & $2310.5605812^{+0.0000785}_{-0.0000763}$ & $5.1857539^{+0.0000006}_{-0.0000005}$ & $10.28^{+0.27}_{-0.19}$ & $20.19^{+1.04}_{-1.35}$ & $0.34^{+0.15}_{-0.22}$ & $738^{+7}_{-19}$ & $0.0629^{+0.0033}_{-0.0043}$ & $7.52^{+0.22}_{-0.18}$ & Known & VP \\
			211822797.01 & K2-103 b & $2311.4052567^{+0.0022025}_{-0.0022530}$ & $21.1701963^{+0.0000662}_{-0.0000642}$ & $3.03^{+0.19}_{-0.10}$ & $44.37^{+3.58}_{-9.35}$ & $0.38^{+0.30}_{-0.27}$ & $431^{+4}_{-17}$ & $0.1195^{+0.0100}_{-0.0251}$ & $1.92^{+0.12}_{-0.07}$ & Known & VP \\
			211904310.01 &   & $3432.6937079^{+0.0012236}_{-0.0012639}$ & $24.3998690^{+0.0019024}_{-0.0019239}$ & $10.29^{+1.32}_{-1.18}$ & $15.59^{+0.65}_{-0.47}$ & $0.97^{+0.02}_{-0.02}$ & $1144^{+56}_{-54}$ & $0.1201^{+0.0148}_{-0.0142}$ & $18.43^{+3.17}_{-2.76}$ & New & PC \\
			211913977.01 & K2-101 b & $2319.6849877^{+0.0011205}_{-0.0011677}$ & $14.6762429^{+0.0000201}_{-0.0000203}$ & $2.40^{+0.17}_{-0.07}$ & $31.82^{+2.67}_{-6.79}$ & $0.41^{+0.29}_{-0.28}$ & $618^{+9}_{-25}$ & $0.1124^{+0.0095}_{-0.0239}$ & $2.00^{+0.14}_{-0.07}$ & Known & VP \\
			211914998.01 & K2-358 b & $3426.0094216^{+0.0031628}_{-0.0029238}$ & $11.2510285^{+0.0018188}_{-0.0017892}$ & $2.69^{+0.12}_{-0.07}$ & $20.61^{+1.41}_{-3.89}$ & $0.35^{+0.30}_{-0.24}$ & $888^{+0}_{-47}$ & $0.0863^{+0.0079}_{-0.0159}$ & $2.68^{+0.17}_{-0.14}$ & New & VP \\
			211914998.02 &   & $3435.5995863^{+0.0051114}_{-0.0047980}$ & $24.6237203^{+0.0090182}_{-0.0079234}$ & $2.35^{+0.20}_{-0.12}$ & $47.18^{+6.26}_{-14.07}$ & $0.43^{+0.35}_{-0.30}$ & $585^{+12}_{-43}$ & $0.1984^{+0.0291}_{-0.0587}$ & $2.33^{+0.23}_{-0.16}$ & New & PC \\
			211916756.01 & K2-95 b & $2307.7422121^{+0.0007600}_{-0.0007697}$ & $10.1346454^{+0.0000121}_{-0.0000123}$ & $7.56^{+0.45}_{-0.22}$ & $28.28^{+3.06}_{-5.82}$ & $0.46^{+0.25}_{-0.31}$ & $474^{+8}_{-24}$ & $0.0525^{+0.0059}_{-0.0108}$ & $3.31^{+0.20}_{-0.14}$ & Known & VP \\
			211919004.01 & K2-273 b & $2316.0961619^{+0.0009530}_{-0.0009911}$ & $11.7195663^{+0.0000134}_{-0.0000135}$ & $3.27^{+0.25}_{-0.08}$ & $18.49^{+1.24}_{-3.62}$ & $0.37^{+0.30}_{-0.26}$ & $849^{+7}_{-29}$ & $0.0729^{+0.0053}_{-0.0141}$ & $3.05^{+0.23}_{-0.12}$ & Known & VP \\
			211969807.01 & K2-104 b & $2307.3799790^{+0.0009780}_{-0.0009972}$ & $1.9741954^{+0.0000024}_{-0.0000024}$ & $3.40^{+0.19}_{-0.09}$ & $10.89^{+0.95}_{-2.50}$ & $0.39^{+0.32}_{-0.27}$ & $798^{+9}_{-39}$ & $0.0242^{+0.0022}_{-0.0055}$ & $1.79^{+0.10}_{-0.06}$ & Known & VP \\
			212006344.01 & K2-122 b & $2308.8300469^{+0.0006420}_{-0.0006263}$ & $2.2193023^{+0.0000020}_{-0.0000020}$ & $1.80^{+0.15}_{-0.06}$ & $13.61^{+1.35}_{-3.45}$ & $0.43^{+0.32}_{-0.30}$ & $769^{+20}_{-37}$ & $0.0373^{+0.0038}_{-0.0095}$ & $1.16^{+0.10}_{-0.05}$ & Known & VP \\
			212008766.01 & K2-274 b & $2312.1112611^{+0.0013858}_{-0.0014376}$ & $14.1330314^{+0.0000282}_{-0.0000282}$ & $2.73^{+0.16}_{-0.07}$ & $29.68^{+1.73}_{-5.06}$ & $0.34^{+0.29}_{-0.23}$ & $655^{+0}_{-20}$ & $0.0964^{+0.0059}_{-0.0163}$ & $2.08^{+0.12}_{-0.07}$ & Known & VP \\
			212008766.02 &   & $3443.8646162^{+0.0033673}_{-0.0032444}$ & > 74.8 & $5.69^{+1.06}_{-8.70}$ & > 98.40 & $0.97^{+0.02}_{-0.02}$ & < 86 & > 0.320 & $4.34^{+0.81}_{-0.66}$ & New & PC \\
			212012119.01 & K2-275 b & $2309.1339592^{+0.0003369}_{-0.0003349}$ & $3.2809626^{+0.0000016}_{-0.0000016}$ & $3.11^{+0.32}_{-0.36}$ & $8.61^{+0.42}_{-2.54}$ & $0.77^{+0.13}_{-0.56}$ & $1166^{+0}_{-216}$ & $0.0277^{+0.0141}_{-0.0082}$ & $2.34^{+0.25}_{-0.27}$ & Known & VP \\
			212012119.02 & K2-275 c & $2309.4867161^{+0.0004575}_{-0.0005019}$ & $8.4388385^{+0.0000075}_{-0.0000067}$ & $2.93^{+0.11}_{-0.06}$ & $27.50^{+1.45}_{-3.26}$ & $0.33^{+0.23}_{-0.23}$ & $653^{+3}_{-17}$ & $0.0881^{+0.0049}_{-0.0104}$ & $2.21^{+0.08}_{-0.06}$ & Known & VP \\
			212110888.01 & K2-34 b & $2308.3514834^{+0.0000689}_{-0.0000699}$ & $2.9956348^{+0.0000002}_{-0.0000002}$ & $8.87^{+0.05}_{-0.05}$ & $6.68^{+0.10}_{-0.10}$ & $0.83^{+0.01}_{-0.01}$ & $1687^{+24}_{-25}$ & $0.0442^{+0.0020}_{-0.0020}$ & $13.75^{+0.58}_{-0.59}$ & Known & CP \\
			\hline
		\end{tabular}
	\end{center}
\label{tab:planet_params}
\end{table} \end{landscape} \noindent

\subsection{Transit modelling}
\label{transit_modeling}

\begin{figure*}
    \centering
    \includegraphics[width=4cm]{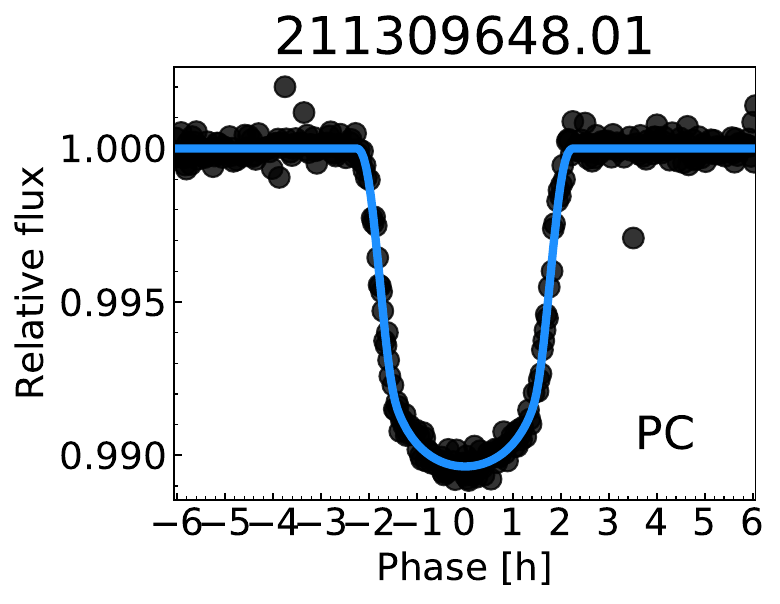}
    \includegraphics[width=4cm]{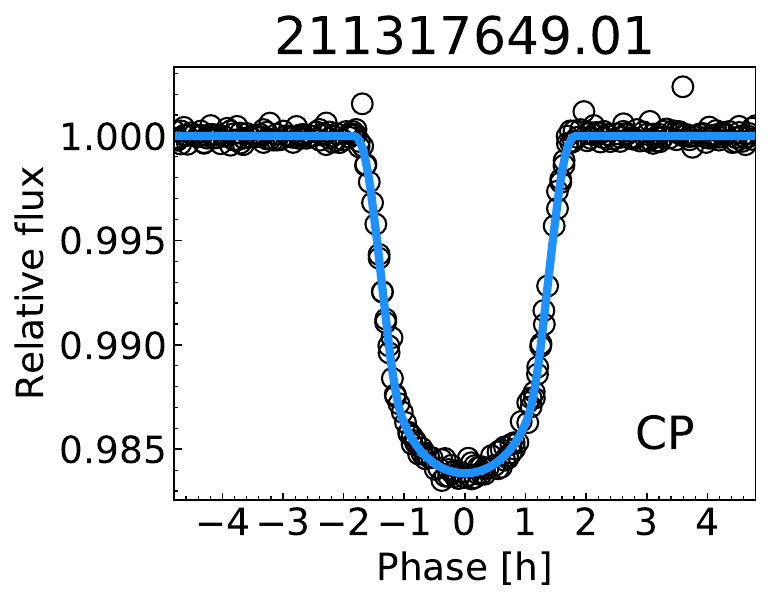}
    \includegraphics[width=4cm]{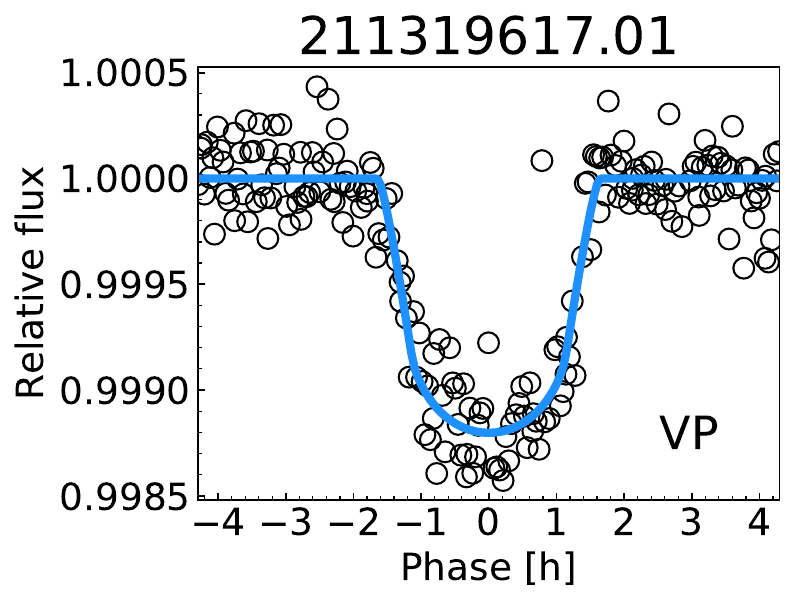}
    \includegraphics[width=4cm]{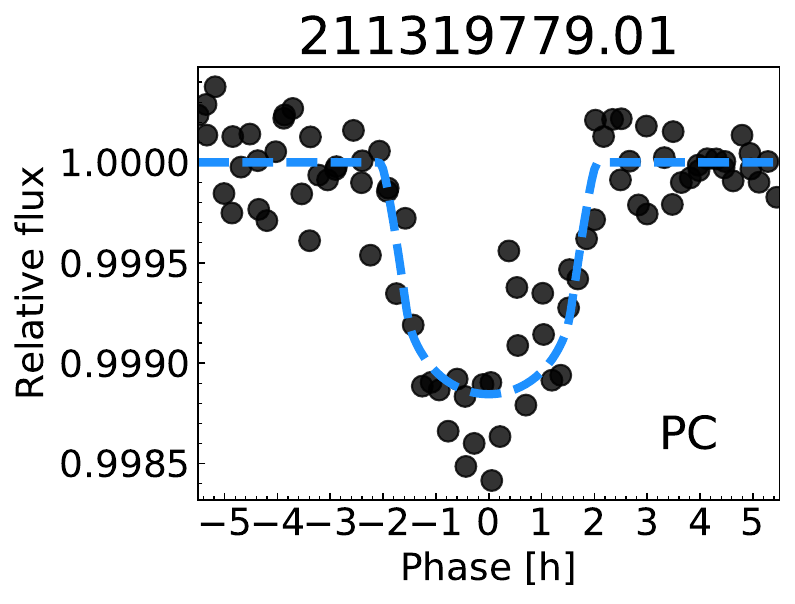}
    \includegraphics[width=4cm]{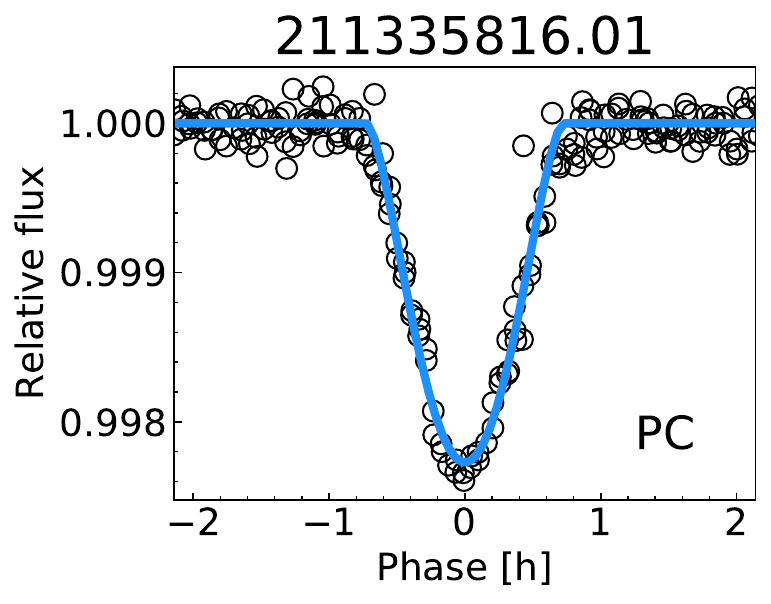}
    \includegraphics[width=4cm]{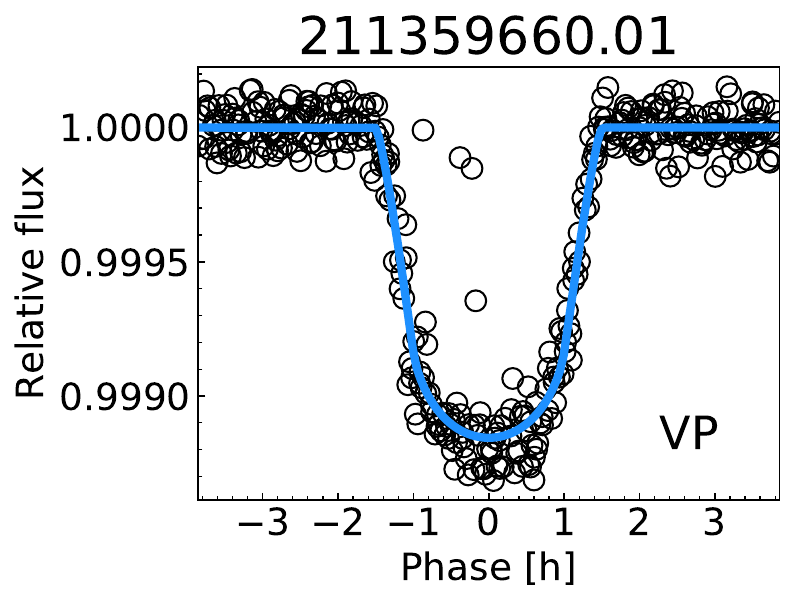}
    \includegraphics[width=4cm]{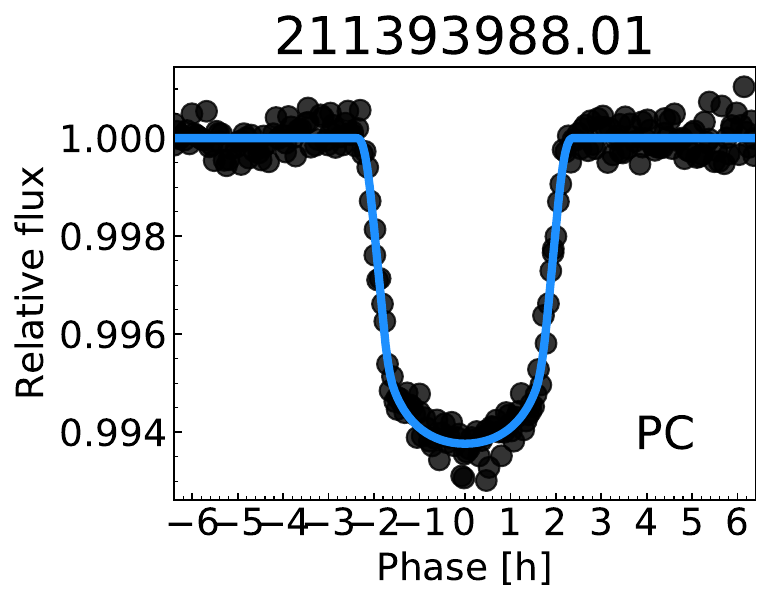}
    \includegraphics[width=4cm]{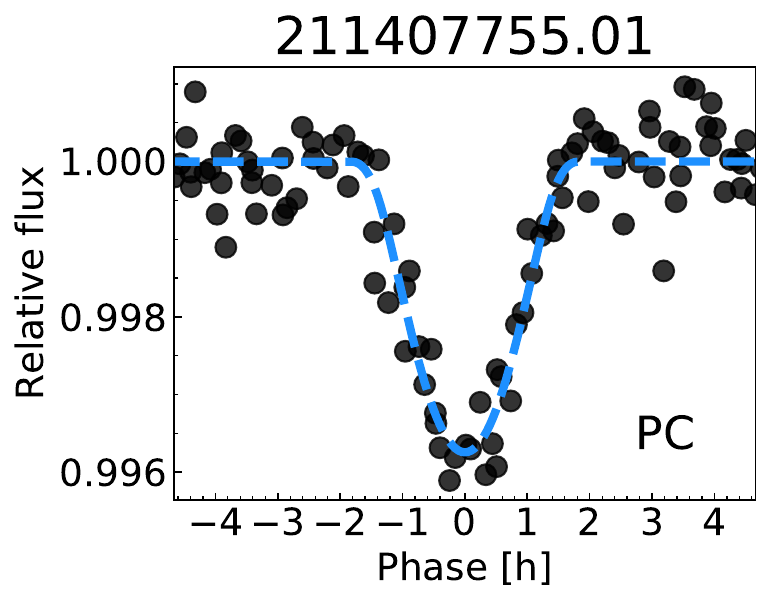}
    \includegraphics[width=4cm]{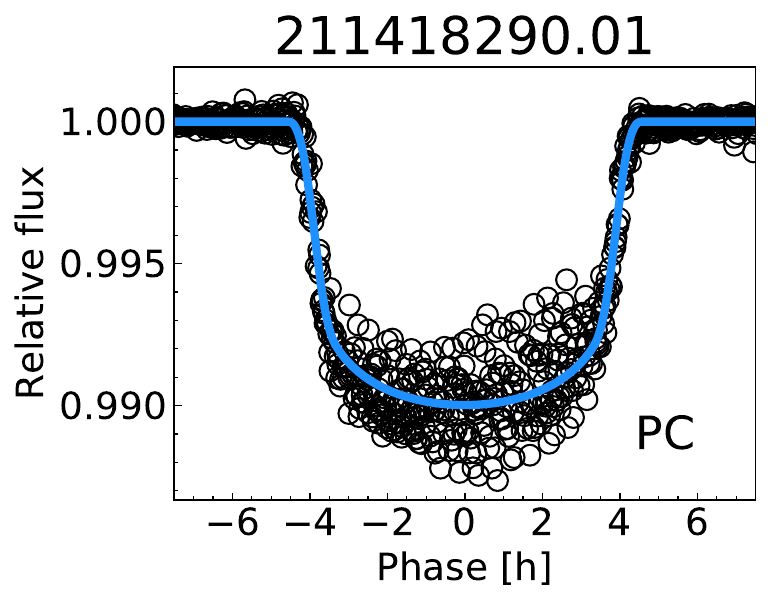}
    \includegraphics[width=4cm]{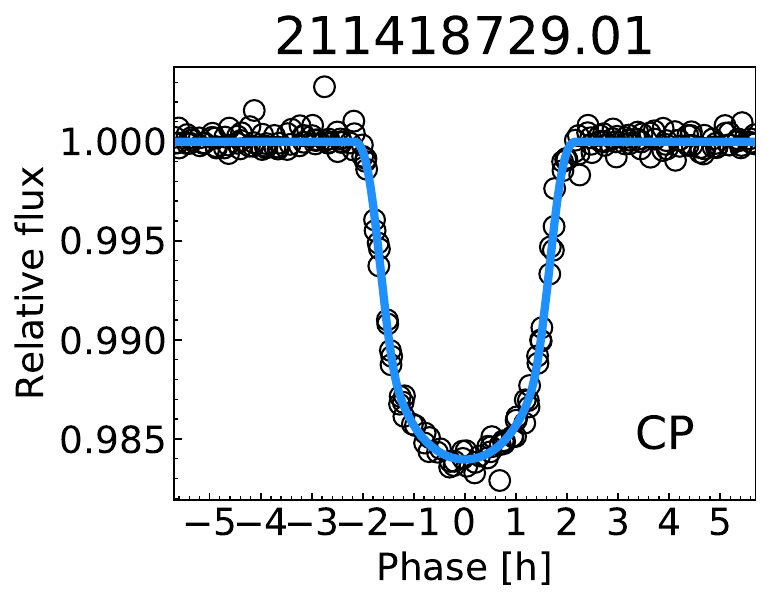}
    \includegraphics[width=4cm]{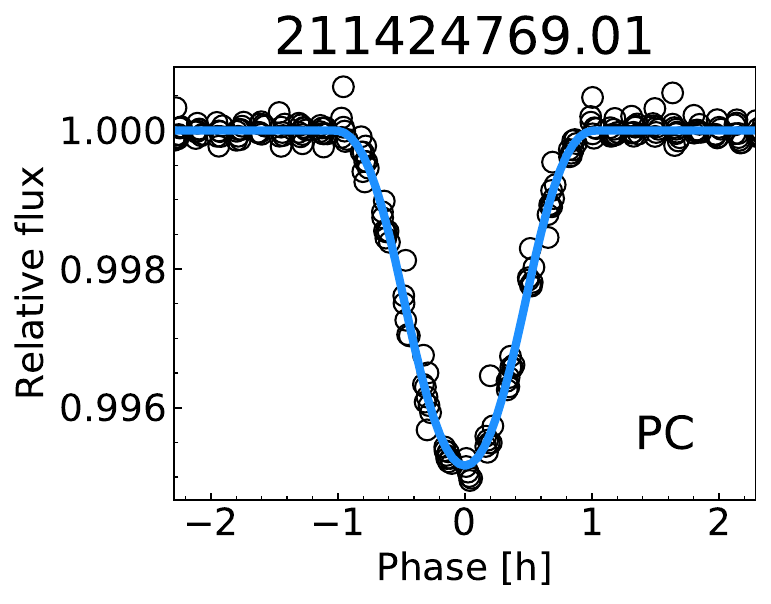}
    \includegraphics[width=4cm]{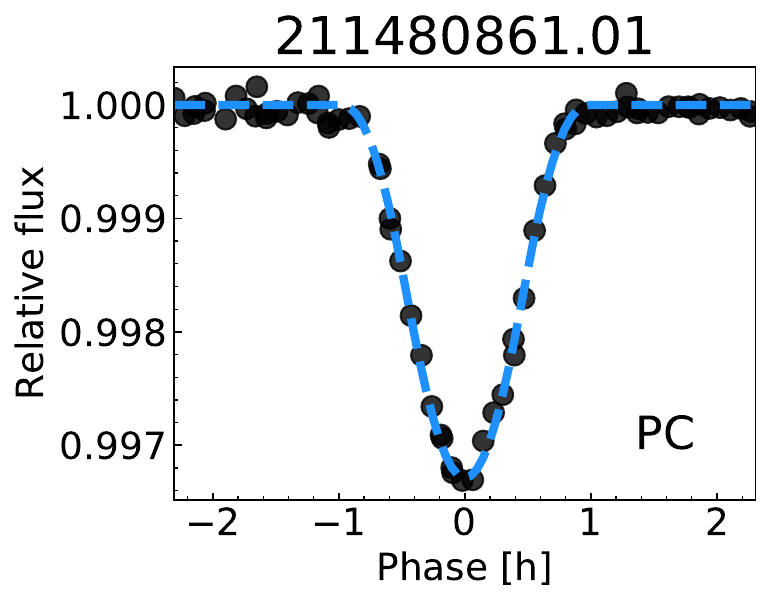}
    \includegraphics[width=4cm]{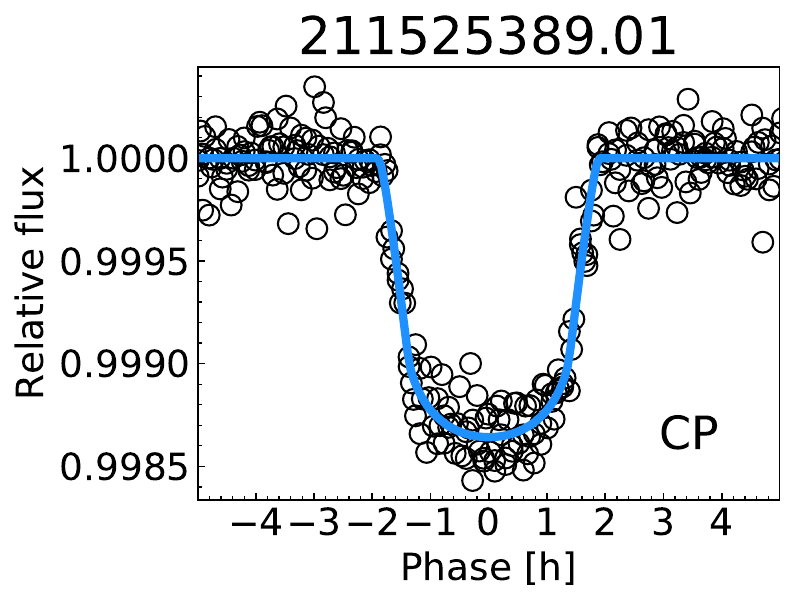}
    \includegraphics[width=4cm]{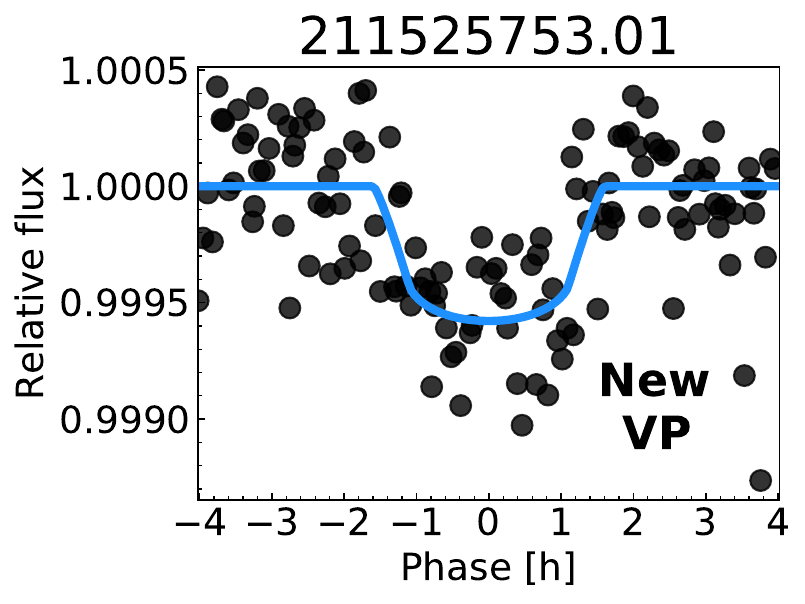}
    \includegraphics[width=4cm]{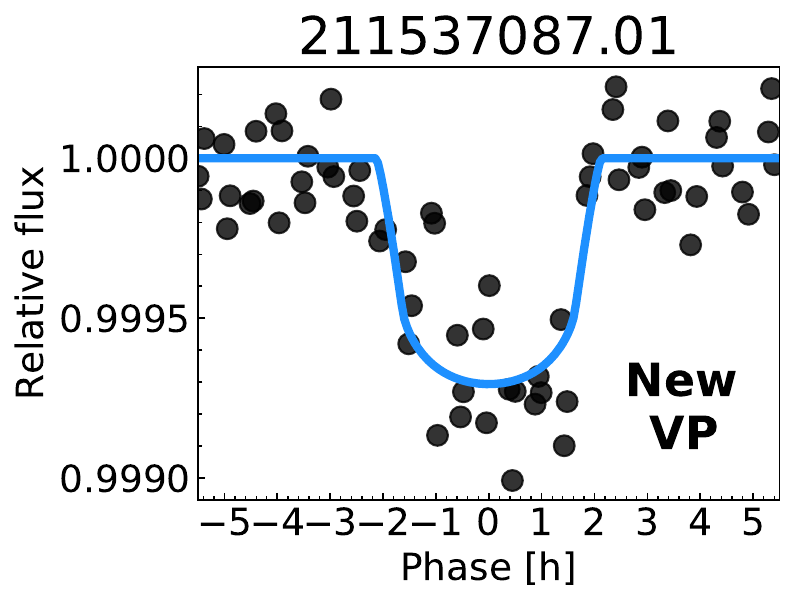}
    \includegraphics[width=4cm]{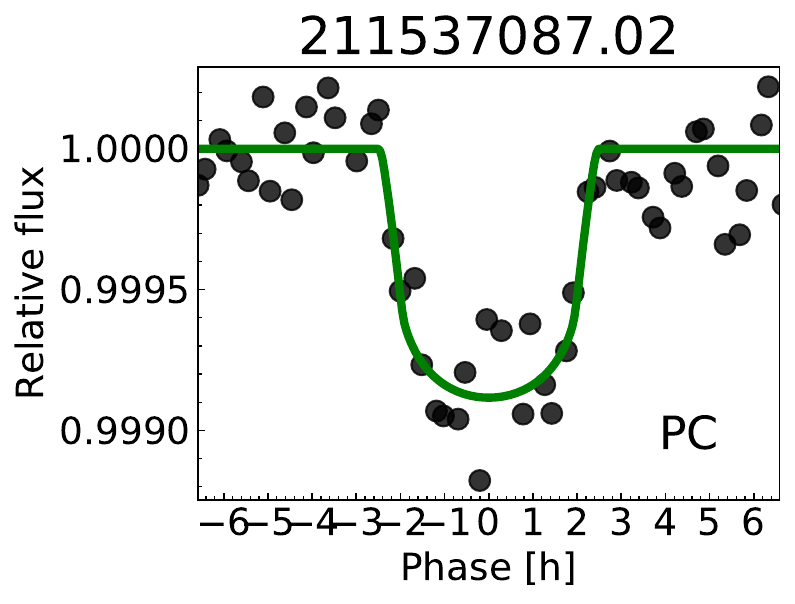}
    \includegraphics[width=4cm]{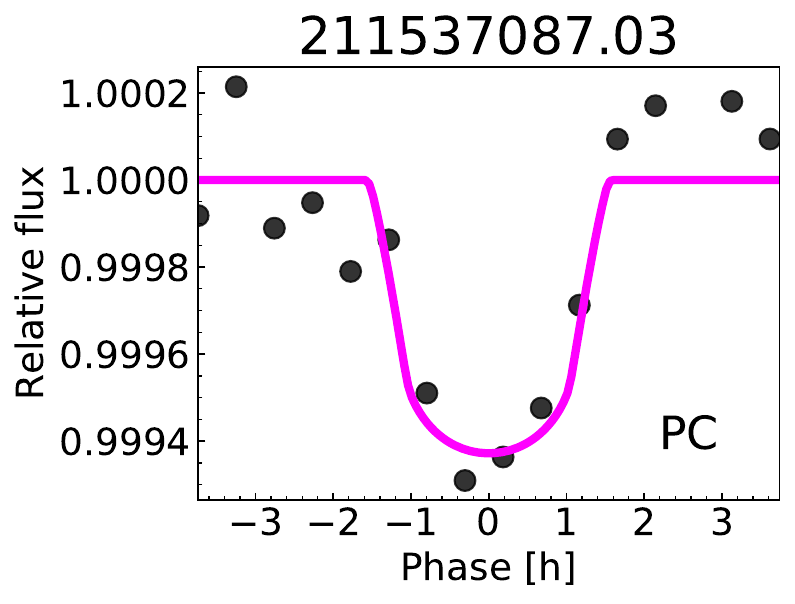}
    \includegraphics[width=4cm]{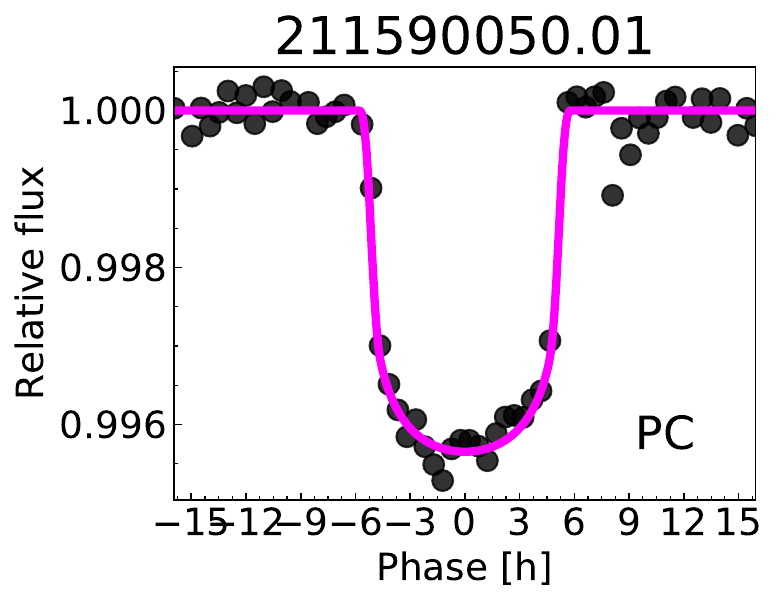}
    \includegraphics[width=4cm]{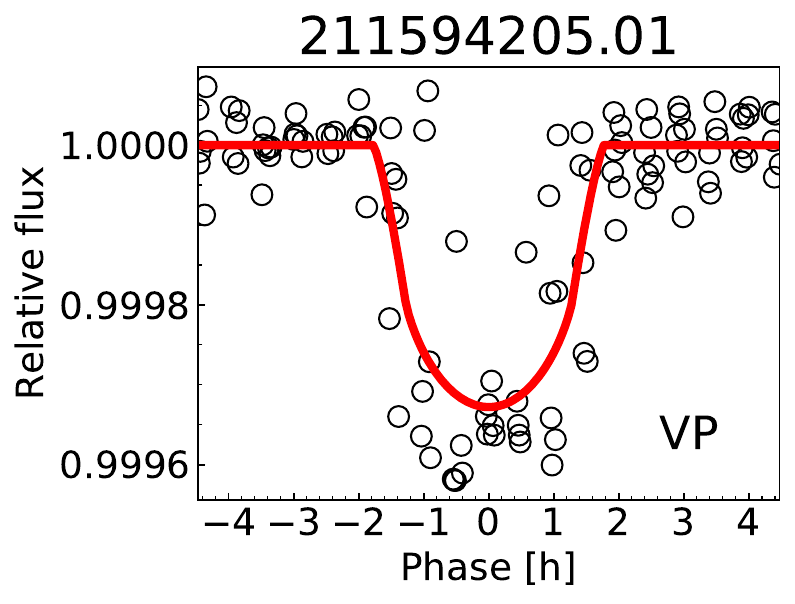}
    \includegraphics[width=4cm]{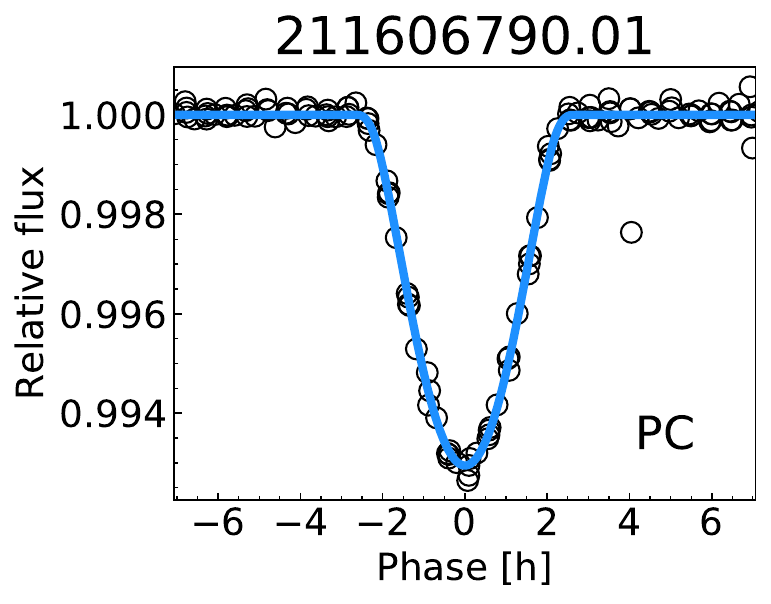}
    \includegraphics[width=4cm]{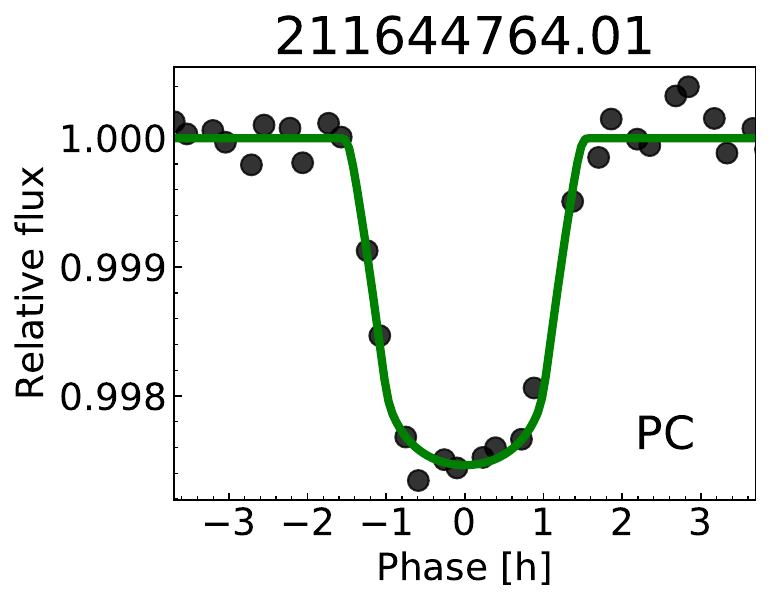}
    \includegraphics[width=4cm]{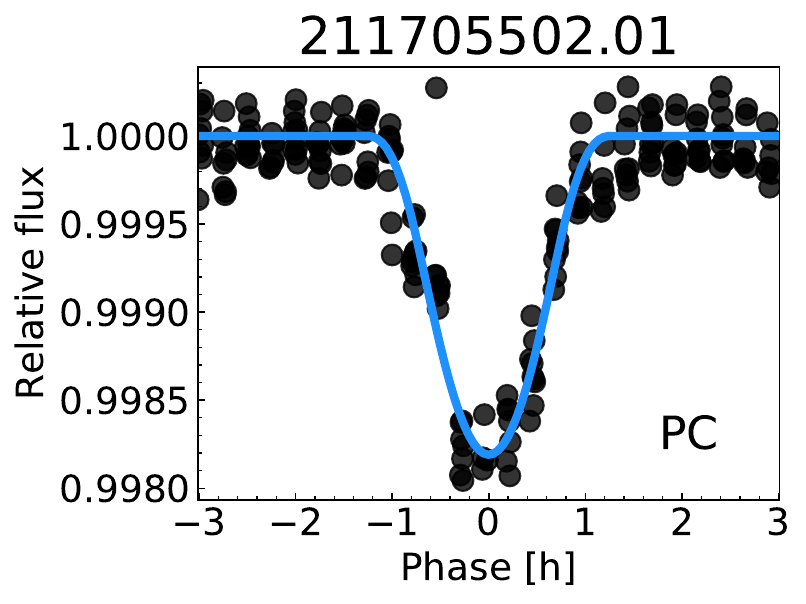}
    \includegraphics[width=4cm]{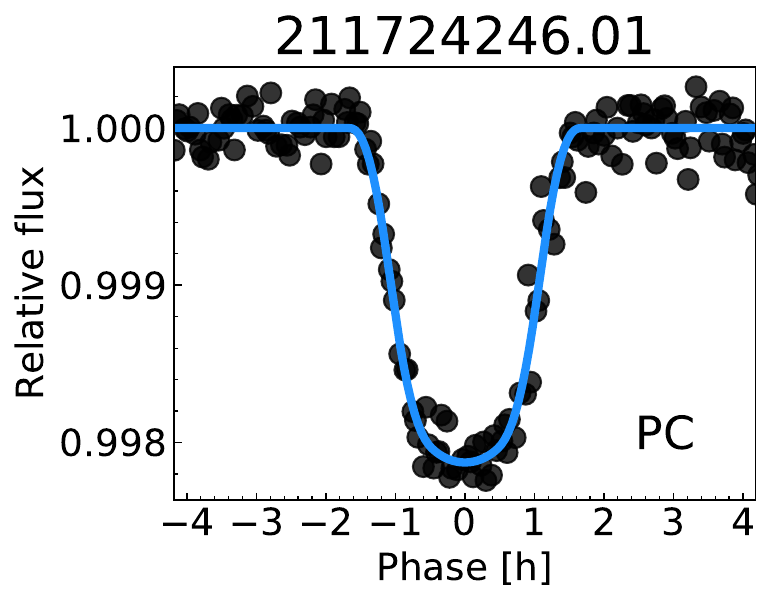}
    \includegraphics[width=4cm]{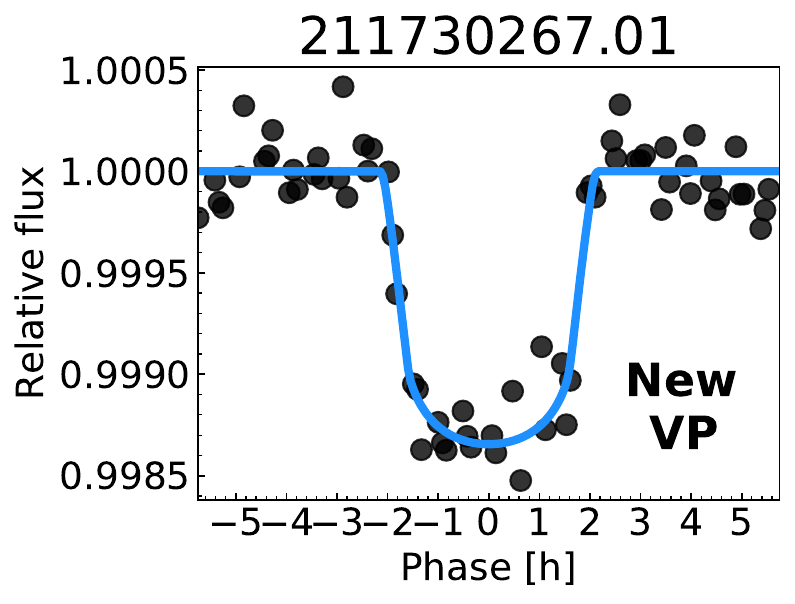}
    \includegraphics[width=4cm]{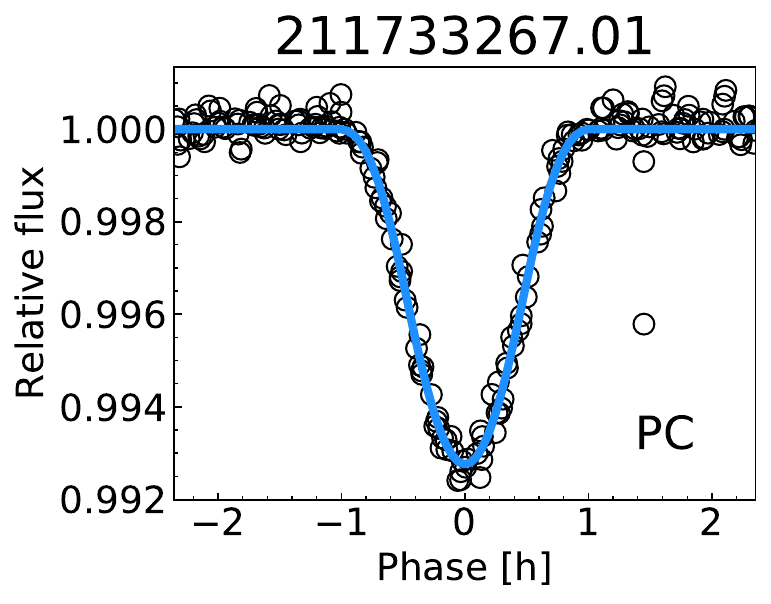}
    \includegraphics[width=4cm]{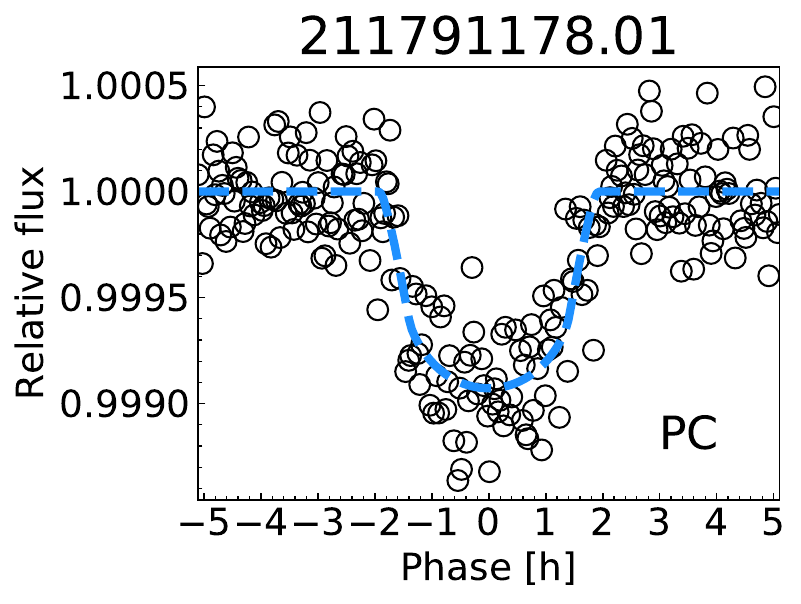}
    \includegraphics[width=4cm]{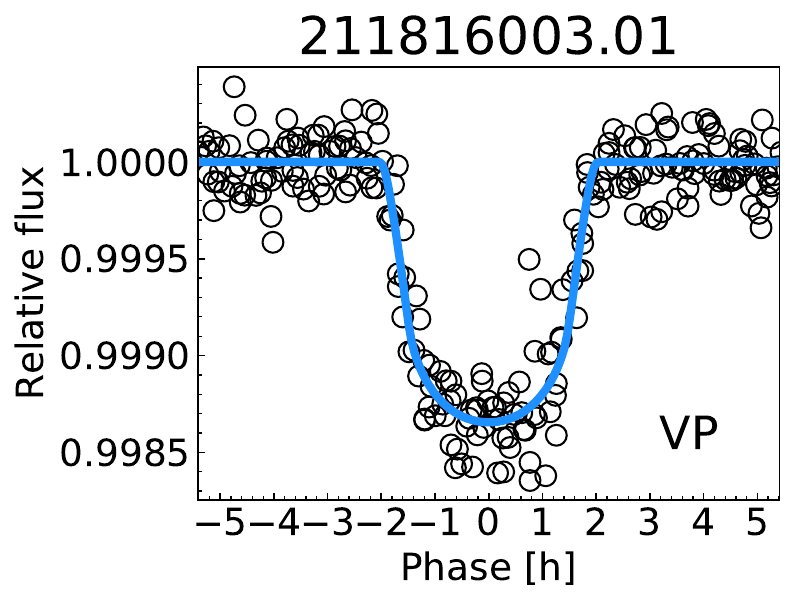}
    \includegraphics[width=4cm]{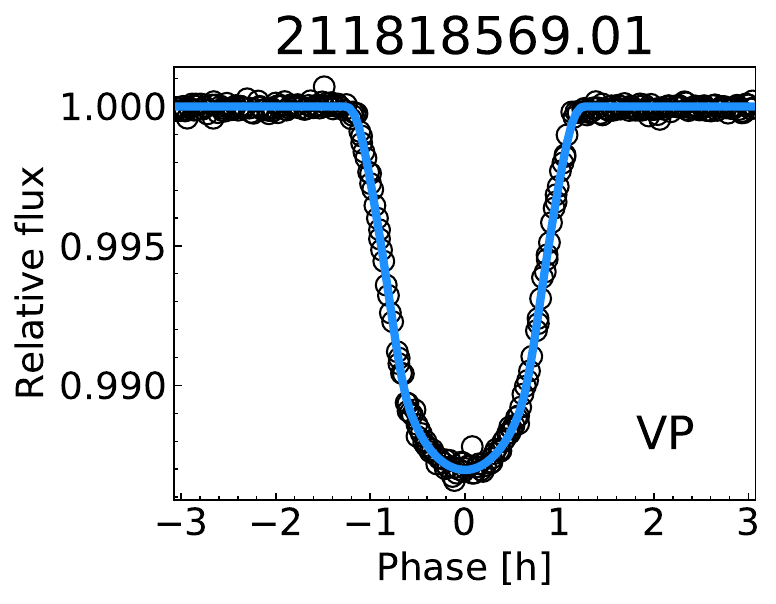}

    \caption{Phase-folded transits of confirmed planets (CP), validated planets (VP), and planet candidates (PC) analysed in this work. Photometric data are plotted with solid symbols for new detections and with open circles for already known planets and candidates. The best-fitting quadratic \citet{2002ApJ...580L.171M} transit models obtained by \textsc{pyaneti} are overplotted in blue for signals with three or more transits, in green for two-transit signals, and in magenta for single transits. The transit model for EPIC 211594205.01 is overplotted in red as it is not suitable to properly fit the data due to its strong TTVs. The transit models for EPICs 211319779.01, 211407755.01, 211480861.01, and 211791178.01 are plotted with dashed lines, indicating that the targets have a contaminant star causing an indeterminacy in the origin of the signal, so the derived planetary parameters are not reliable.}
    
    \label{fig:phase_folded_1}
\end{figure*}

\begin{figure*}
    \centering
    \includegraphics[width=4cm]{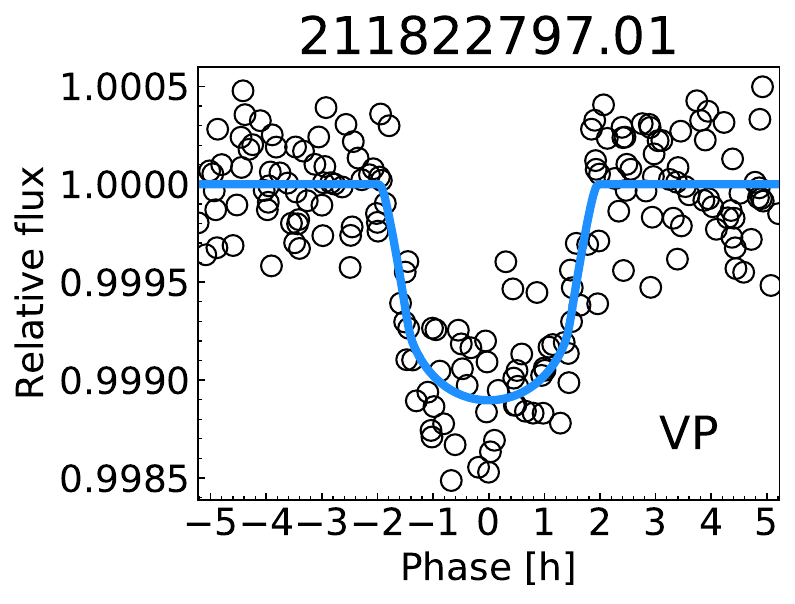}
    \includegraphics[width=4cm]{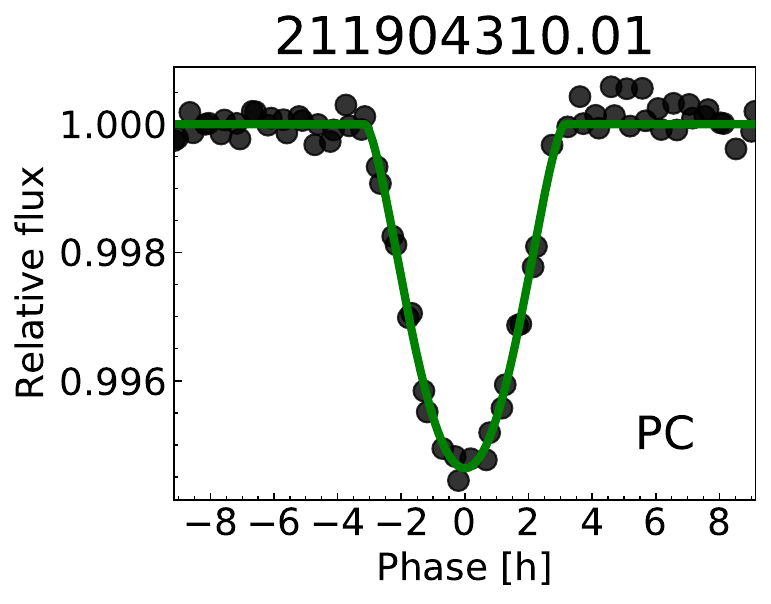}
    \includegraphics[width=4cm]{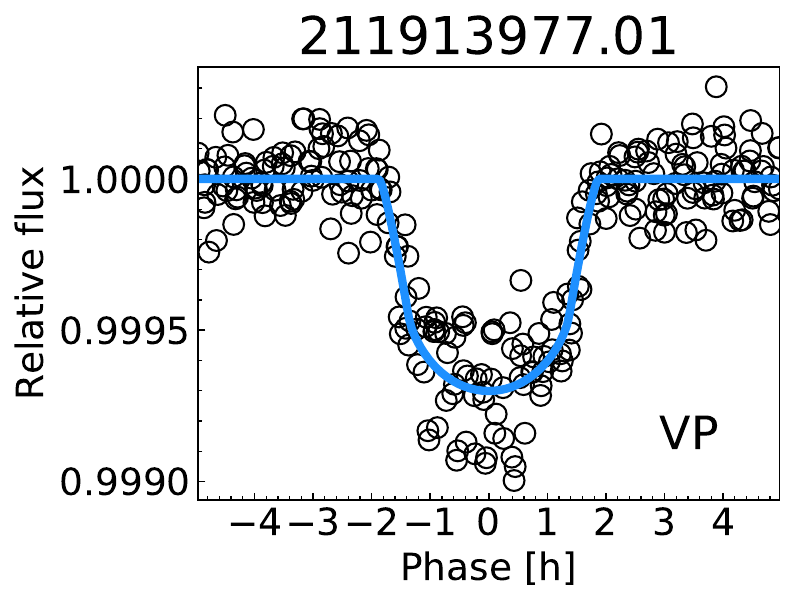}
    \includegraphics[width=4cm]{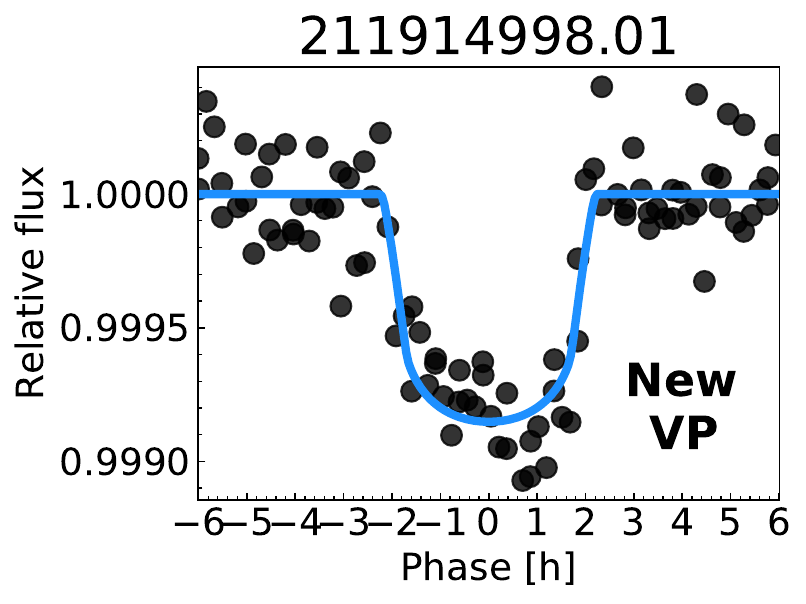}
    \includegraphics[width=4cm]{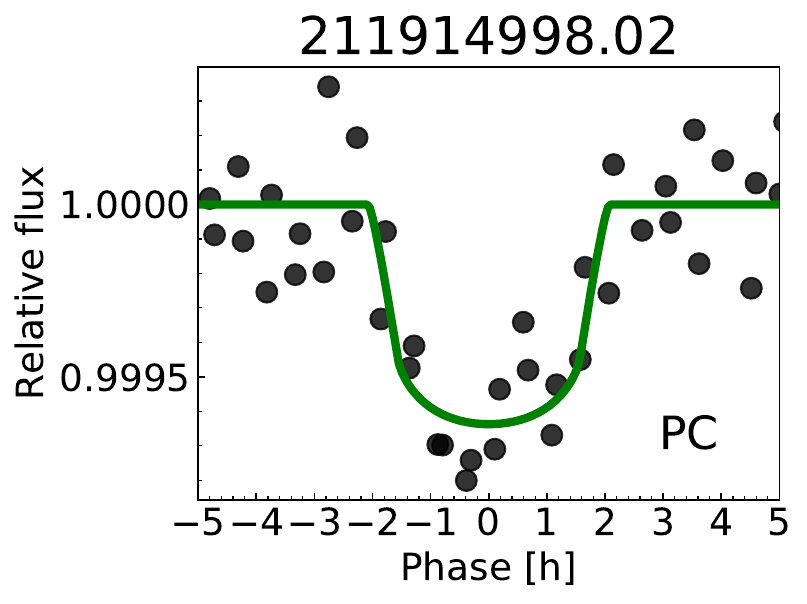}
    \includegraphics[width=4cm]{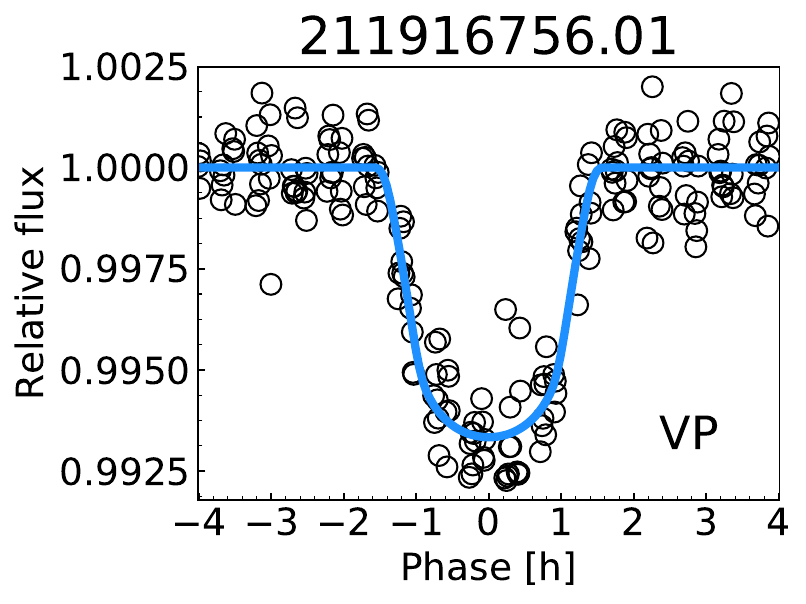}
    \includegraphics[width=4cm]{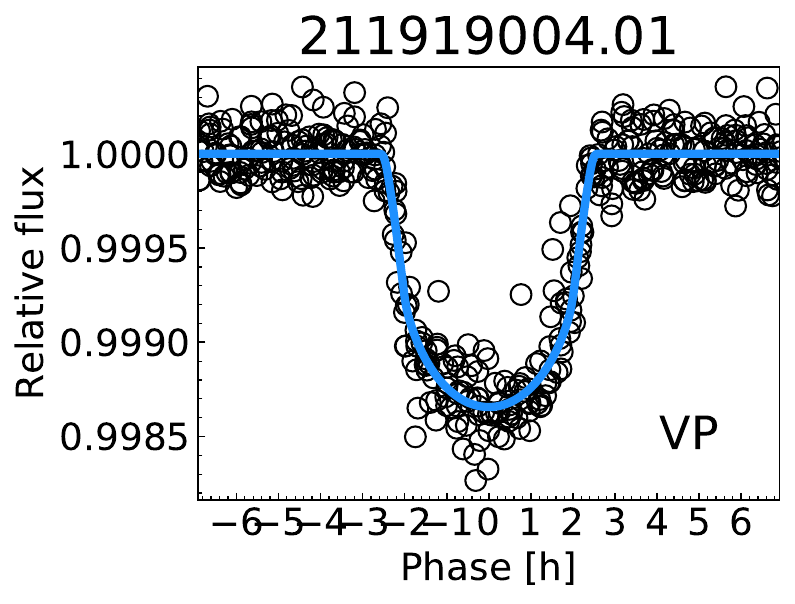}
    \includegraphics[width=4cm]{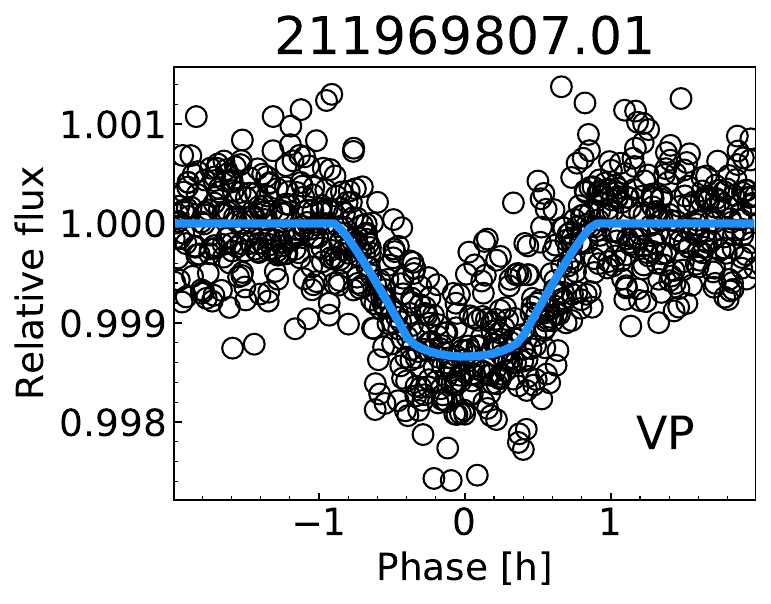}
    \includegraphics[width=4cm]{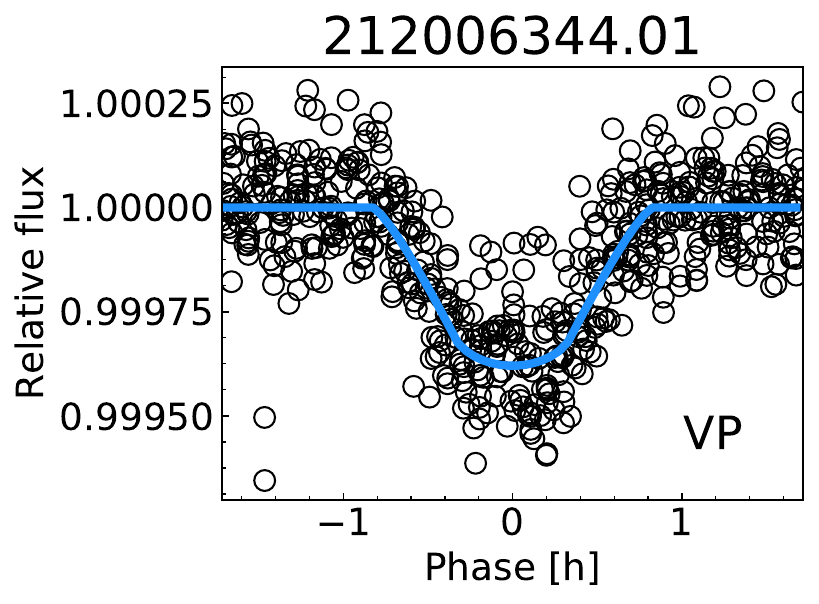}
    \includegraphics[width=4cm]{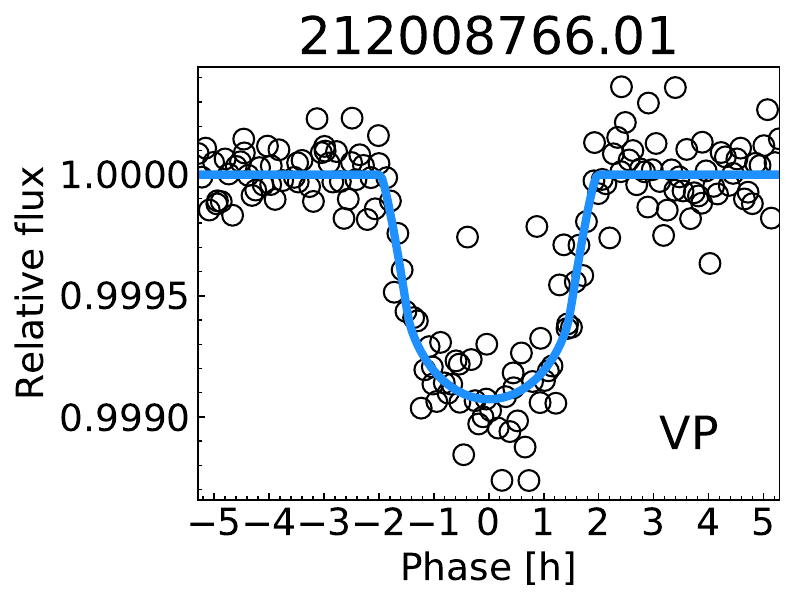}
    \includegraphics[width=4cm]{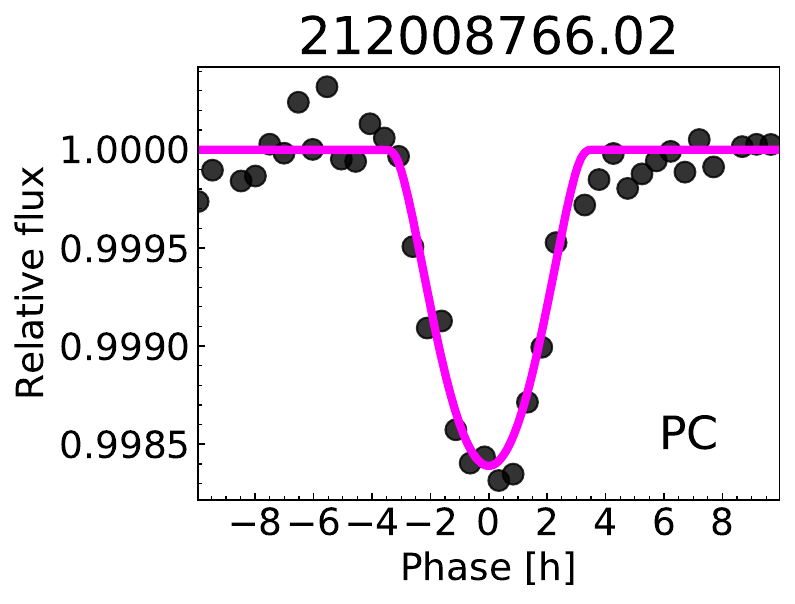}
    \includegraphics[width=4cm]{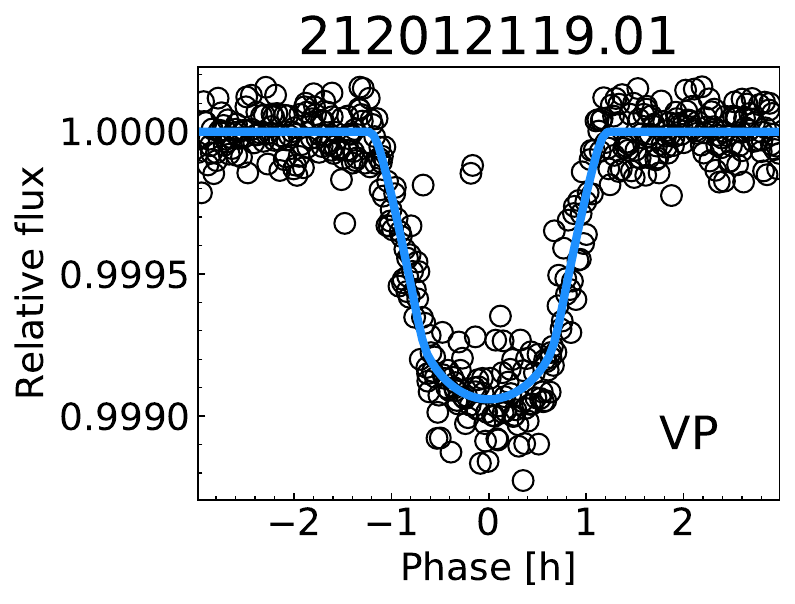}
    \includegraphics[width=4cm]{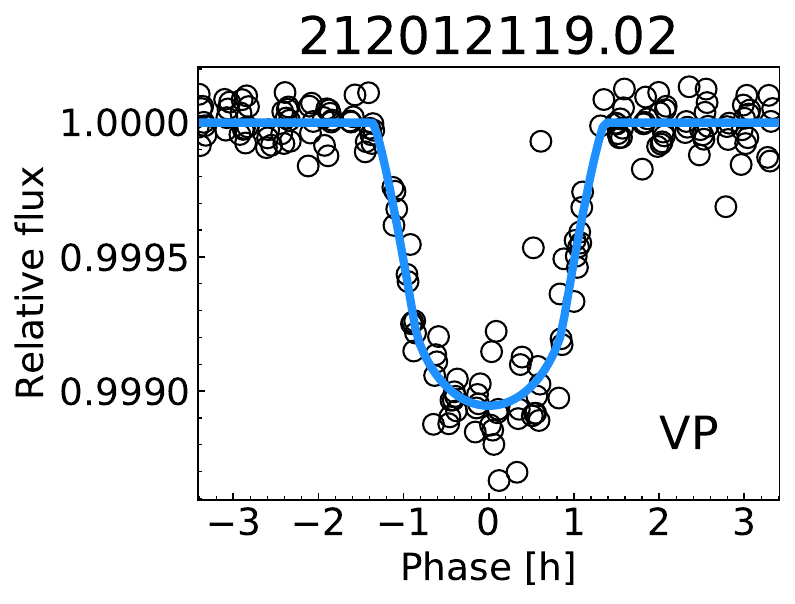}
    \includegraphics[width=4cm]{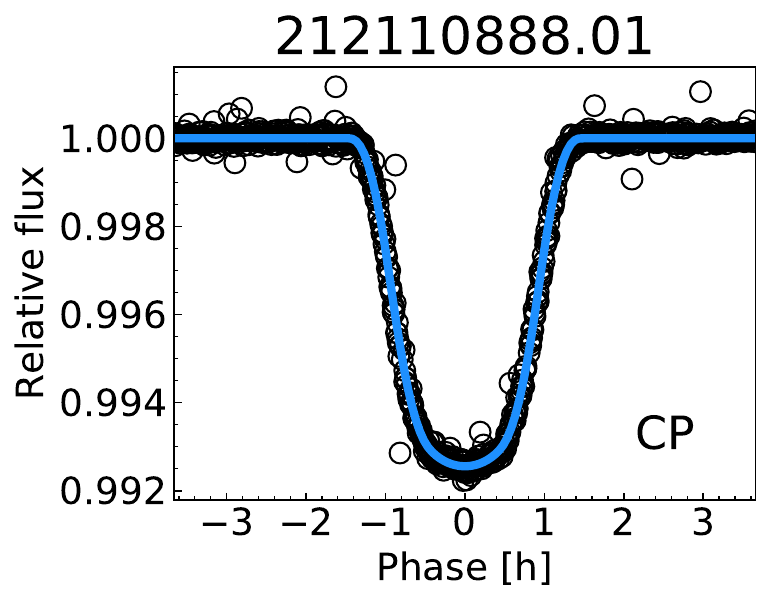}
    
    \caption{\textit{continued} Fig. \ref{fig:phase_folded_1}.}
    \label{fig:phase_folded_2}
\end{figure*}

To derive the planetary parameters, we used the pre-processed light curves as described in Section \ref{K2_photometry} and modelled the transits with the \citet{2002ApJ...580L.171M} quadratic limb darkened transit model. For that, we used  the \textsc{pyaneti} package \citep{2019MNRAS.482.1017B}, accounting for the \textit{K2} 30-min cadence by supersampling the transit model with 10 subsamples per cadence \citep{2010MNRAS.408.1758K}. We assumed circular orbits by fixing the eccentricity to 0. We set wide uniform priors  for the impact parameter ($b$), scaled planet radius ($R_{p} / R_{\star}$) and semimajor axis ($a/R_{\star}$). We set narrow uniform priors for the orbital period ($P$) and the mid-transit time ($T_{0}$) by using the values and uncertainties obtained from the TLS algorithm. We settled Gaussian priors on the quadratic limb darkening coefficients in the q-space \citep{2013MNRAS.435.2152K}, which we estimated from the \textsc{limb-darkening} package \citep{2015MNRAS.450.1879E} by adopting the ATLAS models for the stellar atmospheres \citep{1979ApJS...40....1K}. We also included a photometric jitter term in order to account for underestimated white noise. 


For the single-transit candidates we fit the same parameters as for the multitransit candidates, except for $P$ and $a/R_{\star}$, which cannot be determined. Instead, we obtained a lower bound for $P$ as the time between the transit and the farthest edge of the light curve, and we also estimated a lower bound for $a/R_{\star}$ making use of the bound for $P$ and the Kepler's third law.

Table \ref{tab:planet_params} shows the main derived parameters together with their uncertainties (i.e. the median and 68.3 per cent credible intervals of the posterior distribution) for the new \newVP planets and \newPC candidates presented in this work, as well as for the 24 already published planets and candidates. Figs \ref{fig:phase_folded_1} and \ref{fig:phase_folded_2} show the transit light curves folded to the orbital periods of each planet and candidate, together with the inferred median transit model. 


\subsection{Transit timing variations}
\label{TTVs}

We searched for TTVs produced by additional non-transiting planets in the light curves of our sample. For this, we took for each target the pre-processed and combined light curves as described in Section~\ref{K2_photometry}, and searched for TTVs using the Python Tool for Transit Variations \citep[\textsc{pyttv;}][]{Korth2020}.

The procedure is as follows: the transits from all the planets in a system are fitted together simultaneously by modelling them with the quadratic \citet{2002ApJ...580L.171M} transit model implemented in \textsc{pytransit} \citep{Parviainen2015} via a Taylor-series expansion \citep{2020MNRAS.499.3356P}, and fitted for all the transit centers $t_{c}$, impact parameter $b$, and planet-to-star radius ratio $R_{p}/R_{\star}$ for all planets, for the quadratic limb darkening coefficients $(u,v)$ and mean stellar density {\ensuremath{\rho_{\star}}}. The search for TTVs is carried out by fitting a linear, quadratic or sinusoidal model to the transit times, that is subtracted afterwards and evaluated through the GLS periodogram from \citet{2009A&A...496..577Z}, where best-fitting parameters and their uncertainties are calculated. The model with the lowest Bayesian Information Criterion (BIC) is chosen as the best model and the significance of the other models with respect to the best model is calculated via the $\Delta \mathrm{BIC}$. 

Significant TTVs were detected for EPIC 211594205 that hint to the existence of an additional non-transiting planet (See Section \ref{sec:EPIC_211594205} for a further discussion on the system). Besides, weak TTVs were detected for EPIC 211418290 and EPIC 211816003. The latter ones are most likely produced by stellar activity and spots, which is visible by the high scatter of the in-transit residuals compared to the out-of-transit residuals in their phase-folded transits (Fig.~\ref{fig:phase_folded_1}). 

As TTVs are most sensitive to planets near resonant orbits, we checked the period ratios of the planets in our sample with more than one planet in a system. K2-356 b (EPIC 211537087.01) and EPIC 211537087.02 have a period ratio close to 2, which hints of strong perturbations that can lead to significant TTVs for both of them. In Section \ref{epic_211537087}, we include a further discussion on this system and estimate the TTV periods and amplitudes for both the planet and the candidate.


\subsection{Statistical validation}
\label{validation}

We carried out a statistical validation analysis for the \newsignals new planet candidates found in this work. First, we computed the false positive probabilities (FPPs), which are the probabilities of the signals being astrophysical false positives (Section \ref{sec:FPP_calculation}). Secondly, we assessed the reliability of the FPP calculation in order to obtain the final disposition of each new planet candidate (Section \ref{sec:FPP_reliability}).

\subsubsection{FPP calculation}
\label{sec:FPP_calculation}

We obtained the FPPs by using the \textsc{vespa} package \citep{2012ApJ...761....6M,2015ascl.soft03011M}, which computes the likelihood of the main astrophysical false positives scenarios: eclipsing binaries (EBs), background eclipsing binaries (BEBs), and hierarchical triple systems (HEBs), taking into account the target coordinates and relying on simulations of the Galaxy from the \textsc{trilegal} population synthesis code \citep{2005A&A...436..895G}. Briefly, to assign the probability for each scenario, \textsc{vespa} starts from \textsc{isochrones} to carry out single-, binary-, and triple-star model fits to the observed photometric,  spectroscopic and parallax constraints. Then \textsc{vespa} simulates thousands of planetary and non-planetary scenarios to be compared with the observed phase-folded light curve, which is modelled through a trapezoidal transit fit. Finally the FPP is computed as the posterior probability of the non-planetary scenarios.

We ran \textsc{vespa} starting from the aforementioned constraints and several additional constraints that help to assess the different scenarios. We used the orbital period and the planet-to-star radius ratio as derived from \textsc{pyaneti}. We computed the maximum aperture radius for which the signal is expected to come from (\texttt{maxrad} parameter) as $\sqrt{A \pi^{-1}}$, being $A$ the area of the aperture. We also constrained the maximum allowed depth of a potential secondary eclipse (\texttt{secthresh}) to be thrice the standard deviation of the out-of-transit region. This constraint is quite conservative, as any secondary eclipse with that depth would be noticed clearly even with the naked eye. We show the obtained FPP broken down by scenario in Table \ref{tab:FPPs}.

\subsubsection{Reliability of FPPs and final dispositions}
\label{sec:FPP_reliability}

The most commonly adopted criteria to consider a candidate as statistically validated planet (VP) is to have a false positive probability lower than 1 per cent  \citep[FPP $<$ 0.01; e.g.][]{2014ApJ...784...45R, 2015ApJ...809...25M, 2016ApJ...822...86M, 2019A&A...627A..66H}, while a candidate with a false positive probability greater than 90 per cent (FPP $>$ 0.9) is considered as a false positive \citep[FP; e.g.][]{2015ApJ...809...25M, 2016ApJ...822...86M}. For the rest of cases (0.01 < FPP < 0.9), the planet candidate disposition (PC) prevails. However, it has been widely discussed \citep[e.g.][]{2018AJ....156..277L,2018AJ....155..136M} and proved \citep[e.g.][]{2017A&A...606A..75C,2017ApJ...847L..18S} that only relying on the FPP can lead to misclassifications related to several factors not being taken into  consideration by the validation packages.   In the next paragraphs, we detail the conditions considered in this work that any target and candidate needs to meet before being able to be assigned a disposition based on its FPP, as well as which crucial factors  can prevent validation, independently of the FPP.

As planetary signals must be periodic, we do not validate any candidate with less than three consecutive transits within the light curve. Therefore, the signals found from EPICs 211537087.03 (1 dip), 211590050.01 (1 dip), 211644764.01 (2 dips), 211904310.01 (2 dips), and 212008766.02 (1 dip) are considered as planetary candidates. For those targets with three or more transit signals, we searched for odd-even transit depth mismatches in order to identify possible secondary transits. For that, we modelled separately the odd and even transit events and compared their depths, avoiding to validate any signal with a transit depth mismatch higher than 3-$\sigma$.  We also avoid validating noisy signals in order to discard possible non-physical origins. Quantitatively, we do not validate any signal  with a signal-to-noise ratio (SNR)\footnote{We computed the signal-to-noise ratio as SNR = $ d  \sqrt{N_{p}} \sigma^{-1}$, where $d$ is the  transit depth, $N_{p}$ the number of data points in transit, and $\sigma$ the standard deviation of the detrended light curve.} lower than 10, which is a conservative threshold adopted by several authors \citep[e.g.][]{2012ApJS..201...15H,2016ApJ...822...86M,2020MNRAS.499.5416C}.  Besides, similarly to previous works \citep[e.g.][]{2018AJ....155..136M,2021AJ....161...24G}, we do not validate  any signal with $R_{\rm p}$ > 8 $\rm R_{\oplus}$ in order to avoid validating brown dwarfs and low-mass eclipsing stellar companions.   In Table \ref{tab:FPPs} we include the number of observed transits, odd-even mismatches, SNRs and $R_{\rm p}$ of each new signal. 

\renewcommand{\arraystretch}{1.2}
\begin{table*}
\scriptsize
\caption{Summary of the statistical validation analysis for the new targets reported in this work. From left to right: ID, final probabilities (computed from \textsc{vespa}) that the signal is due to a BEB, EB, HEB, probability that the signal comes from a planet, FPP, number of measured transits, signal-to-noise ratio, odd-even mismatch, Astrometric Goodness of Fit of the astrometric solution for the star in the Along-Scan direction, Astrometric Excess Noise significance, candidate radius, fulfillment of the condition $\delta'$ > $\gamma^{-1}$, and final disposition assigned (VP = validated planet, PC = planet candidate).}
	\begin{center}
		\begin{tabular}{c|c|c|c|c|c|c|c|c|c|c|c|c|c}
		    \hline
			ID & P(BEB) & P(EB) & P(HEB) & P(Pl) & FPP & $\#$tr & SNR & mismatch [$\sigma$] & GOF\_AL & D & $R_{p}$ [$R_{\oplus}$] & $\delta'$ > $\gamma^{-1}$ & Disp \\
			\hline
			211309648.01 & $\rm 2.8 \times 10^{-25}$ & $\rm 2.1 \times 10^{-2}$ & $\rm 4.1 \times 10^{-4}$ & $\rm 9.8 \times 10^{-1}$ & $\rm 2.1 \times 10^{-2}$ & $\geq 3$ & 101 & {0.43} & {-3.41} & {0.00} & {$\geq 8$} & Yes & PC \\
			211319779.01 & $\rm 1.0 \times 10^{-2}$ & $\rm 2.8 \times 10^{-6}$ & $\rm 1.9 \times 10^{-10}$ & $\rm 9.9 \times 10^{-1}$ & $\rm 1.0 \times 10^{-2}$ & $\geq 3$ & 18 & {0.25} & {1.43} & {0.00} & {2.4} & No & PC \\
			211393988.01 & $\rm 5.2 \times 10^{-15}$ & $\rm 5.7 \times 10^{-2}$ & $\rm 7.9 \times 10^{-4}$ & $\rm 9.4 \times 10^{-1}$ & $\rm 5.8 \times 10^{-2}$ & $\geq 3$ & 62 & {1.20} & {0.97} & {0.00} & {$\geq 8$} & Yes & PC \\
			211407755.01 & $\rm 7.6 \times 10^{-1}$ & $\rm 2.1 \times 10^{-1}$ & $\rm 1.1 \times 10^{-3}$ & $\rm 2.5 \times 10^{-2}$ & $\rm 9.7 \times 10^{-1}$ & $\geq 3$ & 22 & {0.42} & {-2.62} & {0.00} & {6.1} & No & PC \\
			211480861.01 & $\rm 2.8 \times 10^{-2}$ & $\rm 8.3 \times 10^{-1}$ & $\rm 7.0 \times 10^{-2}$ & $\rm 6.9 \times 10^{-2}$ & $\rm 9.3 \times 10^{-1}$ & $\geq 3$  & 48 & {0.04} & {96.54} & {185.60} & {$\geq 8$} & Yes & PC \\
			211525753.01 & $\rm 7.6 \times 10^{-3}$ & $\rm 4.5 \times 10^{-5}$ & $\rm 1.4 \times 10^{-9}$ & $\rm 9.9 \times 10^{-1}$ & $\rm 7.7 \times 10^{-3}$ & $\geq 3$ & 13 & {0.15} & {-5.52} & {1.04} & {5.7} & Yes & \textbf{VP}  \\
			211537087.01 & $\rm 2.8 \times 10^{-3}$ & $\rm 7.2 \times 10^{-5}$ & $\rm 2.5 \times 10^{-9}$ & $\rm 1.0 \times 10^{0}$ & $\rm 2.9 \times 10^{-3}$ & $\geq 3$ & 15 & {0.63} & {-3.51} & {0.00} & {2.3} & Yes & \textbf{VP} \\
			211537087.02 & $\rm 7.8 \times 10^{-3}$ & $\rm 2.8 \times 10^{-3}$ & $\rm 4.5 \times 10^{-9}$ & $\rm 9.9 \times 10^{-1}$ & $\rm 1.1 \times 10^{-2}$ & 2 & 20 & {-} & {-3.51} & {0.00} & {2.6} & Yes & PC \\
			211537087.03 & - & - & - & - & - & 1 & 11 & {-} & {-3.51} & {0.00} & {2.3} & Yes & PC \\
			211590050.01 & - & - & - & - & - & 1 & 105 & {-} & {-0.60} & {0.00} & {$\geq 8$} & Yes & PC  \\
			211644764.01 & $\rm 5.9 \times 10^{-2}$ & $\rm 3.6 \times 10^{-1}$ & $\rm 7.7 \times 10^{-2}$ & $\rm 5.1 \times 10^{-1}$ & $\rm 4.9 \times 10^{-1}$ & 2 & 56 & {-} & {4.64} & {0.64} & {$\geq 8$} & Yes & PC \\
			211705502.01 & $\rm 7.9 \times 10^{-1}$ & $\rm 2.0 \times 10^{-1}$ & $\rm 2.8 \times 10^{-3}$ & $\rm 4.9 \times 10^{-3}$ & $\rm 9.9 \times 10^{-1}$ & $\geq 3$ & 46 & {3.70} & {0.57} & {0.00} & {$\geq 8$} & Yes & PC \\
			211724246.01 & $\rm 4.1 \times 10^{-2}$ & $\rm 2.1 \times 10^{-2}$ & $\rm 1.2 \times 10^{-3}$ & $\rm 9.4 \times 10^{-1}$ & $\rm 6.3 \times 10^{-2}$ & $\geq 3$ & 58 & {0.33} & {-3.50} & {0.00} & {$\geq 8$} & Yes & PC \\
			211730267.01 & $\rm 2.9 \times 10^{-4}$ & $\rm 3.5 \times 10^{-4}$ & $\rm 1.1 \times 10^{-16}$ & $\rm 1.0 \times 10^{0}$ & $\rm 6.3 \times 10^{-4}$ & $\geq 3$ & 28 & {0.33} & {-4.86} & {0.00} & {3.7} & Yes & \textbf{VP} \\
			211904310.01 & $\rm 3.3 \times 10^{-1}$ & $\rm 2.1 \times 10^{-1}$ & $\rm 2.0 \times 10^{-2}$ & $\rm 4.4 \times 10^{-1}$ & $\rm 5.6 \times 10^{-1}$ & 2 & 42 & {-} & {-2.13} & {0.00} & {$\geq 8$} & Yes & PC  \\
			211914998.01 & $\rm 8.9 \times 10^{-4}$ & $\rm 4.7 \times 10^{-5}$ & $\rm 1.3 \times 10^{-11}$ & $\rm 1.0 \times 10^{0}$ & $\rm 9.3 \times 10^{-4}$ & $\geq 3$ & 20 & {1.58} & {-6.55} & {0.00} & {2.5} & Yes & \textbf{VP} \\
			211914998.02 & $\rm 1.8 \times 10^{-2}$ & $\rm 4.0 \times 10^{-3}$ & $\rm 3.5 \times 10^{-6}$ & $\rm 9.8 \times 10^{-1}$ & $\rm 2.2 \times 10^{-2}$ & 2 & 8 & {-} & {-6.55} & {0.00} & {2.1} & Yes & PC \\
			212008766.02 & - & - & - & - & - & 1 & 44 & {-} & {-1.15} & {0.00} & {2.0} & Yes & PC \\
			\hline
		\end{tabular}
	\end{center}
	\label{tab:FPPs}
\end{table*}

\begin{table}
\caption{Dilution factors and magnitudes differences for each nearby \textit{Gaia} DR2 star located inside or outside the aperture at a distance $r$ < 40 arcsec. }

\begin{tabular}{cccccc}
\hline
EPIC & Aperture & $r$ (arcsec) & $\Delta_{m}$ & $\gamma_{\rm pri}$ & $\gamma_{\rm sec}$ \\ \hline
211309648 & Outside & 31.86 & 5.93 & 1.004 & 236.548 \\
 & Outside & 39.52 & 5.85 & 1.005 & 219.675 \\ 
211319779 & Inside & 3.41 & 6.24 & 1.003 & 312.918 \\
 & Outside & 11.22 & 2.88 & 1.070 & 15.232 \\ 
211393988 & Outside & 32.18 & -0.96 & 3.424 & 1.412 \\
 & Outside & 34.51 & 3.40 & 1.044 & 23.985 \\ 
\multicolumn{1}{l}{211407755} & Inside & 4.73 & 3.50 & 1.040 & 26.223 \\
\multicolumn{1}{l}{} & Inside & 11.04 & 6.62 & 1.002 & 446.985 \\
\multicolumn{1}{l}{} & Outside & 21.97 & 2.17 & 1.135 & 8.410 \\
\multicolumn{1}{l}{} & Outside & 35.39 & 3.91 & 1.027 & 37.749 \\ 
211480861 & Inside & 12.87 & 6.52 & 1.002 & 407.481 \\
 & Outside & 28.77 & 3.91 & 1.027 & 37.789 \\
 & Outside & 32.41 & 7.27 & 1.001 & 807.789 \\
 & Outside & 34.93 & 6.94 & 1.002 & 598.97 \\ 
211525753 & Outside & 28.45 & 2.72 & 1.082 & 13.242 \\ 
211537087 & Outside & 17.66 & 7.36 & 1.001 & 881.968 \\ 
211590050 & $-$ & $-$ & $-$ & $-$ & $-$ \\
211644764 & Outside & 16.85 & 7.23 & 1.001 & 779.825 \\
 & Outside & 35.27 & 5.62 & 1.006 & 178.812 \\ 
\multicolumn{1}{l}{211705502} & Inside & 5.88 & 6.47 & 1.003 & 389.830 \\
\multicolumn{1}{l}{} & Outside & 31.91 & 4.41 & 1.017 & 58.932 \\
\multicolumn{1}{l}{} & Outside & 35.12 & 0.12 & 1.898 & 2.114 \\ 
211724246 & Outside & 21.68 & 7.24 & 1.001 & 789.497 \\
 & Outside & 26.78 & 5.68 & 1.005 & 188.241 \\
 & Outside & 30.04 & 7.07 & 1.001 & 673.419 \\
 & Outside & 32.11 & 5.76 & 1.005 & 202.131 \\
 & Outside & 36.57 & 5.11 & 1.009 & 111.235 \\ 
211730267 & Outside & 16.56 & 6.47 & 1.003 & 386.549 \\
 & Outside & 30.35 & 6.09 & 1.004 & 272.894 \\
 & Outside & 32.65 & 6.85 & 1.002 & 552.47 \\ 
211904310 & Outside & 30.95 & 5.57 & 1.006 & 169.749 \\ 
211914998 & Outside & 34.94 & 4.40 & 1.017 & 58.671 \\ 
\hline
\end{tabular}
\label{tab:dilution_factors}
\end{table}

\begin{figure*}
    \includegraphics[scale = 0.35]{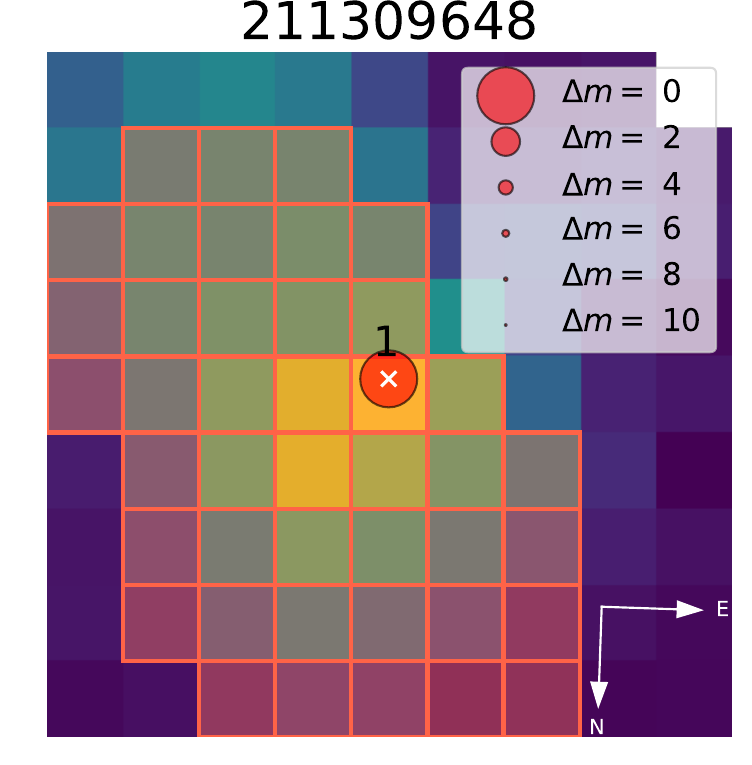}
    \includegraphics[scale=0.35]{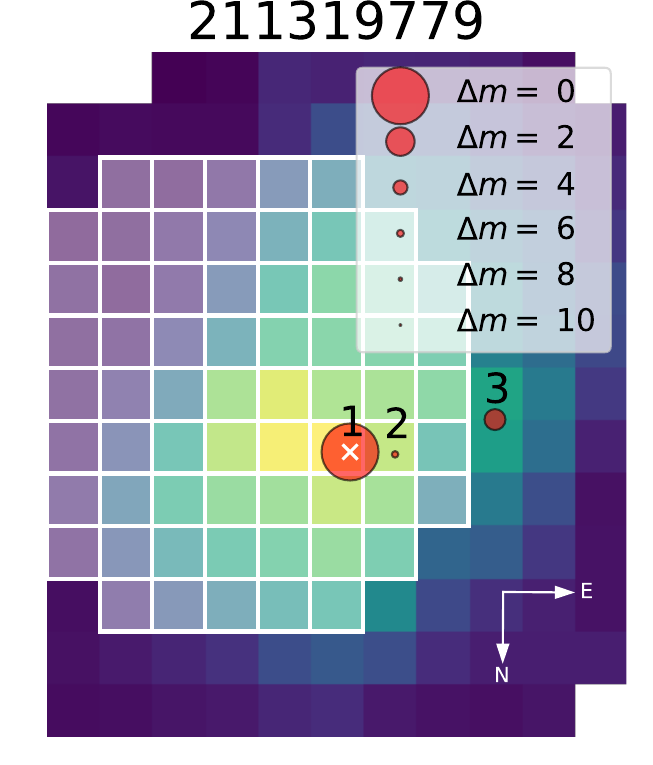}
    \includegraphics[scale=0.34]{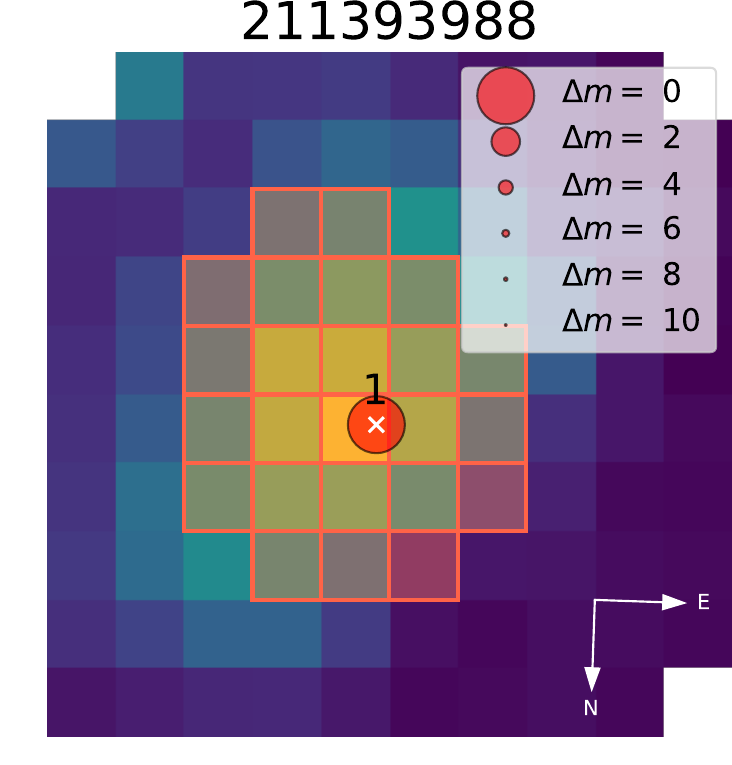}
    \includegraphics[scale=0.34]{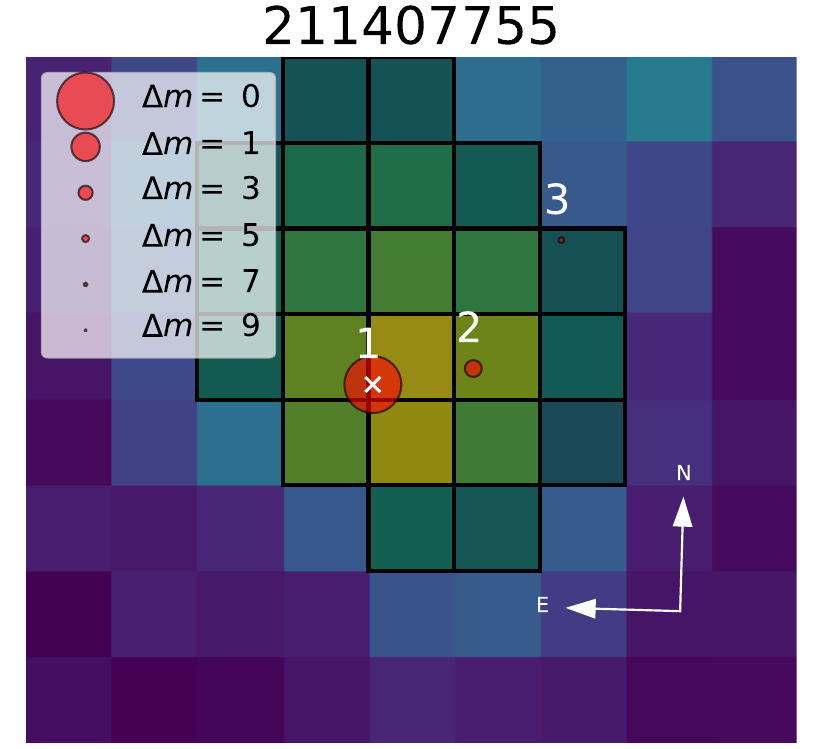}
    \includegraphics[scale=0.33]{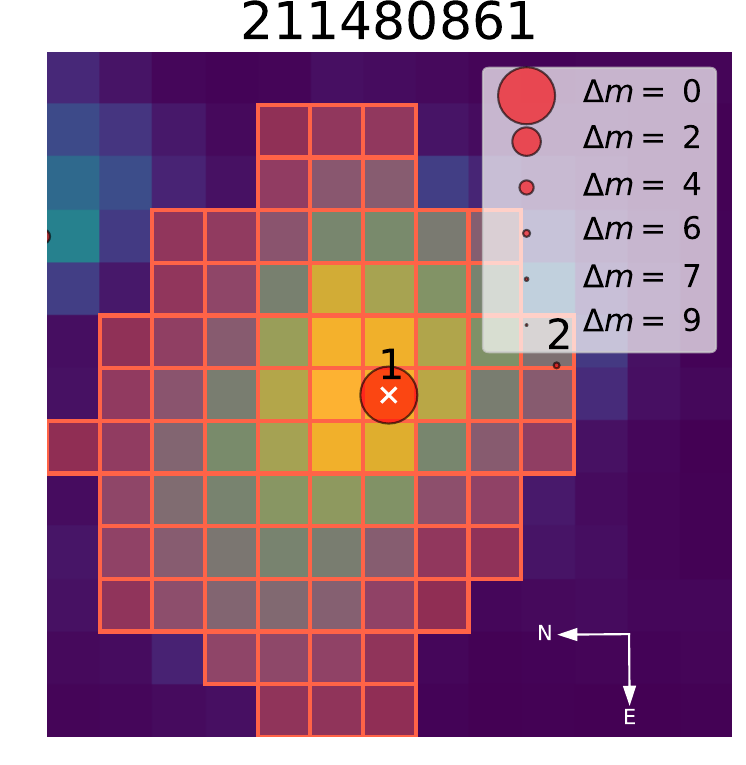}
    \includegraphics[scale=0.33]{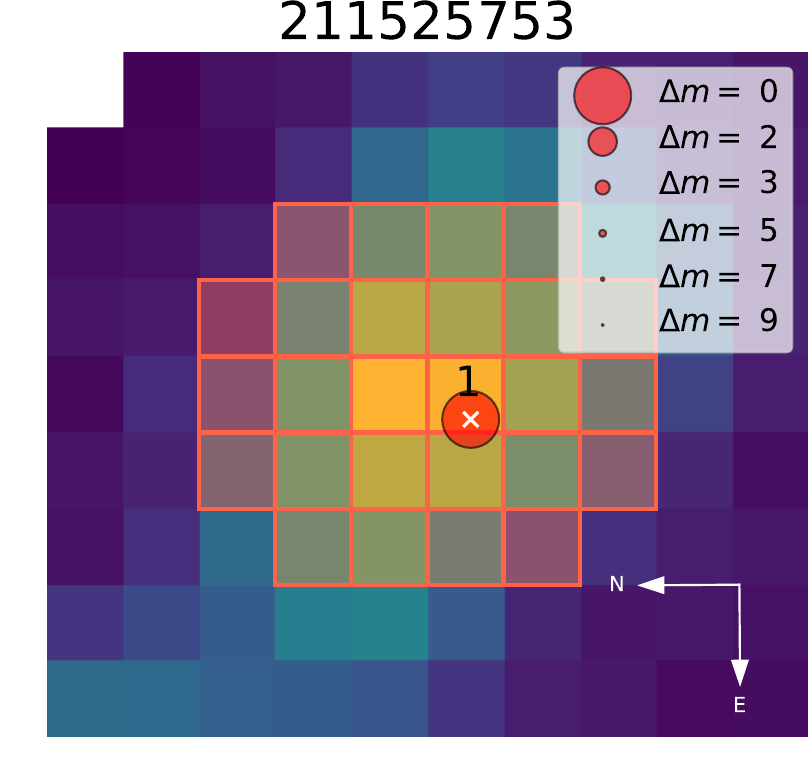}
    \includegraphics[scale=0.33]{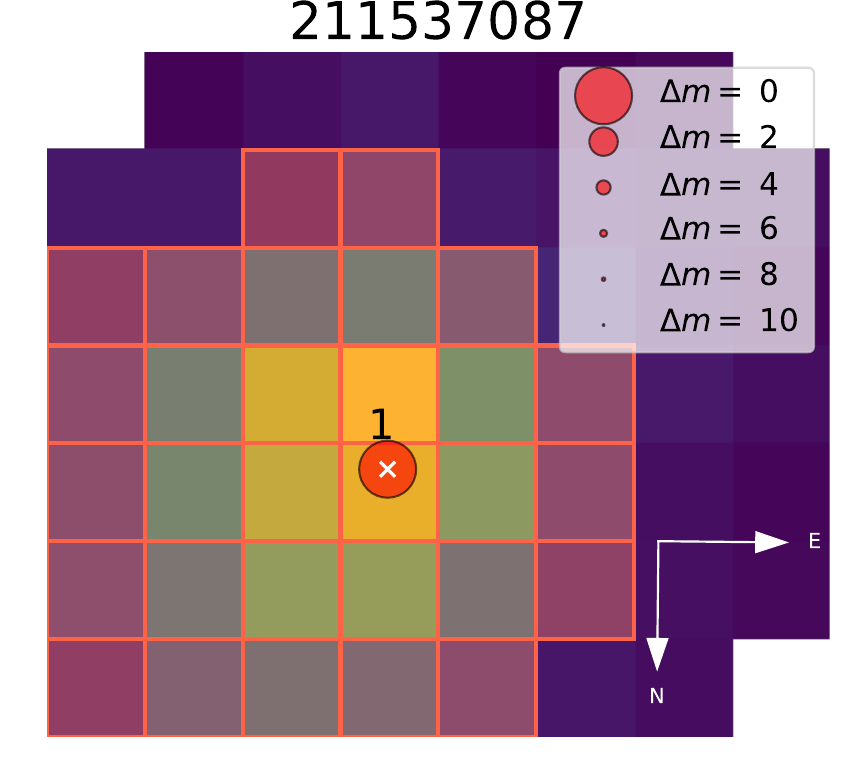}
    \includegraphics[scale=0.33]{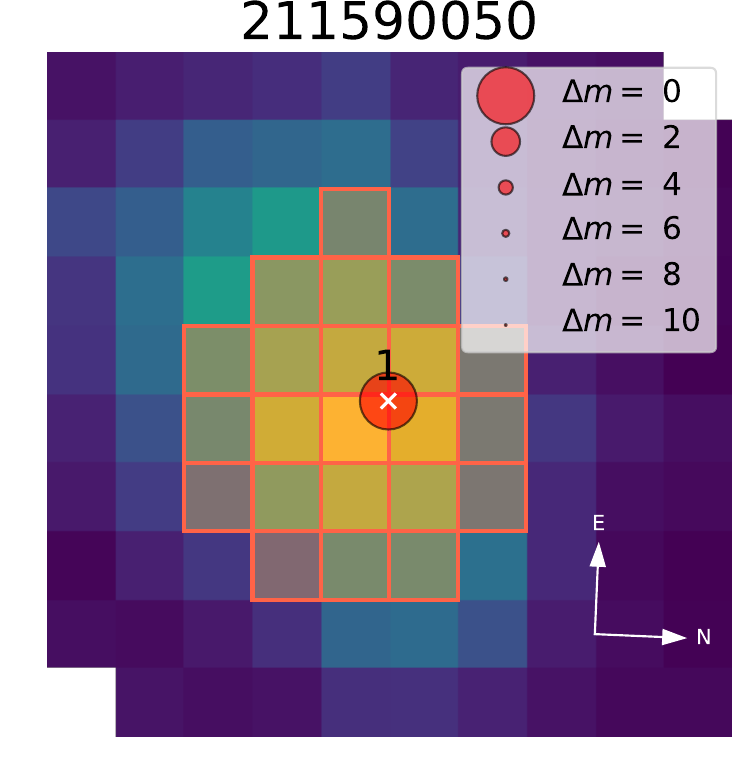}
    \includegraphics[scale=0.33]{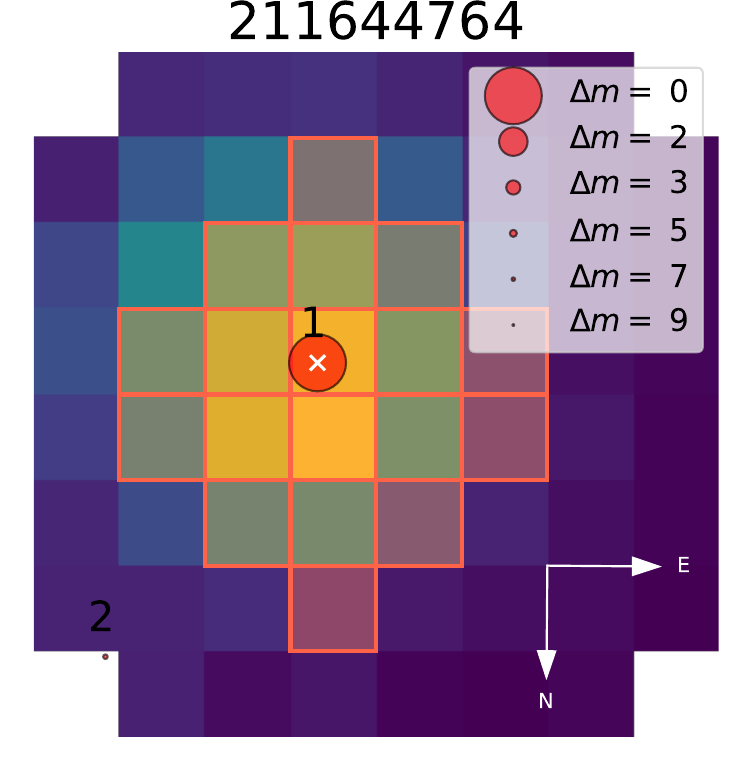}
    \includegraphics[scale=0.33]{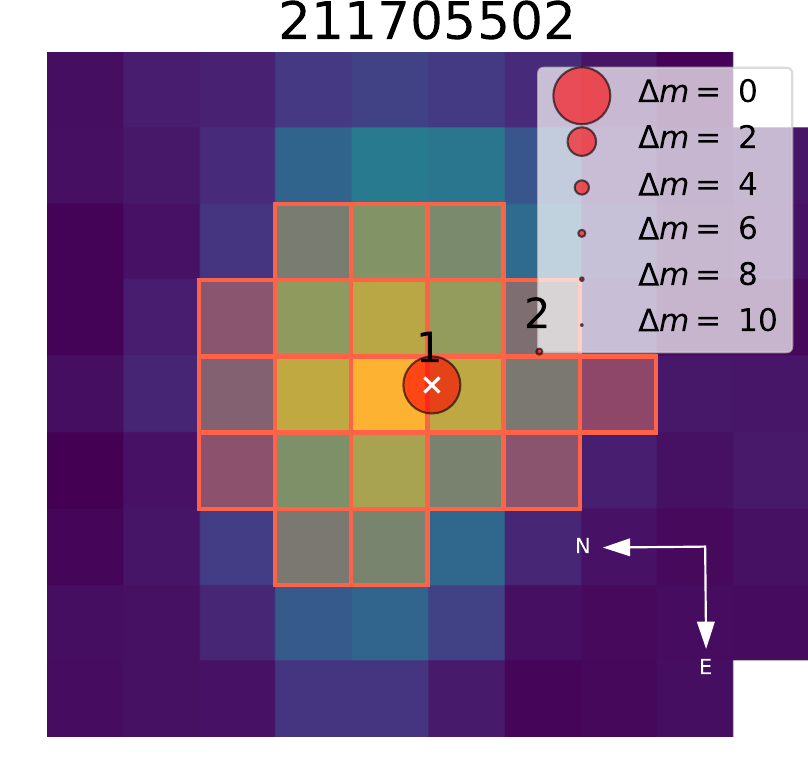}
    \includegraphics[scale=0.33]{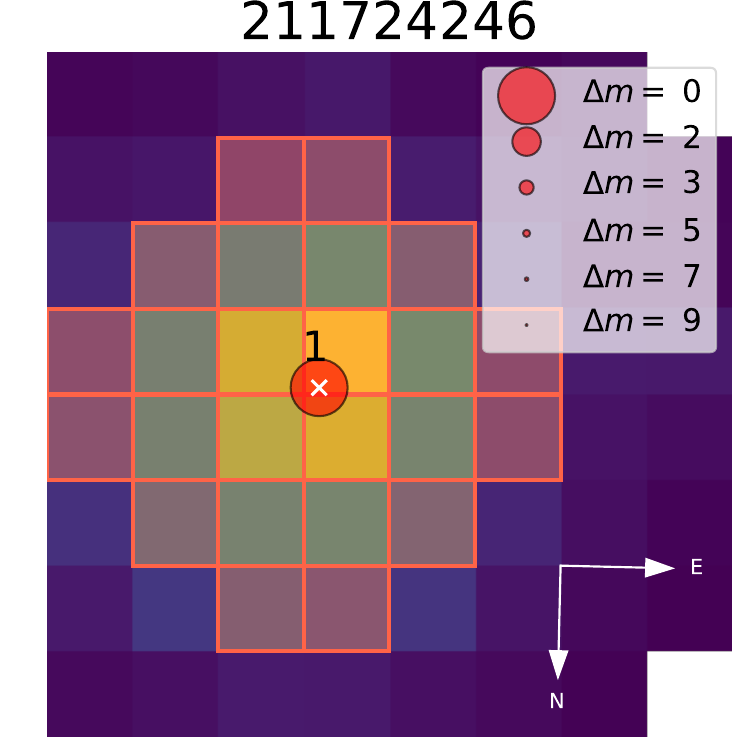}
    \includegraphics[scale=0.33]{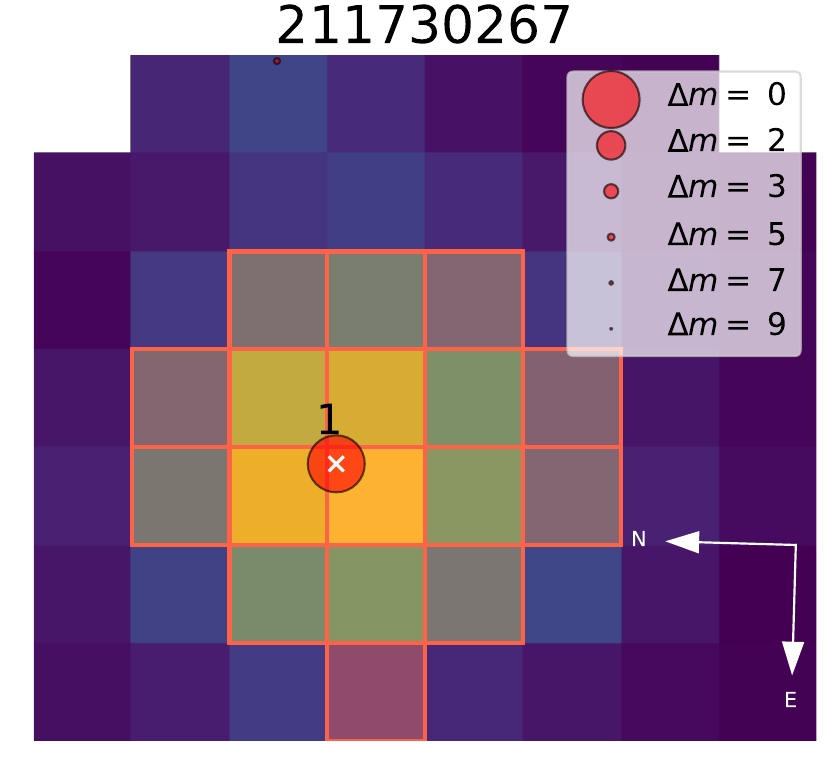}
    \includegraphics[scale=0.34]{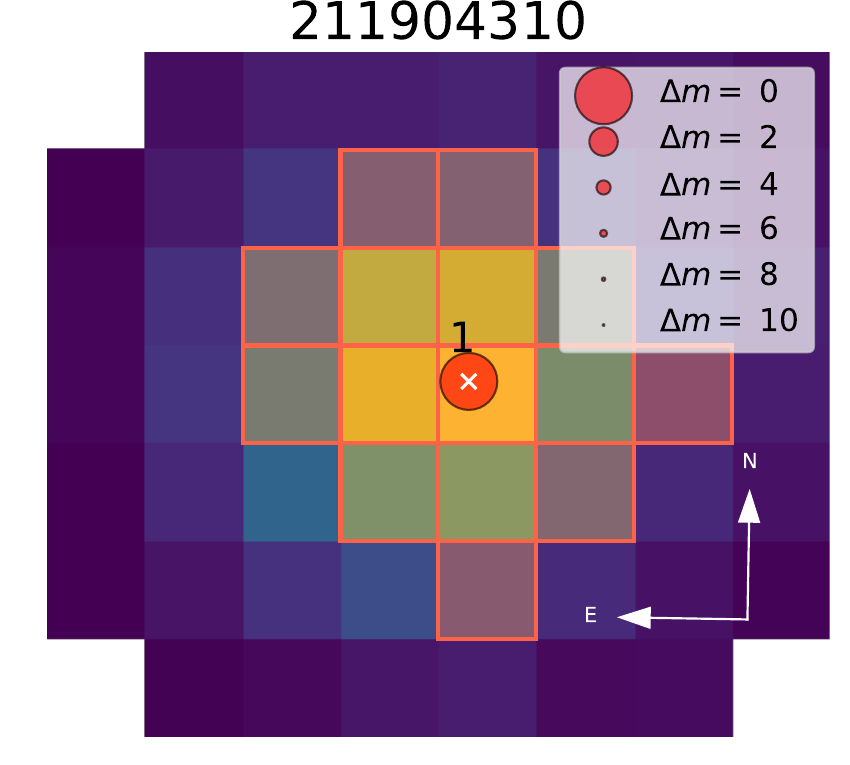}
    \includegraphics[scale=0.34]{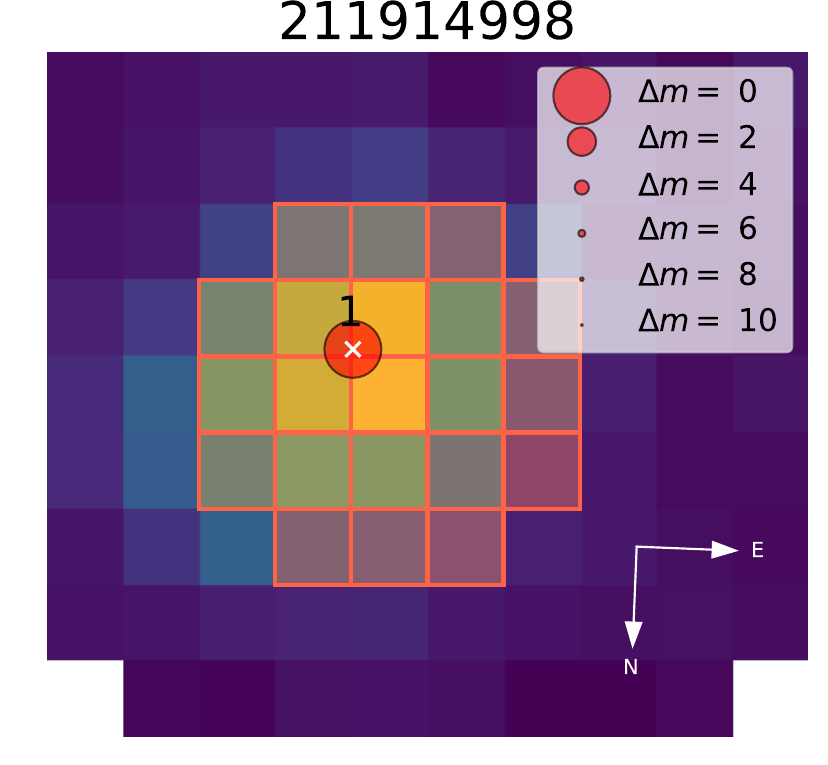}

    \caption{Target pixel files (TPFs) of each new target reported in this work. The red circles are the \textit{Gaia} DR2 sources in the field, which are scaled according to the magnitude difference between the target (highlighted with a white cross) and each nearby star. The overplotted apertures are those considered to obtain the photometry, being the red ones corresponding to the EVEREST pipeline, and the white one to the K2SFF pipeline. The black aperture was defined manually. The \textit{K2} pixel scale is 3.98 arcsec pixel$^{-1}$. }
    \label{fig:tpfplotter_1}
\end{figure*}


\newpage
We searched for hints of binarity by using \textit{Gaia} DR2. Systems with large Astrometric Goodness of Fit of the astrometric solution for the source in the Along-Scan direction (\texttt{GOF\_AL} > 20) and Astrometric Excess Noise significance (\texttt{D} > 5) are plausibly poorly-resolved binaries \citep{2018RNAAS...2...20E}. There is one target in our sample that meets these two conditions (EPIC 211480861.01). Given the possible presence of multiple stars in the system, we designate this target as PC regardless of its FPP, and we do not report its planetary parameters. We include \texttt{GOF\_AL} and \texttt{D} in Table \ref{tab:FPPs} for each target.

Another important consideration before relying on the FPP consists of searching for nearby stars inside or surrounding the aperture, which could be contaminating the photometry. A fairly common false positive scenario involves the presence of a fainter contaminant eclipsing binary, whose deep dips are diluted by the target star, so what we observe appears to be more similar to the typical shallower planetary transits. In another possible scenario, if a planetary signal comes from the target star and there is also a bright nearby contaminant star, the transit depths will be shallower, and thus systematically causing an underestimation of the planet radius. In any case in which we find contaminant stars causing that the origin of the signal cannot be determined, we prevent the candidate from validation, and we do not report the candidate parameters. In order to search for nearby contaminant stars in our new targets, we updated the \textsc{tpfplotter} package \citep{2020A&A...635A.128A}, which in addition to \textit{TESS}, it is now compatible with \textit{Kepler} and \textit{K2} data. The package overlaps the \textit{Gaia} DR2 catalogue to the Target Pixel Files (TPFs), computing and plotting the location of potential contaminant sources relative to the photometric aperture (see Fig. \ref{fig:tpfplotter_1}). Note that the apertures of some targets occupy almost the entire TPF, so in these cases there can be stars outside the TPF but still contaminating the photometry due to the broad point spread function (PSF) of the \textit{Kepler} telescope, which has a typical full-width at half-maximum of FWHM $\approx$ 6 arcsec. For this reason, we conservatively looked for all the \textit{Gaia} DR2 sources within a search radius of 40 arcsec and with up to 10 magnitudes fainter than the target star. We quantified the photometric contamination by computing the dilution factor as $\gamma = 1+10^{0.4\Delta m}$ \citep[equation 1,][]{2018AJ....156..277L}, which defines the relationship between the observed transit depth ($\delta'$) and the true transit depth ($\delta$) as $\delta'$ = $\gamma^{-1}$ $\delta$, being $\Delta m$ the magnitude of the contaminant star minus the magnitude of the star where the signal comes from, in the \textit{Kepler} bandpass. We use the notation $\gamma_{\rm pri}$ and $\gamma_{\rm sec}$ to indicate that the dilution factor is computed considering that the signal comes from the target (primary) star with a true transit depth $\delta_{\rm pri}$ or from a nearby (secondary) star with a true transit depth $\delta_{\rm sec}$. We show in Table \ref{tab:dilution_factors} both $\delta_{\rm pri}$ and $\delta_{\rm sec}$ for all the sources found at a distance < 40 arcsec of our newly detected targets, as well as their separation and magnitude differences. We do not know a priori where the signal comes from, so we followed a procedure to assess if we can discard the nearby star origin. The procedure consisted of assuming that the signal comes from any of the nearby stars located inside the aperture or outside but separated up to 6 arcsec from the nearest edge. So, as their hypothetical eclipses cannot be greater than 100 per cent (i.e. $\delta_{\rm sec}$ $\leq$ 1), if the condition $\delta'$ > $\gamma_{\rm sec}^{-1}$ is met, we can ensure that the observed depth $\delta'$ is too deep to be caused by the nearby secondary star. Otherwise, the origin of the signal is uncertain, so the computed FPP is not reliable, and we consider that signal as a PC until its origin is ascertained.

The condition $\delta'$ > $\gamma_{\rm sec}^{-1}$ is met for all our new targets except for EPIC 211319779 and EPIC 211407755. The dip observed in EPIC 211319779 could be caused by a $\sim$30 per cent dip coming from star $\#2$, and by a $\sim$1.5 per cent dip coming from star $\#3$. Similarly, the dip observed in EPIC 211407755 could be caused by a  $\sim$1 per cent dip coming from star $\#2$ (see Fig. \ref{fig:tpfplotter_1} and Table \ref{tab:dilution_factors}). For these two cases, we performed pixel level multi-aperture analysis in order to figure out the actual origin of the signals found. In some cases, when the target star and the potential contaminant faint star are located several pixels apart, assessing the photometry created with different photometric apertures can solve the signal origin uncertainty, being decisive to unveil possible FP scenarios \citep[e.g. ][found that two \textit{K2} validated planets were in fact background eclipsing binaries, and hence FPs]{2017A&A...606A..75C}. Unfortunately, our targets are not suitable to reach such decisive conclusions by means of multi-aperture analysis, because of the great closeness between the stars. Even though there are clear hints of that the signal from EPIC 211319779 does not come from star $\#3$ (e.g. different apertures both enclosing and excluding it do not alter transit depths, and the transits are still present considering the EVEREST aperture, which is 14 arcsec away from star $\#3$ as shown in Fig. \ref{fig:EVEREST_vs_K2SFF}), the stars $\#1$ and $\#2$ for both  EPIC 211319779 and EPIC 211407755 are 1 pixel away, causing the PSFs to be completely blended. As the signal origin cannot be ascertained for these two targets, we consider them as PCs.

Of the \newsignals new candidates found, four satisfy all the aforementioned conditions. Besides, these candidates have FPP < 0.01. Before relying on the FPP to assign the final disposition, we searched for close sources by observing these stars with the high-spatial resolution camera AstraLux, located at the 2.2 m telescope of the Calar Alto Observatory. The presence of close contaminant sources, identically to nearby sources, implies a potential misidentification in the origin of the signal. Besides, the \textsc{isochrones} stellar characterization would be unreliable because of the photometric contamination between both sources. For each target we found no evidence of additional sources within a 6 × 6 arcsec field of view and within the computed sensitivity limits (see Section \ref{sec:AstraLux} for further details).  After meeting all the conditions imposed, we consider these four candidates as validated planets. To sum up, the statistical validation analysis carried out over the \newsignals new signals found resulted in four validated planets and 14 planet candidates, three of which we report without planetary parameters due to the presence of photometric contamination that prevents the determination of the origin of the signals.



\section{R\,e\,s\,u\,l\,t\,s \, a\,n\,d \, D\,i\,s\,c\,u\,s\,s\,i\,o\,n}
\label{results_discussion}

The K2-OjOS search in \textit{K2} C18 gave rise to 42 planet candidates, of which 24 were published in previous works (four confirmed planets, 14 validated planets, and six planet candidates), and 18 are new detections (four validated planets, and 14 planet candidates). In Section \ref{sec:refining_ephemeris}, we quantify the refinement of transit ephemeris and planetary parameters that we achieve by modelling C5, C16 (when available) and C18 photometric data jointly for the previously known planets and candidates. In Section \ref{sec:detection_efficiency}, we compute and compare the detection efficiencies of both the K2-OjOS and BLS searches.  In Sections \ref{sec:charact_stellar_sample} and \ref{sec:charact_planet_sample}, we contextualize our results by comparing the K2-OjOS detections to the population of known host stars and exoplanets,\footnote{All data for known host stars, planets, and candidates were obtained from the Nasa Exoplanet Archive (NEA).} and discuss about the possible internal structure of the K2-OjOS planets and candidates. We also compute the  habitable zone (HZ) boundaries for our target stars based on both conservative and more optimistic climate models. Finally, in Section \ref{sec:highlights_individual_systems} we highlight interesting features of five individual systems.

\subsection{Refining transit ephemeris and planetary parameters}
\label{sec:refining_ephemeris}

Obtaining long temporal baselines of photometric data for targets hosting transiting planets allows us to measure transit ephemeris very precisely. Besides, increasing the number of observed transits allows us to better constrain the planetary parameters due to a greater in-transit coverage. The latter is especially important for long-period planets observed by \textit{K2}, whose long 30-min cadence corresponds to few data points per transit (typically between 4 and 10). For deep enough dips, transit follow-up can be carried out from ground-based facilities, but for many interesting targets these observations need to be done from space. 

\textit{K2} C18 observed a field that covers 30 per cent of C16 and 95 per cent of C5. The targets observed in both C5 and C18 have a 3-yr temporal baseline with a 4-month duty cycle, while for those targets observed in C5, C16, and C18, their duty cycle increases up to 7 months. In our sample of targets with planets and/or candidates with a full characterization, 12 of them were observed in C18 alone, 13 were observed in both C5 and C18, and 9 were observed in C5, C16, and C18. All the 22 targets with observations in multiple campaigns host published planets or candidates, whose planetary parameters were derived starting from C5 data alone, since when their corresponding papers were in preparation, C16 and C18 had not started yet. In this work, we modelled for the first time the light curves of these \targetstoimprove \, targets by joining photometric data from C5, C16 (when available), and C18, managing to refine their published transit ephemeris and planetary parameters. For the orbital period, we obtain uncertainty improvement factors between 10 and 88, with a median value of \medianfactorP. We also obtain more precise $T_{0}$ and $R_{p}/R_{\star}$, decreasing their uncertainties by a median factor of \medianfactorTo and \medianfactorrprs respectively. The significant orbital period refinement is to be expected given previous similar studies, in which for example \citet[][]{2018AJ....156..277L} decreased the period uncertainty by a median value of 26 for a subset of targets observed in both C5 and C16, and \citet[][]{2021MNRAS.508..195D} obtained a maximum improvement factor of 80 for a planetary signal modelled with C5, C16 and C18 data jointly. The great ephemeris refinement obtained for the 22 targets facilitates future planning of follow-up observations by the new telescope generation. A representative example is K2-274 b, for which we compute, for the year 2028 (the second year of the scheduled \textit{PLATO} mission), a propagated uncertainty in the mid-transit time of 4 min, while the uncertainty obtained from the currently published parameters is 1 h and 45 min.

\subsection{Detection efficiency}
\label{sec:detection_efficiency}

\begin{figure}
    \includegraphics[scale=0.44]{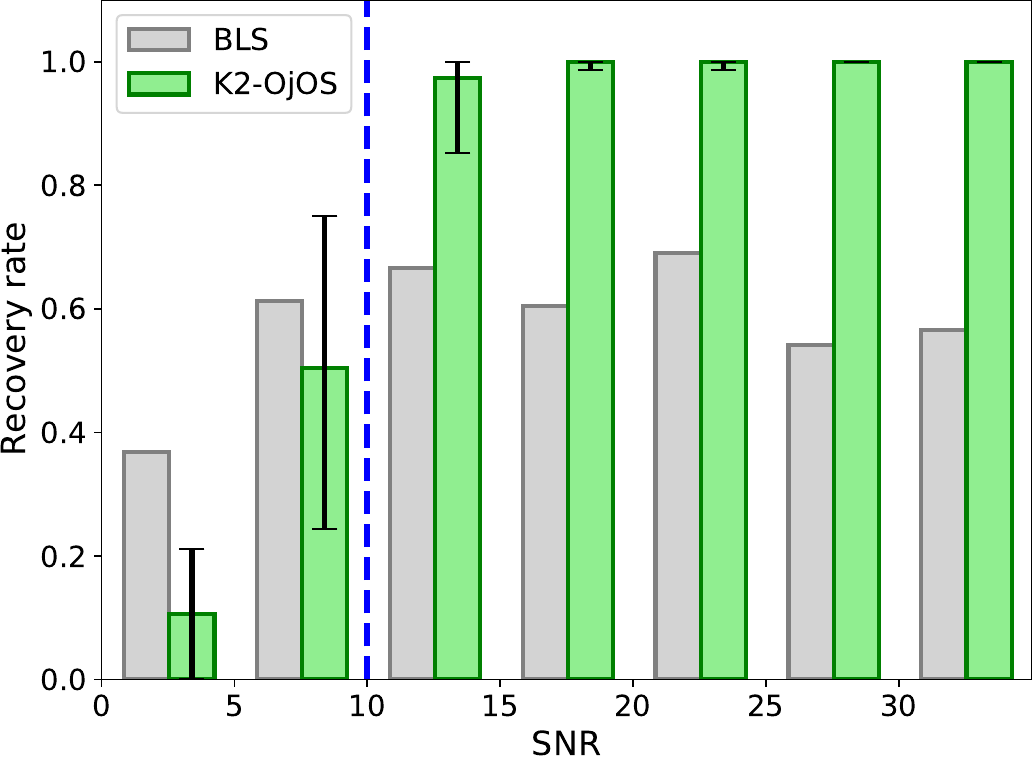}
    \caption{BLS and K2-OjOS recovery rates for different ranges of SNR. The vertical dashed line indicates the minimum required SNR for a signal to be subjected to validation in this work.}
    \label{fig:BLS_K2OJOS_recovered_vs_Rp_Rs}
\end{figure}

\begin{figure}
    \includegraphics[scale=0.44]{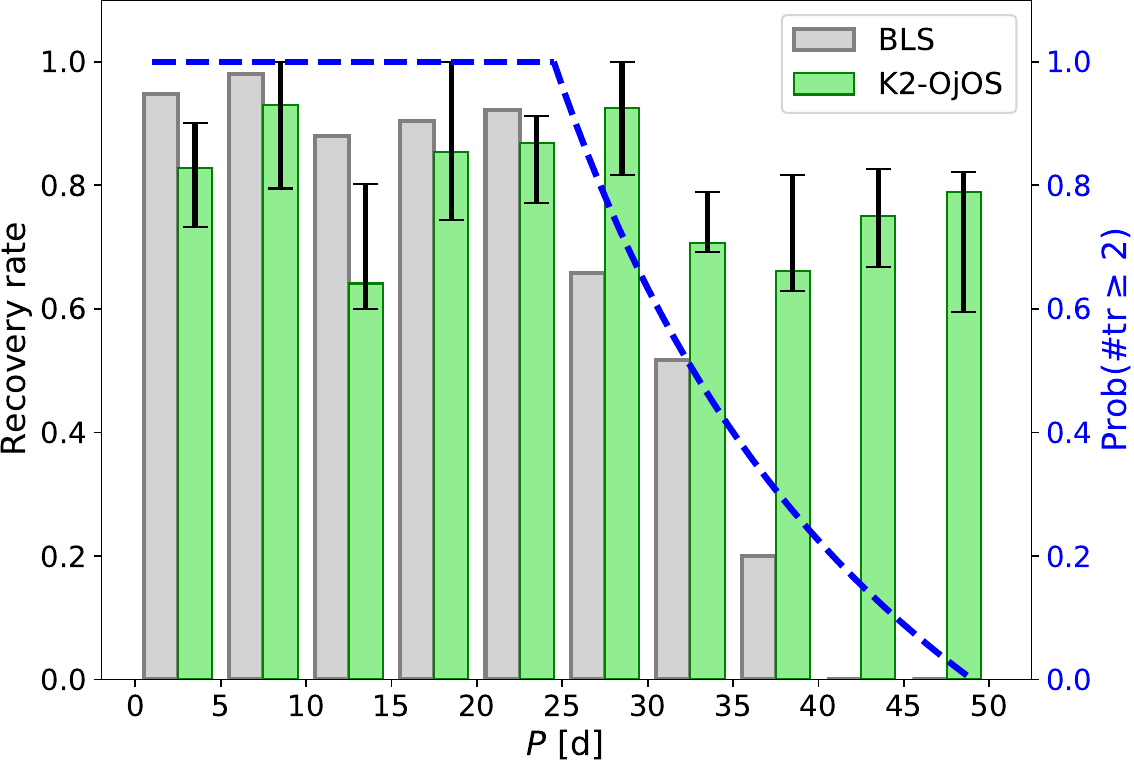}
    \caption{BLS and K2-OjOS recovery rates for different ranges of $P$. Superimposed we plot the probability function of having two or more injected transits per signal. The function takes the value 1 for the interval 1 d < $P$ < 24.5 d, and (49-$P$)/$P$  for 24.5 d < $P$ < 49 d.}
    \label{fig:BLS_K2OJOS_recovered_vs_Porb}
\end{figure}

The K2-OjOS members and the BLS algorithm analysed light curves with simulated transit signals in order to quantify the detection efficiencies of both search methods (Section \ref{sec:injection_recovery}). We first computed the recovery rates of each method by dividing the total number of recovered transits by the total number of injected transits. Given the injection and recovery conditions explained in Section \ref{sec:injection_recovery}, we obtain an overall recovery rate of $\sim$78 per cent for the K2-OjOS members and of $\sim$58 per cent for the BLS algorithm. We also computed the recovery rate of the K2-OjOS team without being biased by the number of batches analysed by each member in this particular work; that is, we calculated the mean value of the recovery rates of each member. As a result, we obtain $\sim$78 per cent as well.  In the following we discuss the K2-OjOS and BLS recovery rates broken down by SNR and $P$.

Fig. \ref{fig:BLS_K2OJOS_recovered_vs_Rp_Rs} is a histogram in which we plot the BLS recovery rates as well as the median and 68.3 per cent credible intervals of the K2-OjOS recovery rates for different ranges of SNR. The K2-OjOS members retrieved 99.5 per cent of the injected signals with SNR > 10, which is the condition that any signal must meet before being subjected to validation in this work. The 0.5 per cent not retrieved typically corresponds to short-period signals with many transits at noise level. As for the BLS, the algorithm retrieved 62 per cent of the injected signals with SNR > 10. These small recovery rates obtained even for high SNRs are related to the inability of BLS to recover single transits, unlike the visual inspection.

In Fig. \ref{fig:BLS_K2OJOS_recovered_vs_Porb}, we plot the recovery rates as a function of the orbital period for both the BLS and K2-OjOS searches. We also plot the probability function of having two or more injected transits per signal (i.e. of not having a single transit) within the C18 light curve, computed from the injection features explained in Section \ref{sec:injection_recovery}. Similar to previous works, we find that the K2-OjOS detection efficiency keeps insensitive to the orbital period. However, for $P>24.5$ d (half of the C18 temporal baseline), the recovery rate of the BLS method drops to zero as expected given the decreasing probability of multitransit signals being injected. 


From the comparison between both methods we can draw two main conclusions. First, the K2-OjOS visual search itself shows great completeness in the search for potentially validatable signals (SNR > 10). Secondly, for signals with SNR < 10, although we obtain higher recovery rates for the BLS algorithm, we highlight the visual inspection as a good complementary method to detect single transits, which are undetectable by the widely used periodicities-based automated transit searchers as BLS.



\subsection{Characteristics of our sample: The host stars}
\label{sec:charact_stellar_sample}

The \textit{K2} stars known to host planets have a median magnitude of $K_{\rm p}$  = 12.6, which is two magnitudes brighter than that of the host stars of the primary \textit{Kepler} mission. Thereby, \textit{K2} targets can be excellent for precise RV follow-up and atmospheric characterizations, allowing us to unveil planetary masses, densities, and atmospheric properties of the planets found. Our sample has a median magnitude of $K_{\rm p}$ = 13.1, which is slightly fainter than that of \textit{K2} host stars. However, we highlight the presence of 3 bright targets ($K_{\rm p}$ < 11): EPICs 211424769 ($K_{\rm p}$ = 9.4), 211480861 ($K_{\rm p}$ = 10.0), and 211594205 ($K_{\rm p}$ = 10.7). Regarding effective temperatures, most of the planets and candidates in our sample orbit stars that are clustered around 5200 K. If we compare the relative occurrence of stars in our sample with that of \textit{K2} hosts, we find a deficit of solar-type and cool-dwarf stars, and a surplus of stars of early K and late G spectral types.




\subsection{Characteristics of our sample: Planets and candidates}
\label{sec:charact_planet_sample}

\subsubsection{Planet radius and orbital period distribution}

\begin{figure}
    \centering
    \includegraphics[scale=0.32]{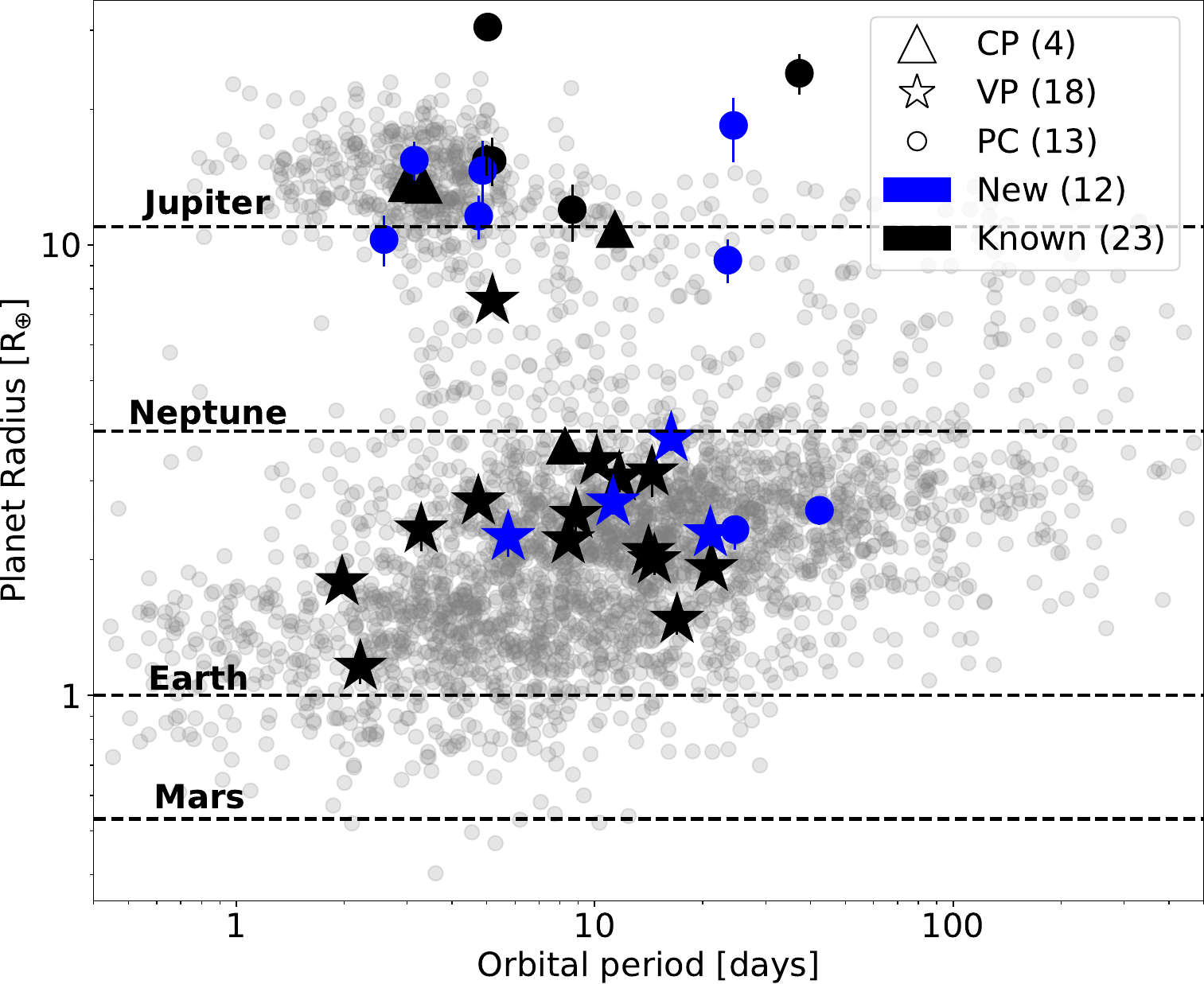}
    \caption{Planet radius as a function of the orbital period for the confirmed planets (CP), validated planets (VP), and planet candidates (PC) analysed in this work with measured orbital period and planetary parameters. The black markers correspond to previously published detections, while blue markers correspond to the new K2-OjOS detections. The grey data points correspond to the population of known confirmed and validated exoplanets.}
    \label{fig:Rp_vs_P}
\end{figure}

\begin{figure}
    \centering
    \includegraphics[scale=0.255]{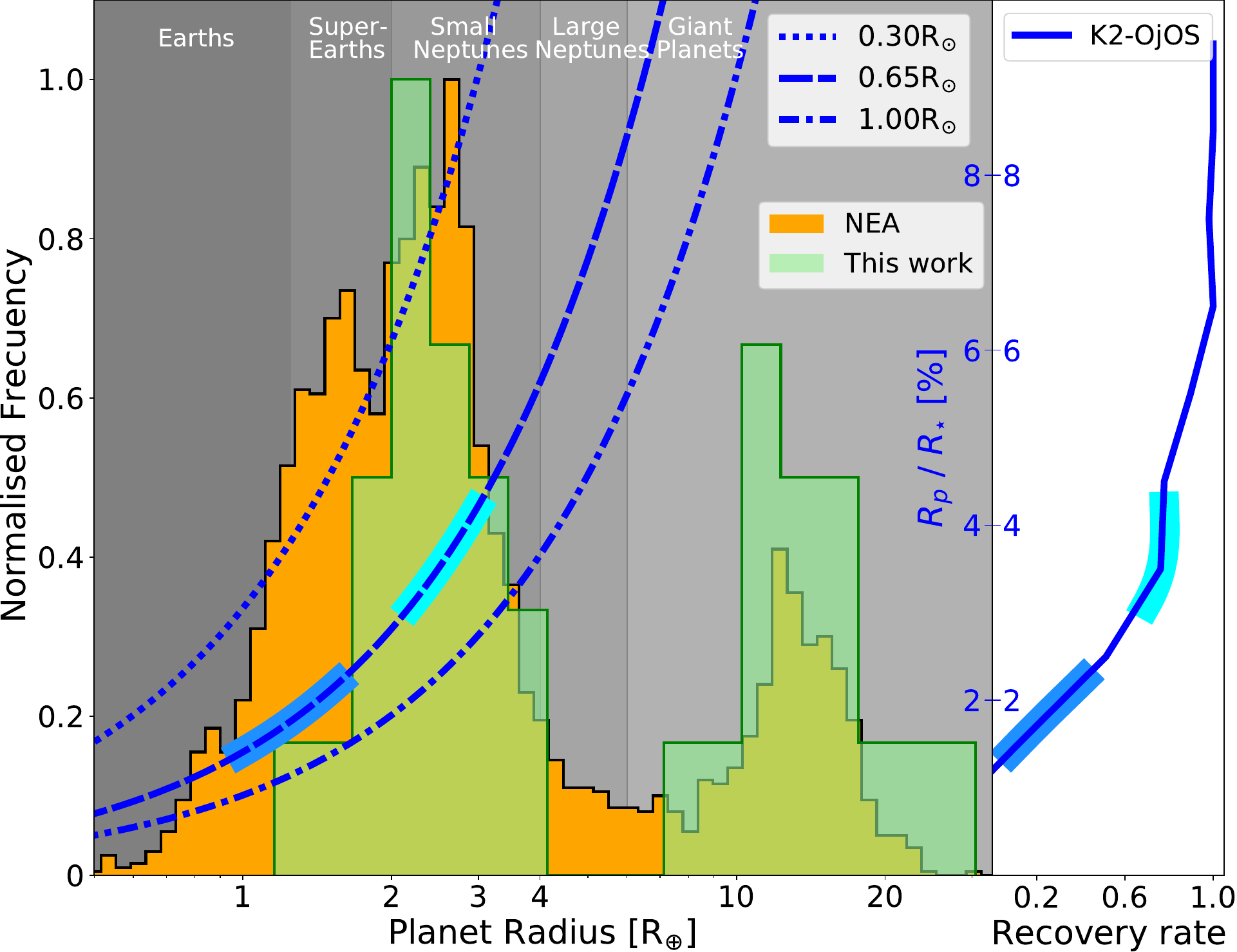}
    \caption{Left-hand panel: Histogram of planet radii for both the planets and candidates analysed in this work and the population of known planets. Superimposed we plot $R_{p} / R_{\star}$ (right y-axis) versus $R_{p}$ for three stellar radii. Right-hand panel: K2-OjOS recovery rates as a function of $R_{p} / R_{\star}$. In the left plot we highlight two sections of the  $R_{p} / R_{\star}$ versus $R_{p}$ curve of a 0.65 $\rm R_{\odot}$ star (dodger blue and cyan), each of them being on one side of the radius gap. In the right plot we can see the K2-OjOS recovery rates corresponding to the highlighted sections.}
    \label{fig:Rp_histogram}
    
\end{figure}

In Fig. \ref{fig:Rp_vs_P}, we plot the planet radius as a function of the orbital period for the planets and candidates analysed in this work, as well as for the current population of known planets. The latter are mainly grouped into two well differentiated clusters: the small planets cluster, which embraces those planets with $R_{p}$ < 4 $\rm R_{\rm \oplus}$ and orbital periods ranging from less than a day to hundreds of days, and the hot Jupiters cluster, which is composed of large planets ($R_{\rm p}$ > 10 $\rm R_{\rm \oplus}$) with short orbital periods ($P$ < 10 d). Despite the relative small size of our sample, the K2-OjOS findings alone match quite well with both clusters, especially if we look at the subsample of confirmed and validated planets. In Fig. \ref{fig:Rp_histogram} (left-hand panel) we plot the distribution of planet radii for both our planet and candidate sample and the population of known planets. The \textit{Kepler} and \textit{K2} findings showed that the most common type of planets belong to the small planets cluster \citep[e.g.][]{2012ApJS..201...15H,2013ApJS..204...24B,2013PNAS..11019273P,2015ApJ...809....8B}. Besides, their findings allowed to unveil a bimodality in the small planets distribution \citep{2017AJ....154..109F}, which shows a lower mode at $\sim$ 1.3 $\rm R_{\oplus}$ and a higher mode at $\sim$ 2.6 $\rm R_{\oplus}$, being both of them separated by the so-called radius gap ($\sim$ 1.9 $\rm R_{\oplus}$). Adopting the same planet size categories as \citet{2013ApJ...766...81F}, our sample of fully characterized planets and candidates contains one Earth (0.8 $\rm R_{\oplus}$ < $R_{\rm p}$ < 1.25 $\rm R_{\oplus}$), four super-Earths (1.25 $\rm R_{\oplus}$ < $R_{\rm p}$ < 2 $\rm R_{\oplus}$), 15 small Neptunes (2 $\rm R_{\oplus}$ < $R_{\rm p}$ < 4 $\rm R_{\oplus}$), and 15 giant planets ($R_{\rm p}$ > 6 $\rm R_{\oplus}$). Focusing on the small planet regime,  when compared to the population of known planets, we find a deficit of planets and candidates with radii smaller than that of the radius gap. We explain this deficit as a consequence of the great difficulty to detect in \textit{K2} data such small planets around stars as large as those in our sample. To illustrate this, in Fig. \ref{fig:Rp_histogram} (left-hand panel) we plot  the $R_{p}$/$R_{\star}$ ratios versus $R_{p}$ for different stellar radii. In the right-hand panel we plot the obtained K2-OjOS recovery rates as a function of $R_{p}$/$R_{\star}$ (see Sections \ref{sec:injection_recovery} and \ref{sec:detection_efficiency} for further details). We highlight two sections of the $R_{p}$/$R_{\star}$ vs $R_{p}$ curve of a 0.65 $\rm R_{\odot}$ star, which corresponds to the typical stellar radius in our sample. The section located in the lower mode of the small planets distribution corresponds to recovery rates between 8 and 40 per cent, while the section located in the higher mode corresponds to recovery rates between 70 and 80 per cent.



\begin{figure*}
    \centering
    \includegraphics[scale = 1]{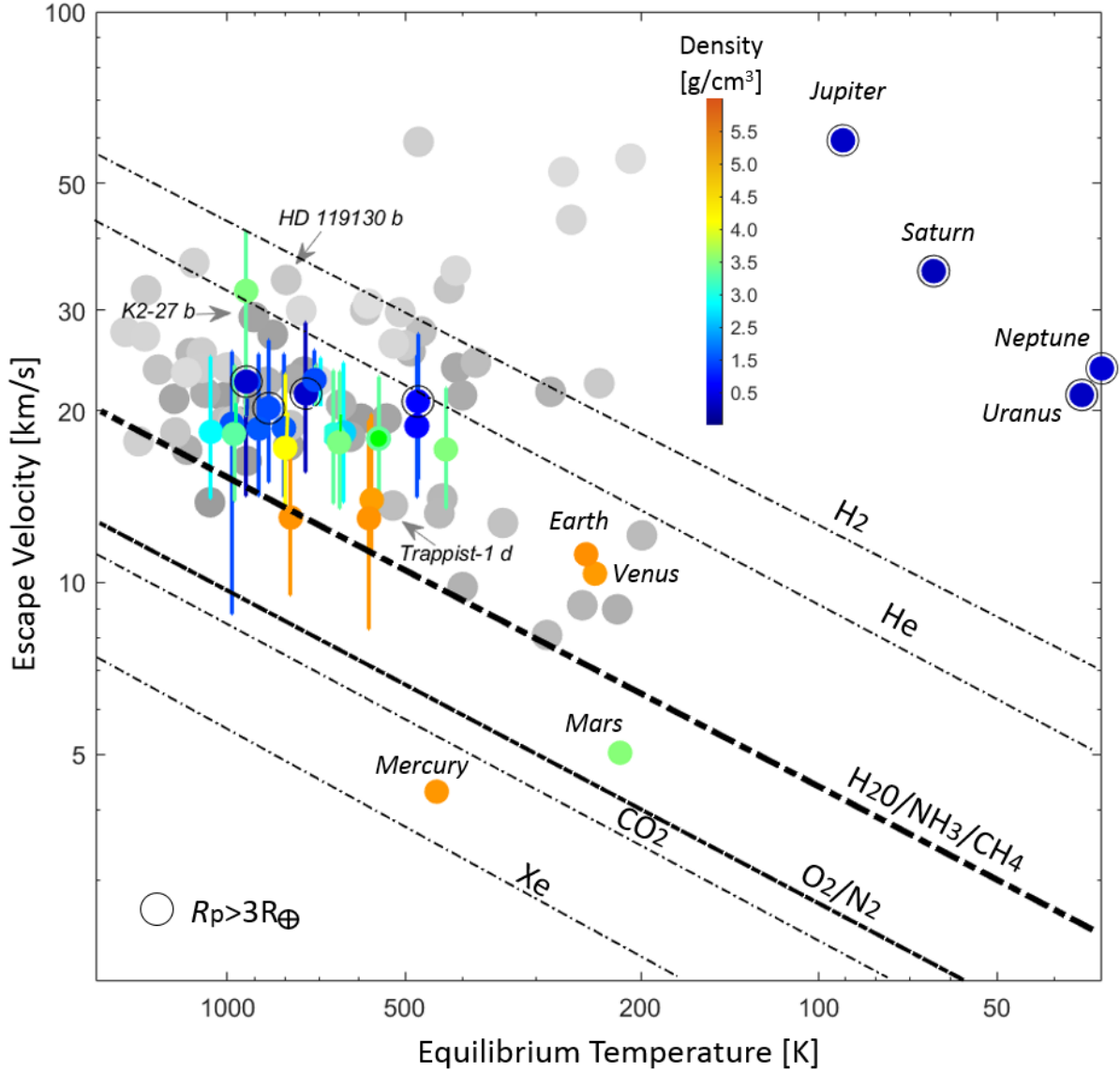}
    \caption{Atmospheric escape velocities as a function of the surface equilibrium temperatures for the small planets and candidates ($R_{\rm p}$ < 4 $\rm R_{\oplus}$) analysed in this work. Colour coding corresponds to the bulk estimated densities of the small exoplanets of this work and Solar System planets. 66 NEA confirmed and candidate planets with 0.5 $\rm R_{\oplus}$ < $R_{\rm p}$ < 4.0 $\rm R_{\oplus}$ with uncertainties lower than ±10 per cent (1-$\sigma$, the average error is about ±7 per cent) and masses determined by the radial velocity method are represented in grey ramp (low bulk density in light grey; high bulk density in dark grey). The average quadratic error of escape velocity is 22.6 per cent of the escape velocity of the Earth. Dot-dashed straight lines of 0.5 slope stand for threshold velocities of chemicals labelling each line.}
    \label{fig:esc_velo}
\end{figure*}

\subsubsection{Gas dwarfs versus water worlds: atmospheric escape velocities and retention of volatiles}

The puzzling bimodality in the population of planets with radii smaller than 4 $\rm R_{\oplus}$ is still a matter of debate \citep[see for example][and references therein]{Zeng9723}. This bimodality is consistent with the existence of two different types of planets. In the range 1-2 $\rm R_{\oplus}$, planets are known to be most likely rocky, whereas the internal compositions of planets between 2 and 4 $\rm R_{\oplus}$ is still an open question. They may either be gas dwarfs or water worlds, being the former planets with a rocky core and a prominent $\rm H_{2}$-He gaseous envelope, and the latter planets with significant amount of multicomponent, $\rm H_{2}O$-dominated ices/fluids in addition to rock and gas \citep[][]{Zeng9723}. Within our dataset, planets below 2 $\rm R_{\oplus}$ are too few to probe the radius gap. However, we can provide some insight into the possible composition of the planets we have detected in the higher mode. We briefly recall two recent models that explain the observed bimodal distribution leading to very different proposals on the composition and evolution of planets between 2 and 4 $\rm R_{\oplus}$. 

In the photoevaporation model, the bimodality is consistent with the theoretical valley predicted by evaporation numerical analysis \citep[][]{2013ApJ...775..105O,2014ApJ...795...65J,2014ApJ...792....1L,2016ApJ...831..180C}. Along the first 100 Myr of the star lifetime, high energy radiation (EUV and X-ray) would have completely stripped the primordial atmosphere of planets that we observe at the lower mode. As a result, gas dwarfs are proposed for planets within the higher mode. On the other side, \citet[][]{Zeng9723} were able to reproduce the two radii subpopulations of small exoplanets by means of a pebble accretion model independent on the planet growth mechanism. This model involves similar ice and rock contributions to the planet composition leading to water worlds rather than gas dwarfs for planets in the higher mode. The authors used Monte Carlo simulations to show that the radii bimodal distribution could arise from the dichotomy of rocky and icy cores.

\begin{figure*}
    \centering
    \includegraphics[scale=0.44]{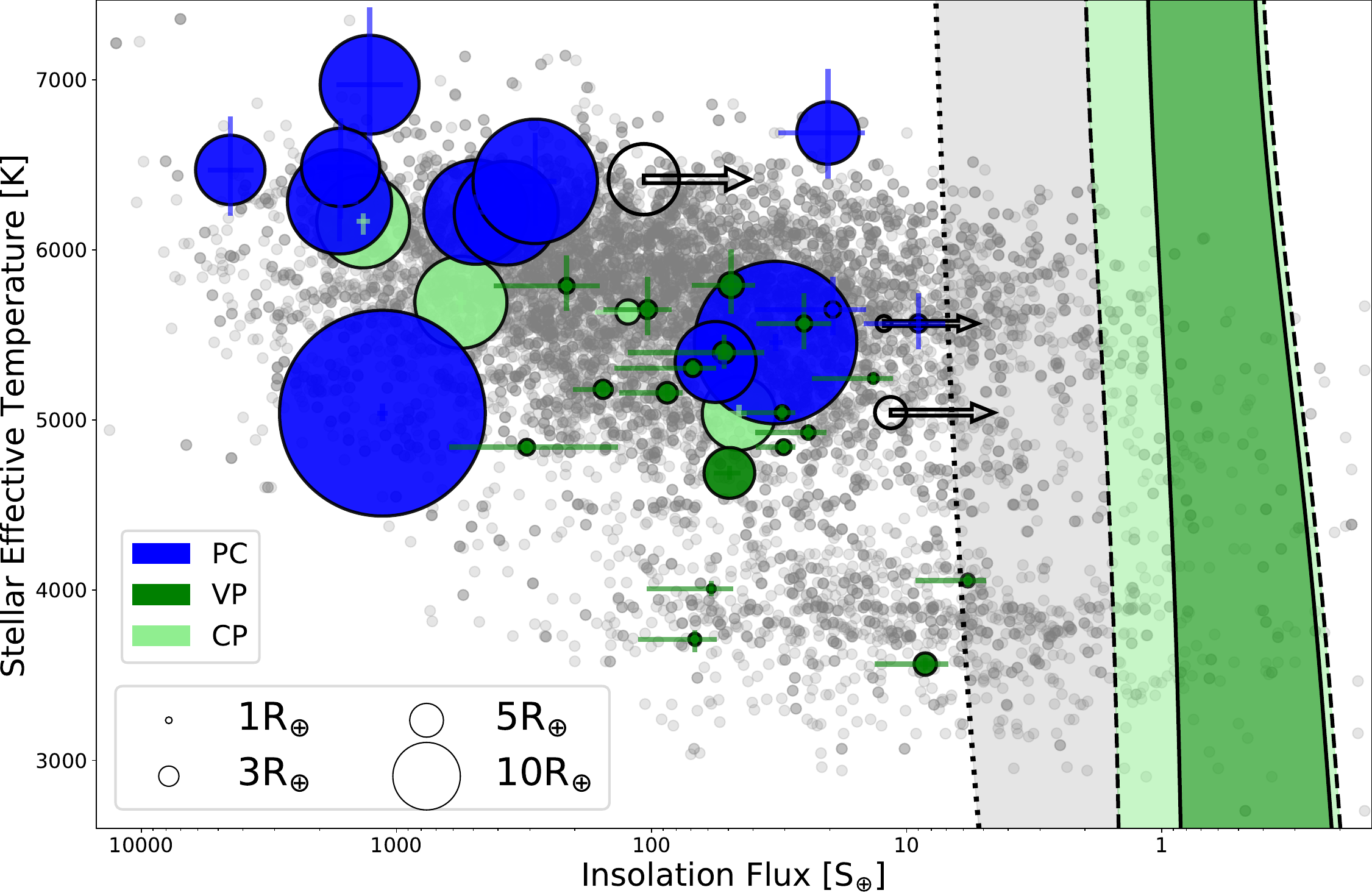}
    \caption{Stellar effective temperatures as a function of the insolation fluxes received by the corresponding planets and candidates. The data points are scaled to the planet radius and the colour coding indicates the dispositions. Open circles correspond to the maximum insolation fluxes for single transits, computed from their minimum orbital periods. The grey data points correspond to the population of known planets and candidates. The dark green region is bounded by two solid lines corresponding to the moist greenhouse inner edge and the maximum greenhouse outer edge. The light green region is bounded by two dashed lines corresponding to the recent Venus inner limit and early Mars outer limit. The \citet{2013ApJ...778..109Z} inner edge corresponds to the dotted line. }
    \label{fig:Teff_vs_S}
\end{figure*}

We now study whether the planets and candidates with radii between 2 and 4 $\rm R_{\oplus}$ analysed in this work can keep an atmospheric $\rm H_{2}$-He envelope over a billion-year timescale. Estimations of gas envelopes can be obtained following the strong correlation that atmospheric escape has with the escape velocities of planetary bodies and their atmospheric compositions in the Solar System. The same equations can be applied to estimate which gaseous species a planetary atmosphere can hold \citep[][]{Zeng9723}.

In Fig. \ref{fig:esc_velo}, we plot the escape velocities ($V_{\rm esc}$) of the planets and candidates with $R$ < 4 $\rm R_{\oplus}$ in our sample, as a function of their equilibrium temperatures ($T_{\rm eq}$). The markers are coloured as a function of the planet densities. Given the difficulty of measuring the masses of our faint host stars via radial velocity measurements, we estimate them from a mass-radius probabilistic algorithm implemented in the widely used program  \textsc{forecaster} \citep[][]{2017ApJ...834...17C}. We use these masses to derive both the escape velocities ($V_{\rm esc}=\sqrt{2GMR^{-1}}$) and planet bulk densities ($\rho = M/V$; $V= 4/3 \pi R^{3}$). The grey dashed dotted lines indicate the thermal escape thresholds of different molecular species. It is interesting to notice that the majority of our planets lie within a specific region with escape velocities of 20 $\pm$ 5 km $\rm s^{-1}$ and equilibrium temperatures between 500 and 1000 K. This particular region of the diagram is characterized by the thermal loss of H$_2$-He gas while the other components are retained. For escape velocities around 20 km $\rm s^{-1}$, one would expect rocky planets smaller and denser while gas dwarfs would be much bigger and with smaller densities than the planets in our sample. Altogether, the escape velocities, density and radii, allow us to tentatively propose, within the uncertainty derived by the use of estimated masses from \textsc{forecaster}, that our subsample of planets and candidates in the higher mode would be composed of water worlds since they would not be able to retain their primordial H$_2$/He envelopes.



\subsubsection{Potentially habitable systems}
\label{sec:habitability}

In order to assess whether any of the low-insolated planets and candidates analysed in this work could be inside the Habitable Zone (HZ) we computed the insolation flux boundaries derived from the \citet{2013ApJ...765..131K} climate model, as well as from the more optimistic \citet{2013ApJ...778..109Z} model. For the Solar System, the former define a conservative HZ between 0.99 AU and 1.70 AU, whereas the latter argues that in certain particular conditions ($\rm N_{2}$-dominated atmosphere, surface gravity of $g_{\rm surf}$ = 25 $\rm m\,s^{-2}$, surface pressure of $P_{\rm surf}$ = 1 bar, relative humidity of $\Phi$ = 1 per cent, surface albedo of $A$ = 0.8, and $\rm C0_{2}$ mixing ratio of $X_{\rm C0_{2}}$ = $\rm 10^{-4}$), the HZ inner edge (IHZ) can be as close as 0.38 AU. We used the polynomial relations from \citet{2013ApJ...765..131K} to determine for a wide range of effective temperatures (2600  K < $T_{\rm eff}$ < 7200 $\rm K$) the moist greenhouse inner edge and the maximum greenhouse outer edge, as well as the more optimistic limits of recent Venus inner limit and early Mars outer limit (see Fig. \ref{fig:Teff_vs_S}). The \citet{2013ApJ...778..109Z} model analytical expression is defined within the distance-luminosity parameter space. To transform the model into the $S_{\rm eff}$-$T_{\rm eff}$ parameter space we used the semi-analytical formulas for the Hertzsprung-Russell diagram from \citet{2008SerAJ.177...73Z}.  None of the analysed planets or candidates in this work belong to the \citet{2013ApJ...765..131K} HZ, but there is one validated planet (K2-103 b, EPIC 211822797.01) whose orbit is located slightly further than \citet{2013ApJ...778..109Z} IHZ. We discuss the habitability of this planet in Section \ref{sec:EPIC_211822797}.

\subsection{Highlights of five individual systems}
\label{sec:highlights_individual_systems}

\subsubsection{A 2:1 Period commensurability on the new planetary system K2-356 (EPIC 211537087)}
\label{epic_211537087}

\begin{figure}
    \includegraphics[scale=0.50]{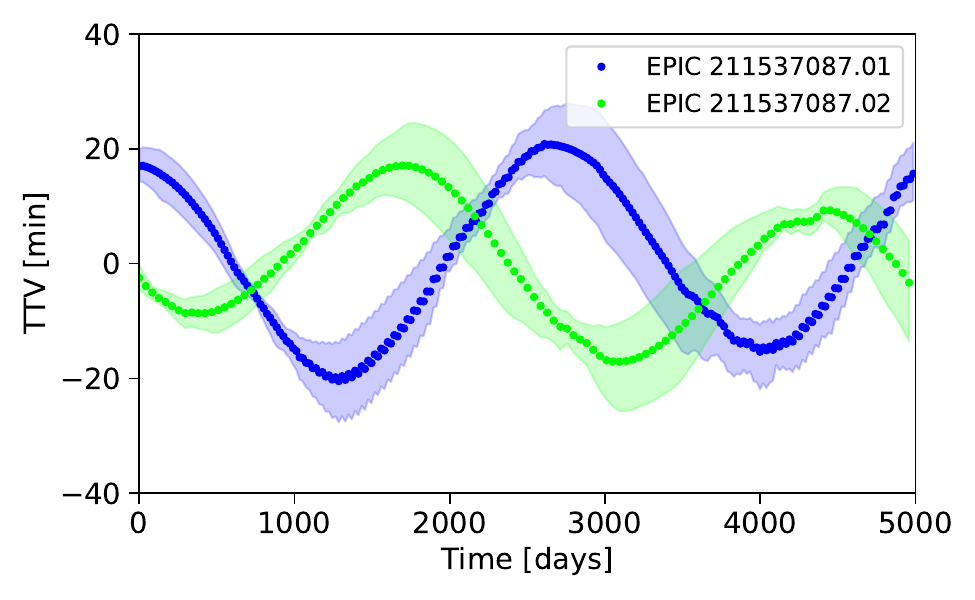}
    \caption{Predicted TTVs ranges with an amplitude of around 20 min for EPIC 211537087.01 (K2-356~b) and EPIC 211537087.02 estimated via forward modelling with \textsc{rebound} assuming circular orbits and planet masses of $M_\mathrm{01}$\,=\,6.1$^{+4.6}_{-2.7}\,\mathrm{M_{\oplus}}$ and $M_\mathrm{02}$\,=\,7.3$^{+5.5}_{-3.2}\,\mathrm{M_{\oplus}}$ derived from the probabilistic mass-radius relation from \textsc{forecaster}. The period of the interaction, the so-called TTV-period or cycle period, is of $\sim$2780\,d which agrees with the theoretically estimated value of $\sim$2700\,d. The coloured shaded area mark the uncertainties in the expected TTVs based on uncertainties in the planetary masses, orbital periods, and $T_{\rm 0}$.}
    \label{fig:ttv_sim}
\end{figure}

The K2-OjOS team detected a new planetary system of three planet candidates transiting around K2-356 (EPIC 211537087), a G-type star with $K_{p}$ = 13.44, $T_{\rm eff}$ = $5568$ K, $R$ = $0.86$ $\rm R_{\odot}$, and $M$ = $0.90$ $\rm M_{\odot}$. The three candidates are small Neptunes; EPIC 211537087.01 has $R_{\rm p}$ = 2.29 $\rm R_{\oplus}$ with a 21.03-d orbital period, EPIC 211537087.02 has $R_{\rm p}$ = 2.58 $\rm R_{\oplus}$ with a 42.38-d orbital period, and EPIC 211537087.03 is a single transit with $R_{\rm p}$ = 2.28 $\rm R_{\oplus}$ and orbital period greater than 41.9 d. EPIC 211537087.01 and EPIC 211537087.02 orbit near a 2:1 period commensurability. The analysis of the AstraLux High resolution image results in a very low probability of the target having a BEB: 0.15 per cent (see Section \ref{sec:AstraLux} for further details). Although both EPIC 211537087.01 and EPIC 211537087.02 show false positive probabilities lower than the required threshold to validate a planet, we only validate EPIC 211537087.01 (K2-356 b) as EPIC 211537087.02 shows only two transits within the light curve. However, we argue that the origin of EPIC 211537087.02 as well as EPIC 211537087.03 must be planetary, given the extremely low probability of finding multiple false positive signals \citep[][]{2010arXiv1006.3727R,2011ApJS..197....8L}. Besides, for planet candidates which have a period ratio near a first-order mean motion resonance the probability of both signals being true planets is even higher, since such resonances would not be seen for random eclipsing binaries \citep{2011ApJS..197....8L}.

As TTVs are more sensitive to planets near resonant orbits and the number of transits (three and two for EPIC 211537087.01 and EPIC 211537087.02, respectively) is not sufficient to detect any TTVs, we computed the theoretical TTVs for this system. We estimated a TTV period of $\sim$2700 d using the analytical formula described in \citet{2012ApJ...761..122L}. To have an idea of the expected TTV amplitude, we carried out n-body simulations using \textsc{rebound} \citep{2012A&A...537A.128R}, assuming circular orbits and adopting the values for the orbital periods and mid-transit times from Table \ref{tab:planet_params}. The values for the masses ($M_\mathrm{01}$\,=\,6.1$^{+4.6}_{-2.7}\,\mathrm{M_{\oplus}}$ and $M_\mathrm{02}$\,=\,7.3$^{+5.5}_{-3.2}\,\mathrm{M_{\oplus}}$) were estimated using the mass-radius relation implemented in \textsc{forecaster} \citep{2017ApJ...834...17C}, starting from the planet radii (2.29\,$\pm$\,0.2\,$\mathrm{R_{\oplus}}$ and 2.58\,$\pm$\,0.2\,$\mathrm{R_{\oplus}}$) for EPIC 211537087.01 (K2-356~b) and EPIC 211537087.02.
The simulations predict a TTV period of $\sim$2780 d, confirming the analytical estimations, and TTV amplitudes of $\sim$20 min (Fig.~\ref{fig:ttv_sim}). The influence of the third planet candidate EPIC 211537087.03 (for which the orbital period is unknown) was not considered, which could affect our predictions of both the TTV amplitude and the TTV period.

\begin{figure}
    \includegraphics[scale=0.5]{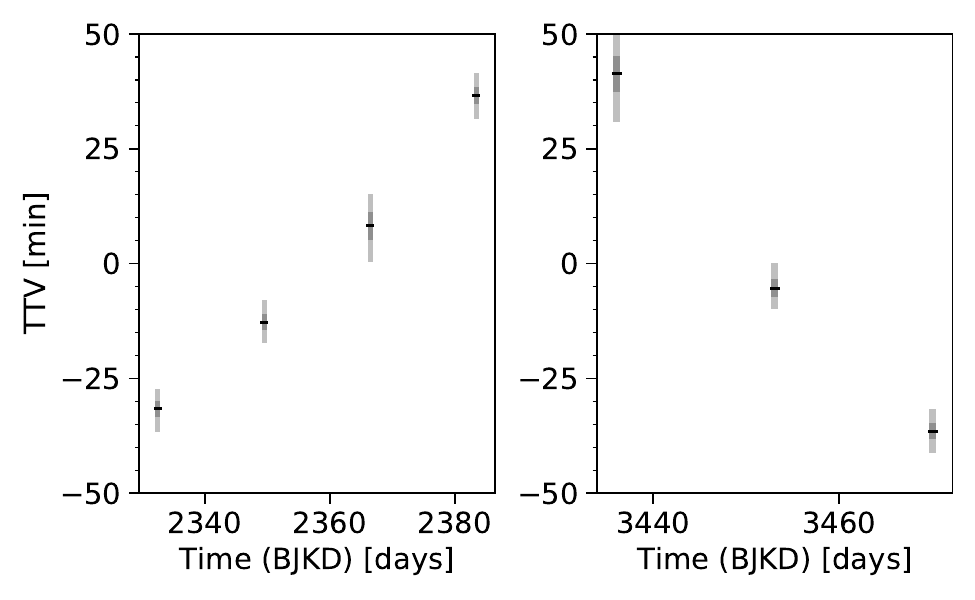}
    \caption{Transit timing variations for K2-184~b observed in C5 (left-hand panel) and C18 (right-hand panel). The black lines mark the median values and the 68 per cent and 99 per cent central posterior percentiles are indicated by the dark and light shaded area, respectively.}
    \label{fig:ttv}
\end{figure}

\begin{figure}
    \includegraphics[scale=0.5]{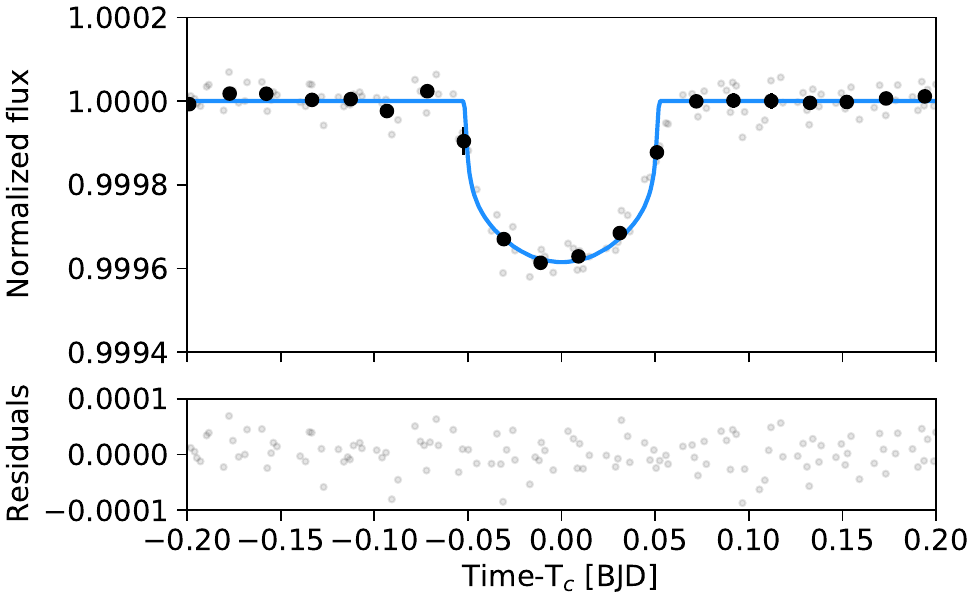}
    \caption{Upper panel: Phase-folded transits of K2-184 b (EPIC 211594205.01) with its orbital period of 16.97 d. Each transit is shifted by their mid-transit time estimated by \textsc{pyttv}. The black points are the binned data points with a 30-min binning. The blue line is the transit model. Lower panel: fit residuals. Note that in-transit and out-of-transit data show the same noise properties, indicating that the data are well fitted.}
    \label{fig:phase_folded_K2184}
\end{figure}

\subsubsection{Transit timing variations on the bright star (V = 10.35) K2-184 (EPIC 211594205)}

\label{sec:EPIC_211594205}

K2-184 b (EPIC 211594205.01)  is a super-Earth with $R_{\rm p}$ = 1.5 $\rm R_{\oplus}$, which orbits a late G-type star ($K_{p}$ = 10.68, $T_{\rm eff}$ = $5245$ K, $R$ = $0.75$ $\rm R_{\odot}$, and $M$ = $0.84$ $\rm M_{\odot}$) with an orbital period of 16.98 d. It was published as a candidate by \citet{2016A&A...594A.100B} and \citet{2016MNRAS.461.3399P}, and latter validated by \citet{2018AJ....155..136M} and \citet{2018AJ....156..277L}. Other works that studied this planet are \citet{2018AJ....155...21P} and \citet{2019ApJS..244...11K}. In this work, the TTV analysis on the joined C5 and C18 data found significant sinusoidal TTVs (Fig.~\ref{fig:ttv}) that hint to the existence of an additional non-transiting planet, since no other planet was detected in the light curve which could produce the detected TTVs. Unfortunately, there is no chance to derive characteristics of the TTV signal, e.g. TTV amplitude or TTV period, because of an insufficient coverage of the TTV period due to the lack of transit observations. The phase-folded transit accounting for the TTVs together with the best-fitting model is shown in Fig.~\ref{fig:phase_folded_K2184}. In Table \ref{tab:planet_params} we report the median and 68.3 per cent credible intervals of the main planetary parameters.

The great brightness of the target (V = 10.35) makes it very appropriate for photometric and/or RV follow-up. We predicted the planet mass through \textsc{forecaster} and then estimated the RV semi-amplitude, obtaining $K \sim$ 1.3 m $\rm s^{-1}$, which is achievable by the current precision spectrographs.  

\subsubsection{A new single transit on K2-274 (EPIC 212008766)}

K2-274 b (EPIC 212008766.01) is a planet with $R_{p}$ = 2.1 $\rm R_{\oplus}$, which orbits an early K-type star ($K_{p}$ = 12.80, $T_{\rm eff}$ = $5044$ K, $R$ = $0.70$ $\rm R_{\odot}$, and $M$ = $0.79$ $\rm M_{\odot}$) with an orbital period of 14.13 d. It was published as a candidate in 2016 November by \citet{2016A&A...594A.100B} and \citet{2016MNRAS.461.3399P}, and in 2016 December by \citet{2016MNRAS.463.1780L}. In 2018 it was first studied for validation by \citet{2018AJ....155..136M}, who did not validate the planet with a FFP = 0.15 per cent due to their more conservative considered threshold (FPP < 0.1 per cent). Later, \citet{2018AJ....156..277L} validated the planet with a computed FPP of 0.03 per cent. Interestingly, the authors found that the photometric pipeline they used (k2phot) includes a nearby contaminant star within the aperture, the same that we found for the EVEREST pipeline, which made us choose the smaller K2SFF aperture for the subsequent analysis. Other works that studied this planet are \citet{2018AJ....155...21P} and \citet{2019ApJS..244...11K}. 

All the aforementioned works detected and analysed the planet starting from C5 data alone. For this work, the K2-OjOS team retrieved the signal in C18 allowing us to model the photometry by joining both C5 and C18 datasets. Besides, the team detected a new single-transit event at 3443.86 BKJD with a 0.15 per cent dip, hinting the existence of an additional long-period planet in the system. Our transit modelling corresponds to a planet candidate of 4.3 $\rm R_{\oplus}$. It is remarkable the V-shape of the transit, which is consistent with a grazing transit. This is to be expected for planets with large semimajor axis, as for a given orbital inclination $b$ $\propto$ $a$. Given the non-presence of such deep dip within the C5 continuous photometry, we constrain the orbital period for EPIC 212008766.02 to be greater than 74.8 days. If confirmed, it would be the second longest period planet detected by \textit{K2}.

\subsubsection{Habitability of K2-103 b (EPIC 211822797.01)}
\label{sec:EPIC_211822797}

K2-103 b (EPIC 211822797.01)  was first detected as a candidate by \citet{2016A&A...594A.100B} and latter validated by \citet{2017AJ....153...64M} and \citet{2017AJ....154..207D} starting from C5 data alone. Other works that analysed the planet are \citet{2017AJ....154..224R} and \citet{2019ApJS..244...11K}. The planet orbits around EPIC 211822797, a late K-type dwarf with $K_{p}$ = 14.57,  $T_{\rm eff}$ = $4057$ K, $R$ = $0.58$ $\rm R_{\odot}$, and $M$ = $0.62$ $\rm M_{\odot}$. Our planet transit modelling of C5, C16, and C18 photometry results in an orbital period of 21.17 d and a radius of $R_{\rm p}$ = 1.92 $\rm R_{\oplus}$, which locates the planet inside the radius gap. The late spectral type of the host star together with the relative long orbital period, causes that this planet receives the least amount of insolation flux of the planets and candidates in our sample: 5.75 $\rm S_{\oplus}$. Although this flux is not low enough to consider the planet within the HZ according to \citet{2013ApJ...765..131K} model, if we consider the optimistic conditions proposed by \citet{2013ApJ...778..109Z} (Section \ref{sec:habitability}), this planet would be within the Habitable Zone of its star (see Fig. \ref{fig:Teff_vs_S}). In terms of distances, the \citet{2013ApJ...778..109Z} inner edge is located at 0.116 AU, while the semimajor axis of the planet is $a$ = 0.120 AU.

\subsubsection{Disposition of K2-120 b (EPIC 211791178.01)}

\begin{figure}
    \centering
    \includegraphics[scale=0.52]{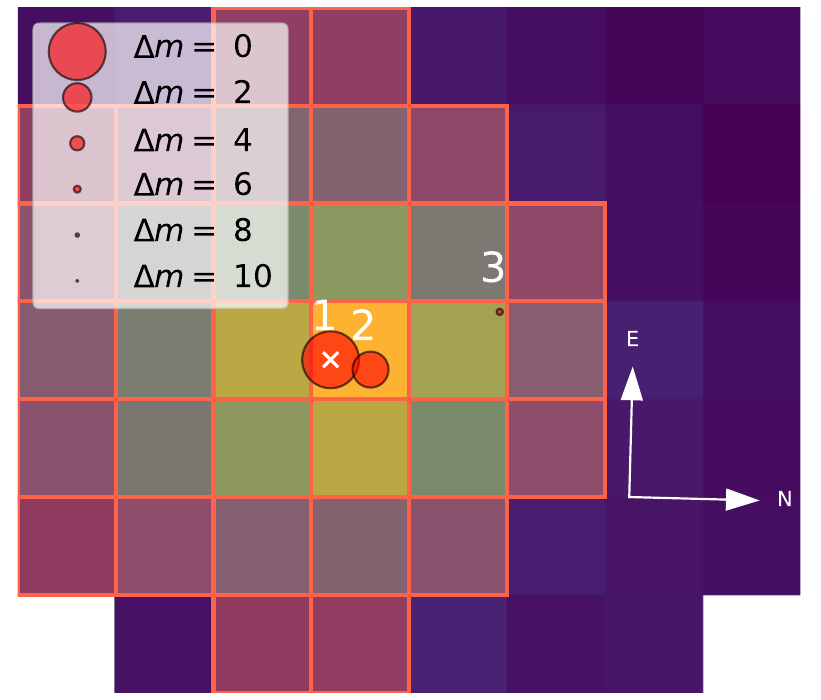}
    \caption{Target pixel file of EPIC 211791178 (star $\#$1) together with the EVEREST aperture and nearby \textit{Gaia} DR2 sources.}
    \label{fig:tpfplotter_K2_120}
\end{figure}

K2-120 b (EPIC 211791178.01) was a validated planet retrieved by the K2-OjOS visual searching. As detailed in Section \ref{validation}, we carried out a complete statistical validation analysis for our newly detected candidates, but not for the retrieved known planets and candidates, which were already subjected to validation. This is mainly motivated by the fact that simply adding photometry does not affect crucially the transit shape as for perturbing \textsc{vespa} dispositions. However, we performed some routine checks for all the 37 analysed targets (e.g. revisit photometry with different apertures and/or pipelines, search for nearby contaminant stars, and search for secondary eclipses). As a result, we detected for this target a bright contaminant star within the photometric aperture. In Fig. \ref{fig:tpfplotter_K2_120} we show the locations of EPIC 211791178 (source $\#$1, $G_{mag}$ = 13.9) and \textit{Gaia} DR2 659785145072281600 (source $\#$2, $G_{mag}$ = 15.3) within the TPF. These sources are separated 1.67 arcsec from each other, with $\Delta m$ = 1.33, making any type of multi-aperture analysis pointless. Their \textit{Gaia} DR2 measured parallaxes and proper motions are almost identical: $\pi = 3.45 \pm 0.03$ mas, $\mu_{\alpha}$ = $-31.30 \pm 0.06$ $\rm mas \times yr ^{-1}$, $\mu_{\delta}$ = $-20.02 \pm 0.03$ $\rm mas \times yr ^{-1}$ for source $\#$1, and $\pi = 3.28 \pm 0.05$ mas, $\mu_{\alpha}$ = $-31.51 \pm 0.09$ $\rm mas \times yr ^{-1}$, $\mu_{\delta}$ = $-19.19 \pm 0.05$ $\rm mas \times yr ^{-1}$ for source $\#$2, which indicates that both sources form a binary system.

Given the observed transit depth of $\sim$0.077 per cent, if the validated signal comes from source $\#$1, the flux would be diluted by $\gamma_{\rm pri}$ = 1.29, whereas if the signal comes from source $\#$2, the dilution factor would be $\gamma_{\rm sec}$ = 4.40. This yields a dilution-corrected planet radius of 2.80 $\rm R_{\oplus}$ for the first case, and 5.16 $\rm R_{\oplus}$ for the second case, in contrast to the current value of 2.48 $\rm R_{\oplus}$. Therefore, following the criteria in Section \ref{sec:FPP_reliability}, as the origin of the signal cannot be ascertained, we consider EPIC 211791178.01 as a planet candidate. 

\section{S\,u\,m\,m\,a\,r\,y}
\label{summary}

We have presented the first results of K2-OjOS, a Pro-Am project primarily dedicated to search, characterize, and validate new extrasolar planets. In this work, a group of 10 amateur astronomers visually inspected the 20\,427 light curves of \textit{K2} C18 and performed a preliminary vetting of the signals found jointly with professional astronomers, resulting in 42 planet candidates in 37 systems. We characterized homogeneously all the host stars starting from published spectroscopic parameters, photometry, and parallaxes. We modelled the transit signals by joining \textit{K2} photometry from C18 and the overlapping C5 and C16 when available, and searched for TTVs in the joined dataset. An exhaustive search revealed that 24 of the findings had been previously published in several works devoted to analyse C5 data alone, while the remaining 18 are new detections. For the former, we refined their ephemeris by joining C5, C16, and C18 data, managing to decrease their uncertainties by a median factor of \medianfactorP\, for $P$, and \medianfactorTo for $T_{0}$. For the latter, we carried out a careful statistical validation analysis that resulted in four new validated planets (K2-355 b, K2-356 b, K2-357 b, K2-358 b) and 14 planet candidates. For the planet sample with 2 $\rm R_\oplus$ < $R_{p}$ < 4 $\rm R_\oplus$, their escape velocities and densities computed from estimated masses, suggest a composition compatible with water worlds rather than gas dwarfs.  Regarding individual systems, we highlight the presence of a 2:1 period commensurability in the new detected system K2-356, the detection of significant TTVs in the bright star K2-184 (V = 10.35), the location of K2-103 b inside the HZ according to optimistic models, the detection of a new single transit in the known system K2-274, and the disposition reassignment of K2-120 b,
which we consider as a planet candidate as the origin of the signal cannot be ascertained. 

Although exoplanetary research is moving from mass detections towards a more comprehensive characterization and understanding of individual systems, the works aimed to detect large amounts of planets and candidates are of great value. These works greatly increase the statistical information of the population of planets in the Galaxy, and allow the follow-up studies to have a greater diversity of planetary systems from which to choose to invest telescope and economic resources. In this context, well coordinated Pro-Am projects in which amateur and professional astronomers join forces can play an important role.

\section*{A\,c\,k\,n\,o\,w\,l\,e\,d\,g\,e\,m\,e\,n\,t\,s}

AC-G, EDA, SLSG, CGG, FGR, and JCJ would like to acknowledge Spanish ministry project MINECO AYA2017-89121-Pystems. LB and JG-N acknowledge financial support from the PGC 2018 project PGC2018-101948-B-I00 (MICINN and FEDER). JMR acknowledges financial support from project PGC2018-094814-B-C22 (MICINN and FEDER). JL-B acknowledges financial support received from  ‘la Caixa' Foundation (ID 100010434) and from the European Union’s Horizon 2020 research and innovation programme under the Marie Skłodowska-Curie grant agreement No 847648, with fellowship code LCF/BQ/PI20/11760023. JK gratefully acknowledges the support of the Swedish National Space Agency (DNR 2020-00104).


This paper includes data collected by the Kepler mission and obtained from the MAST data archive at the Space Telescope Science Institute (STScI). Funding for the Kepler mission is provided by the NASA Science Mission Directorate. STScI is operated by the Association of Universities for Research in Astronomy, Inc., under NASA contract NAS 5–26555.


This work has made use of data from the European Space Agency (ESA) mission
{\it Gaia} (\url{https://www.cosmos.esa.int/gaia}), processed by the {\it Gaia}
Data Processing and Analysis Consortium (DPAC,
\url{https://www.cosmos.esa.int/web/gaia/dpac/consortium}). Funding for the DPAC
has been provided by national institutions, in particular the institutions
participating in the {\it Gaia} Multilateral Agreement.


This research has made use of the NASA Exoplanet Archive, which is operated by the California Institute of Technology, under contract with the National Aeronautics and Space Administration under the Exoplanet Exploration Program.


This research has made use of the Exoplanet Follow-up Observation Program website, which is operated by the California Institute of Technology, under contract with the National Aeronautics and Space Administration under the Exoplanet Exploration Program.

This work made use of \texttt{tpfplotter} by J. Lillo-Box (publicly available in \url{www.github.com/jlillo/tpfplotter}), which also made use of the python packages \texttt{astropy}, \texttt{lightkurve}, \texttt{matplotlib} and \texttt{numpy}.

We are very grateful to Pia Valentin Sørensen (\url{https://www.artmajeur.com/pia-valentin-sorensen}) for designing the K2-OjOS logo and stunning illustrations for the outreach activities.  

\section*{Data availability}

The data underlying this article will be shared on reasonable request to the corresponding author.




\bibliographystyle{mnras}
\bibliography{refs} 

\begin{thebibliography}{}
\makeatletter
\relax
\def\mn@urlcharsother{\let\do\@makeother \do\$\do\&\do\#\do\^\do\_\do\%\do\~}
\def\mn@doi{\begingroup\mn@urlcharsother \@ifnextchar [ {\mn@doi@}
  {\mn@doi@[]}}
\def\mn@doi@[#1]#2{\def\@tempa{#1}\ifx\@tempa\@empty \href
  {http://dx.doi.org/#2} {doi:#2}\else \href {http://dx.doi.org/#2} {#1}\fi
  \endgroup}
\def\mn@eprint#1#2{\mn@eprint@#1:#2::\@nil}
\def\mn@eprint@arXiv#1{\href {http://arxiv.org/abs/#1} {{\tt arXiv:#1}}}
\def\mn@eprint@dblp#1{\href {http://dblp.uni-trier.de/rec/bibtex/#1.xml}
  {dblp:#1}}
\def\mn@eprint@#1:#2:#3:#4\@nil{\def\@tempa {#1}\def\@tempb {#2}\def\@tempc
  {#3}\ifx \@tempc \@empty \let \@tempc \@tempb \let \@tempb \@tempa \fi \ifx
  \@tempb \@empty \def\@tempb {arXiv}\fi \@ifundefined
  {mn@eprint@\@tempb}{\@tempb:\@tempc}{\expandafter \expandafter \csname
  mn@eprint@\@tempb\endcsname \expandafter{\@tempc}}}

\bibitem[\protect\citeauthoryear{{Akeson} et~al.,}{{Akeson}
  et~al.}{2013}]{2013PASP..125..989A}
{Akeson} R.~L.,  et~al., 2013, \mn@doi [\pasp] {10.1086/672273}, \href
  {https://ui.adsabs.harvard.edu/abs/2013PASP..125..989A} {125, 989}

\bibitem[\protect\citeauthoryear{{Aller}, {Lillo-Box}, {Jones}, {Miranda}  \&
  {Barcel{\'o} Forteza}}{{Aller} et~al.}{2020}]{2020A&A...635A.128A}
{Aller} A.,  {Lillo-Box} J.,  {Jones} D.,  {Miranda} L.~F.,   {Barcel{\'o}
  Forteza} S.,  2020, \mn@doi [\aap] {10.1051/0004-6361/201937118}, \href
  {https://ui.adsabs.harvard.edu/abs/2020A&A...635A.128A} {635, A128}

\bibitem[\protect\citeauthoryear{{Barrag{\'a}n}, {Gandolfi}  \&
  {Antoniciello}}{{Barrag{\'a}n} et~al.}{2019}]{2019MNRAS.482.1017B}
{Barrag{\'a}n} O.,  {Gandolfi} D.,   {Antoniciello} G.,  2019, \mn@doi [\mnras]
  {10.1093/mnras/sty2472}, \href
  {https://ui.adsabs.harvard.edu/abs/2019MNRAS.482.1017B} {482, 1017}

\bibitem[\protect\citeauthoryear{{Barros}, {Demangeon}  \& {Deleuil}}{{Barros}
  et~al.}{2016}]{2016A&A...594A.100B}
{Barros} S.~C.~C.,  {Demangeon} O.,   {Deleuil} M.,  2016, \mn@doi [\aap]
  {10.1051/0004-6361/201628902}, \href
  {https://ui.adsabs.harvard.edu/abs/2016A&A...594A.100B} {594, A100}

\bibitem[\protect\citeauthoryear{{Batalha}}{{Batalha}}{2014}]{2014PNAS..11112647B}
{Batalha} N.~M.,  2014, \mn@doi [Proceedings of the National Academy of
  Science] {10.1073/pnas.1304196111}, \href
  {https://ui.adsabs.harvard.edu/abs/2014PNAS..11112647B} {111, 12647}

\bibitem[\protect\citeauthoryear{{Batalha} et~al.,}{{Batalha}
  et~al.}{2013}]{2013ApJS..204...24B}
{Batalha} N.~M.,  et~al., 2013, \mn@doi [\apjs] {10.1088/0067-0049/204/2/24},
  \href {https://ui.adsabs.harvard.edu/abs/2013ApJS..204...24B} {204, 24}

\bibitem[\protect\citeauthoryear{{Bhatti}, {Bouma}, {Joshua}, {John}  \&
  {Price-Whelan}}{{Bhatti} et~al.}{2018}]{astrobase}
{Bhatti} W.,  {Bouma} L.,  {Joshua} {John}  {Price-Whelan} A.,  2018,
  {Waqasbhatti/Astrobase: Astrobase V0.3.20}, \mn@doi{10.5281/zenodo.1469822}

\bibitem[\protect\citeauthoryear{{Boisse} et~al.,}{{Boisse}
  et~al.}{2013}]{Boisse_2013}
{Boisse} I.,  et~al., 2013, \mn@doi [\aap] {10.1051/0004-6361/201220993}, \href
  {https://ui.adsabs.harvard.edu/abs/2013A&A...558A..86B} {558, A86}

\bibitem[\protect\citeauthoryear{{Borucki} et~al.,}{{Borucki}
  et~al.}{2010}]{2010Sci...327..977B}
{Borucki} W.~J.,  et~al., 2010, \mn@doi [Science] {10.1126/science.1185402},
  \href {https://ui.adsabs.harvard.edu/abs/2010Sci...327..977B} {327, 977}

\bibitem[\protect\citeauthoryear{{Boyajian} et~al.,}{{Boyajian}
  et~al.}{2012}]{2012ApJ...757..112B}
{Boyajian} T.~S.,  et~al., 2012, \mn@doi [\apj] {10.1088/0004-637X/757/2/112},
  \href {https://ui.adsabs.harvard.edu/abs/2012ApJ...757..112B} {757, 112}

\bibitem[\protect\citeauthoryear{{Brahm} et~al.,}{{Brahm}
  et~al.}{2016}]{2016PASP..128l4402B}
{Brahm} R.,  et~al., 2016, \mn@doi [\pasp] {10.1088/1538-3873/128/970/124402},
  \href {https://ui.adsabs.harvard.edu/abs/2016PASP..128l4402B} {128, 124402}

\bibitem[\protect\citeauthoryear{{Burke} et~al.,}{{Burke}
  et~al.}{2015}]{2015ApJ...809....8B}
{Burke} C.~J.,  et~al., 2015, \mn@doi [\apj] {10.1088/0004-637X/809/1/8}, \href
  {https://ui.adsabs.harvard.edu/abs/2015ApJ...809....8B} {809, 8}

\bibitem[\protect\citeauthoryear{{Cabrera} et~al.,}{{Cabrera}
  et~al.}{2017}]{2017A&A...606A..75C}
{Cabrera} J.,  et~al., 2017, \mn@doi [\aap] {10.1051/0004-6361/201731233},
  \href {https://ui.adsabs.harvard.edu/abs/2017A&A...606A..75C} {606, A75}

\bibitem[\protect\citeauthoryear{{Campante} et~al.,}{{Campante}
  et~al.}{2015}]{2015ApJ...799..170C}
{Campante} T.~L.,  et~al., 2015, \mn@doi [\apj] {10.1088/0004-637X/799/2/170},
  \href {https://ui.adsabs.harvard.edu/abs/2015ApJ...799..170C} {799, 170}

\bibitem[\protect\citeauthoryear{{Castro Gonz{\'a}lez} et~al.,}{{Castro
  Gonz{\'a}lez} et~al.}{2020a}]{2020sea..confE..97C}
{Castro Gonz{\'a}lez} A.,  et~al., 2020a, in Contributions to the XIV.0
  Scientific Meeting (virtual) of the Spanish Astronomical Society. p.~97

\bibitem[\protect\citeauthoryear{{Castro Gonz{\'a}lez} et~al.,}{{Castro
  Gonz{\'a}lez} et~al.}{2020b}]{2020MNRAS.499.5416C}
{Castro Gonz{\'a}lez} A.,  et~al., 2020b, \mn@doi [\mnras]
  {10.1093/mnras/staa2353}, \href
  {https://ui.adsabs.harvard.edu/abs/2020MNRAS.499.5416C} {499, 5416}

\bibitem[\protect\citeauthoryear{{Chen} \& {Kipping}}{{Chen} \&
  {Kipping}}{2017}]{2017ApJ...834...17C}
{Chen} J.,  {Kipping} D.,  2017, \mn@doi [\apj] {10.3847/1538-4357/834/1/17},
  \href {https://ui.adsabs.harvard.edu/abs/2017ApJ...834...17C} {834, 17}

\bibitem[\protect\citeauthoryear{{Chen} \& {Rogers}}{{Chen} \&
  {Rogers}}{2016}]{2016ApJ...831..180C}
{Chen} H.,  {Rogers} L.~A.,  2016, \mn@doi [\apj]
  {10.3847/0004-637X/831/2/180}, \href
  {https://ui.adsabs.harvard.edu/abs/2016ApJ...831..180C} {831, 180}

\bibitem[\protect\citeauthoryear{{Choi}, {Dotter}, {Conroy}, {Cantiello},
  {Paxton}  \& {Johnson}}{{Choi} et~al.}{2016}]{2016ApJ...823..102C}
{Choi} J.,  {Dotter} A.,  {Conroy} C.,  {Cantiello} M.,  {Paxton} B.,
  {Johnson} B.~D.,  2016, \mn@doi [\apj] {10.3847/0004-637X/823/2/102}, \href
  {https://ui.adsabs.harvard.edu/abs/2016ApJ...823..102C} {823, 102}

\bibitem[\protect\citeauthoryear{{Christiansen} et~al.,}{{Christiansen}
  et~al.}{2018}]{2018AJ....155...57C}
{Christiansen} J.~L.,  et~al., 2018, \mn@doi [\aj] {10.3847/1538-3881/aa9be0},
  \href {https://ui.adsabs.harvard.edu/abs/2018AJ....155...57C} {155, 57}

\bibitem[\protect\citeauthoryear{{Crossfield} et~al.,}{{Crossfield}
  et~al.}{2016}]{2016ApJS..226....7C}
{Crossfield} I. J.~M.,  et~al., 2016, \mn@doi [\apjs]
  {10.3847/0067-0049/226/1/7}, \href
  {https://ui.adsabs.harvard.edu/abs/2016ApJS..226....7C} {226, 7}

\bibitem[\protect\citeauthoryear{{Cui} et~al.,}{{Cui}
  et~al.}{2012}]{2012RAA....12.1197C}
{Cui} X.-Q.,  et~al., 2012, \mn@doi [Research in Astronomy and Astrophysics]
  {10.1088/1674-4527/12/9/003}, \href
  {https://ui.adsabs.harvard.edu/abs/2012RAA....12.1197C} {12, 1197}

\bibitem[\protect\citeauthoryear{{D{\'\i}az}, {Almenara}, {Santerne}, {Moutou},
  {Lethuillier}  \& {Deleuil}}{{D{\'\i}az} et~al.}{2014}]{2014MNRAS.441..983D}
{D{\'\i}az} R.~F.,  {Almenara} J.~M.,  {Santerne} A.,  {Moutou} C.,
  {Lethuillier} A.,   {Deleuil} M.,  2014, \mn@doi [\mnras]
  {10.1093/mnras/stu601}, \href
  {https://ui.adsabs.harvard.edu/abs/2014MNRAS.441..983D} {441, 983}

\bibitem[\protect\citeauthoryear{{D{\'\i}ez Alonso} et~al.,}{{D{\'\i}ez Alonso}
  et~al.}{2018a}]{2018MNRAS.476L..50D}
{D{\'\i}ez Alonso} E.,  et~al., 2018a, \mn@doi [\mnras]
  {10.1093/mnrasl/sly040}, \href
  {https://ui.adsabs.harvard.edu/abs/2018MNRAS.476L..50D} {476, L50}

\bibitem[\protect\citeauthoryear{{D{\'\i}ez Alonso} et~al.,}{{D{\'\i}ez Alonso}
  et~al.}{2018b}]{2018MNRAS.480L...1D}
{D{\'\i}ez Alonso} E.,  et~al., 2018b, \mn@doi [\mnras]
  {10.1093/mnrasl/sly102}, \href
  {https://ui.adsabs.harvard.edu/abs/2018MNRAS.480L...1D} {480, L1}

\bibitem[\protect\citeauthoryear{{D{\'\i}ez Alonso} et~al.,}{{D{\'\i}ez Alonso}
  et~al.}{2019}]{2019MNRAS.489.5928D}
{D{\'\i}ez Alonso} E.,  et~al., 2019, \mn@doi [\mnras] {10.1093/mnras/sty3467},
  \href {https://ui.adsabs.harvard.edu/abs/2019MNRAS.489.5928D} {489, 5928}

\bibitem[\protect\citeauthoryear{{Dotter}}{{Dotter}}{2016}]{2016ApJS..222....8D}
{Dotter} A.,  2016, \mn@doi [\apjs] {10.3847/0067-0049/222/1/8}, \href
  {https://ui.adsabs.harvard.edu/abs/2016ApJS..222....8D} {222, 8}

\bibitem[\protect\citeauthoryear{{Dressing} et~al.,}{{Dressing}
  et~al.}{2017a}]{2017AJ....154..207D}
{Dressing} C.~D.,  et~al., 2017a, \mn@doi [\aj] {10.3847/1538-3881/aa89f2},
  \href {https://ui.adsabs.harvard.edu/abs/2017AJ....154..207D} {154, 207}

\bibitem[\protect\citeauthoryear{{Dressing}, {Newton}, {Schlieder},
  {Charbonneau}, {Knutson}, {Vanderburg}  \& {Sinukoff}}{{Dressing}
  et~al.}{2017b}]{2017ApJ...836..167D}
{Dressing} C.~D.,  {Newton} E.~R.,  {Schlieder} J.~E.,  {Charbonneau} D.,
  {Knutson} H.~A.,  {Vanderburg} A.,   {Sinukoff} E.,  2017b, \mn@doi [\apj]
  {10.3847/1538-4357/836/2/167}, \href
  {https://ui.adsabs.harvard.edu/abs/2017ApJ...836..167D} {836, 167}

\bibitem[\protect\citeauthoryear{{Eisner} et~al.,}{{Eisner}
  et~al.}{2020a}]{2020arXiv201113944E}
{Eisner} N.~L.,  et~al., 2020a, arXiv e-prints, \href
  {https://ui.adsabs.harvard.edu/abs/2020arXiv201113944E} {p. arXiv:2011.13944}

\bibitem[\protect\citeauthoryear{{Eisner} et~al.,}{{Eisner}
  et~al.}{2020b}]{2020MNRAS.494..750E}
{Eisner} N.~L.,  et~al., 2020b, \mn@doi [\mnras] {10.1093/mnras/staa138}, \href
  {https://ui.adsabs.harvard.edu/abs/2020MNRAS.494..750E} {494, 750}

\bibitem[\protect\citeauthoryear{{Espinoza} \& {Jord{\'a}n}}{{Espinoza} \&
  {Jord{\'a}n}}{2015}]{2015MNRAS.450.1879E}
{Espinoza} N.,  {Jord{\'a}n} A.,  2015, \mn@doi [\mnras]
  {10.1093/mnras/stv744}, \href
  {https://ui.adsabs.harvard.edu/abs/2015MNRAS.450.1879E} {450, 1879}

\bibitem[\protect\citeauthoryear{{Evans}}{{Evans}}{2018}]{2018RNAAS...2...20E}
{Evans} D.~F.,  2018, \mn@doi [Research Notes of the American Astronomical
  Society] {10.3847/2515-5172/aac173}, \href
  {https://ui.adsabs.harvard.edu/abs/2018RNAAS...2...20E} {2, 20}

\bibitem[\protect\citeauthoryear{{Feinstein} et~al.,}{{Feinstein}
  et~al.}{2019}]{2019AJ....157...40F}
{Feinstein} A.~D.,  et~al., 2019, \mn@doi [\aj] {10.3847/1538-3881/aafa70},
  \href {https://ui.adsabs.harvard.edu/abs/2019AJ....157...40F} {157, 40}

\bibitem[\protect\citeauthoryear{{Feroz} \& {Hobson}}{{Feroz} \&
  {Hobson}}{2008}]{2008MNRAS.384..449F}
{Feroz} F.,  {Hobson} M.~P.,  2008, \mn@doi [\mnras]
  {10.1111/j.1365-2966.2007.12353.x}, \href
  {https://ui.adsabs.harvard.edu/abs/2008MNRAS.384..449F} {384, 449}

\bibitem[\protect\citeauthoryear{{Feroz}, {Hobson}  \& {Bridges}}{{Feroz}
  et~al.}{2009}]{2009MNRAS.398.1601F}
{Feroz} F.,  {Hobson} M.~P.,   {Bridges} M.,  2009, \mn@doi [\mnras]
  {10.1111/j.1365-2966.2009.14548.x}, \href
  {https://ui.adsabs.harvard.edu/abs/2009MNRAS.398.1601F} {398, 1601}

\bibitem[\protect\citeauthoryear{{Feroz}, {Hobson}, {Cameron}  \&
  {Pettitt}}{{Feroz} et~al.}{2019}]{2019OJAp....2E..10F}
{Feroz} F.,  {Hobson} M.~P.,  {Cameron} E.,   {Pettitt} A.~N.,  2019, \mn@doi
  [The Open Journal of Astrophysics] {10.21105/astro.1306.2144}, \href
  {https://ui.adsabs.harvard.edu/abs/2019OJAp....2E..10F} {2, 10}

\bibitem[\protect\citeauthoryear{{Fischer} et~al.,}{{Fischer}
  et~al.}{2012}]{2012MNRAS.419.2900F}
{Fischer} D.~A.,  et~al., 2012, \mn@doi [\mnras]
  {10.1111/j.1365-2966.2011.19932.x}, \href
  {https://ui.adsabs.harvard.edu/abs/2012MNRAS.419.2900F} {419, 2900}

\bibitem[\protect\citeauthoryear{{Fressin} et~al.,}{{Fressin}
  et~al.}{2013}]{2013ApJ...766...81F}
{Fressin} F.,  et~al., 2013, \mn@doi [\apj] {10.1088/0004-637X/766/2/81}, \href
  {https://ui.adsabs.harvard.edu/abs/2013ApJ...766...81F} {766, 81}

\bibitem[\protect\citeauthoryear{{Fulton} et~al.,}{{Fulton}
  et~al.}{2017}]{2017AJ....154..109F}
{Fulton} B.~J.,  et~al., 2017, \mn@doi [\aj] {10.3847/1538-3881/aa80eb}, \href
  {https://ui.adsabs.harvard.edu/abs/2017AJ....154..109F} {154, 109}

\bibitem[\protect\citeauthoryear{{Gaia Collaboration} et~al.,}{{Gaia
  Collaboration} et~al.}{2018}]{2018A&A...616A...1G}
{Gaia Collaboration} et~al., 2018, \mn@doi [\aap]
  {10.1051/0004-6361/201833051}, \href
  {http://adsabs.harvard.edu/abs/2018A%26A...616A...1G} {616, A1}

\bibitem[\protect\citeauthoryear{{Giacalone} et~al.,}{{Giacalone}
  et~al.}{2021}]{2021AJ....161...24G}
{Giacalone} S.,  et~al., 2021, \mn@doi [\aj] {10.3847/1538-3881/abc6af}, \href
  {https://ui.adsabs.harvard.edu/abs/2021AJ....161...24G} {161, 24}

\bibitem[\protect\citeauthoryear{{Girardi}, {Groenewegen}, {Hatziminaoglou}  \&
  {da Costa}}{{Girardi} et~al.}{2005}]{2005A&A...436..895G}
{Girardi} L.,  {Groenewegen} M.~A.~T.,  {Hatziminaoglou} E.,   {da Costa} L.,
  2005, \mn@doi [\aap] {10.1051/0004-6361:20042352}, \href
  {https://ui.adsabs.harvard.edu/abs/2005A&A...436..895G} {436, 895}

\bibitem[\protect\citeauthoryear{{Girardi} et~al.,}{{Girardi}
  et~al.}{2012}]{girardi12}
{Girardi} L.,  et~al., 2012, \mn@doi [Astrophysics and Space Science
  Proceedings] {10.1007/978-3-642-18418-5_17}, \href
  {https://ui.adsabs.harvard.edu/abs/2012ASSP...26..165G} {26, 165}

\bibitem[\protect\citeauthoryear{{Grimm} et~al.,}{{Grimm}
  et~al.}{2018}]{2018A&A...613A..68G}
{Grimm} S.~L.,  et~al., 2018, \mn@doi [\aap] {10.1051/0004-6361/201732233},
  \href {https://ui.adsabs.harvard.edu/abs/2018A&A...613A..68G} {613, A68}

\bibitem[\protect\citeauthoryear{{Hardegree-Ullman} \&
  {Christiansen}}{{Hardegree-Ullman} \&
  {Christiansen}}{2019}]{2019AAS...23316407H}
{Hardegree-Ullman} K.,  {Christiansen} J.,  2019, in American Astronomical
  Society Meeting Abstracts \#233. p. 164.07

\bibitem[\protect\citeauthoryear{{Hardegree-Ullman}, {Zink}, {Christiansen},
  {Dressing}, {Ciardi}  \& {Schlieder}}{{Hardegree-Ullman}
  et~al.}{2020}]{2020ApJS..247...28H}
{Hardegree-Ullman} K.~K.,  {Zink} J.~K.,  {Christiansen} J.~L.,  {Dressing}
  C.~D.,  {Ciardi} D.~R.,   {Schlieder} J.~E.,  2020, \mn@doi [\apjs]
  {10.3847/1538-4365/ab7230}, \href
  {https://ui.adsabs.harvard.edu/abs/2020ApJS..247...28H} {247, 28}

\bibitem[\protect\citeauthoryear{{Heller}, {Hippke}  \& {Rodenbeck}}{{Heller}
  et~al.}{2019}]{2019A&A...627A..66H}
{Heller} R.,  {Hippke} M.,   {Rodenbeck} K.,  2019, \mn@doi [\aap]
  {10.1051/0004-6361/201935600}, \href
  {https://ui.adsabs.harvard.edu/abs/2019A&A...627A..66H} {627, A66}

\bibitem[\protect\citeauthoryear{{Hippke} \& {Heller}}{{Hippke} \&
  {Heller}}{2019}]{2019A&A...623A..39H}
{Hippke} M.,  {Heller} R.,  2019, \mn@doi [\aap] {10.1051/0004-6361/201834672},
  \href {https://ui.adsabs.harvard.edu/abs/2019A&A...623A..39H} {623, A39}

\bibitem[\protect\citeauthoryear{{Hippke}, {David}, {Mulders}  \&
  {Heller}}{{Hippke} et~al.}{2019}]{2019AJ....158..143H}
{Hippke} M.,  {David} T.~J.,  {Mulders} G.~D.,   {Heller} R.,  2019, \mn@doi
  [\aj] {10.3847/1538-3881/ab3984}, \href
  {https://ui.adsabs.harvard.edu/abs/2019AJ....158..143 H} {158, 143}

\bibitem[\protect\citeauthoryear{{Hormuth}, {Brandner}, {Hippler}  \&
  {Henning}}{{Hormuth} et~al.}{2008}]{hormuth08}
{Hormuth} F.,  {Brandner} W.,  {Hippler} S.,   {Henning} T.,  2008, \mn@doi
  [Journal of Physics Conference Series] {10.1088/1742-6596/131/1/012051},
  \href {http://adsabs.harvard.edu/abs/2008JPhCS.131a2051H} {131, 012051}

\bibitem[\protect\citeauthoryear{{Howard} et~al.,}{{Howard}
  et~al.}{2012}]{2012ApJS..201...15H}
{Howard} A.~W.,  et~al., 2012, \mn@doi [\apjs] {10.1088/0067-0049/201/2/15},
  \href {https://ui.adsabs.harvard.edu/abs/2012ApJS..201...15H} {201, 15}

\bibitem[\protect\citeauthoryear{{Howell} et~al.,}{{Howell}
  et~al.}{2014}]{2014PASP..126..398H}
{Howell} S.~B.,  et~al., 2014, \mn@doi [\pasp] {10.1086/676406}, \href
  {https://ui.adsabs.harvard.edu/abs/2014PASP..126..398H} {126, 398}

\bibitem[\protect\citeauthoryear{{Huber} et~al.,}{{Huber}
  et~al.}{2016}]{2016ApJS..224....2H}
{Huber} D.,  et~al., 2016, \mn@doi [\apjs] {10.3847/0067-0049/224/1/2}, \href
  {https://ui.adsabs.harvard.edu/abs/2016ApJS..224....2H} {224, 2}

\bibitem[\protect\citeauthoryear{{Jin}, {Mordasini}, {Parmentier}, {van
  Boekel}, {Henning}  \& {Ji}}{{Jin} et~al.}{2014}]{2014ApJ...795...65J}
{Jin} S.,  {Mordasini} C.,  {Parmentier} V.,  {van Boekel} R.,  {Henning} T.,
  {Ji} J.,  2014, \mn@doi [\apj] {10.1088/0004-637X/795/1/65}, \href
  {https://ui.adsabs.harvard.edu/abs/2014ApJ...795...65J} {795, 65}

\bibitem[\protect\citeauthoryear{{Kipping}}{{Kipping}}{2010}]{2010MNRAS.408.1758K}
{Kipping} D.~M.,  2010, \mn@doi [\mnras] {10.1111/j.1365-2966.2010.17242.x},
  \href {https://ui.adsabs.harvard.edu/abs/2010MNRAS.408.1758K} {408, 1758}

\bibitem[\protect\citeauthoryear{{Kipping}}{{Kipping}}{2013}]{2013MNRAS.435.2152K}
{Kipping} D.~M.,  2013, \mn@doi [\mnras] {10.1093/mnras/stt1435}, \href
  {https://ui.adsabs.harvard.edu/abs/2013MNRAS.435.2152K} {435, 2152}

\bibitem[\protect\citeauthoryear{{Kopparapu} et~al.,}{{Kopparapu}
  et~al.}{2013}]{2013ApJ...765..131K}
{Kopparapu} R.~K.,  et~al., 2013, \mn@doi [\apj] {10.1088/0004-637X/765/2/131},
  \href {https://ui.adsabs.harvard.edu/abs/2013ApJ...765..131K} {765, 131}

\bibitem[\protect\citeauthoryear{Korth}{Korth}{2020}]{Korth2020}
Korth J.,  2020, PhD thesis, \url {http://www.uni-koeln.de/}

\bibitem[\protect\citeauthoryear{{Korth} et~al.,}{{Korth}
  et~al.}{2019}]{2019MNRAS.482.1807K}
{Korth} J.,  et~al., 2019, \mn@doi [\mnras] {10.1093/mnras/sty2760}, \href
  {https://ui.adsabs.harvard.edu/abs/2019MNRAS.482.1807K} {482, 1807}

\bibitem[\protect\citeauthoryear{{Kov{\'a}cs}, {Zucker}  \&
  {Mazeh}}{{Kov{\'a}cs} et~al.}{2002}]{2002A&A...391..369K}
{Kov{\'a}cs} G.,  {Zucker} S.,   {Mazeh} T.,  2002, \mn@doi [\aap]
  {10.1051/0004-6361:20020802}, \href
  {https://ui.adsabs.harvard.edu/abs/2002A&A...391..369K} {391, 369}

\bibitem[\protect\citeauthoryear{{Kreidberg}}{{Kreidberg}}{2015}]{2015PASP..127.1161K}
{Kreidberg} L.,  2015, \mn@doi [\pasp] {10.1086/683602}, \href
  {https://ui.adsabs.harvard.edu/abs/2015PASP..127.1161K} {127, 1161}

\bibitem[\protect\citeauthoryear{{Kruse}, {Agol}, {Luger}  \&
  {Foreman-Mackey}}{{Kruse} et~al.}{2019}]{2019ApJS..244...11K}
{Kruse} E.,  {Agol} E.,  {Luger} R.,   {Foreman-Mackey} D.,  2019, \mn@doi
  [\apjs] {10.3847/1538-4365/ab346b}, \href
  {https://ui.adsabs.harvard.edu/abs/2019ApJS..244...11K} {244, 11}

\bibitem[\protect\citeauthoryear{{Kunimoto} \& {Matthews}}{{Kunimoto} \&
  {Matthews}}{2020}]{2020AJ....159..248K}
{Kunimoto} M.,  {Matthews} J.~M.,  2020, \mn@doi [\aj]
  {10.3847/1538-3881/ab88b0}, \href
  {https://ui.adsabs.harvard.edu/abs/2020AJ....159..248K} {159, 248}

\bibitem[\protect\citeauthoryear{{Kurucz}}{{Kurucz}}{1979}]{1979ApJS...40....1K}
{Kurucz} R.~L.,  1979, \mn@doi [\apjs] {10.1086/190589}, \href
  {https://ui.adsabs.harvard.edu/abs/1979ApJS...40....1K} {40, 1}

\bibitem[\protect\citeauthoryear{{Lam} et~al.,}{{Lam}
  et~al.}{2018}]{2018A&A...620A..77L}
{Lam} K.~W.~F.,  et~al., 2018, \mn@doi [\aap] {10.1051/0004-6361/201834073},
  \href {https://ui.adsabs.harvard.edu/abs/2018A&A...620A..77L} {620, A77}

\bibitem[\protect\citeauthoryear{{Libralato} et~al.,}{{Libralato}
  et~al.}{2016}]{2016MNRAS.463.1780L}
{Libralato} M.,  et~al., 2016, \mn@doi [\mnras] {10.1093/mnras/stw1932}, \href
  {https://ui.adsabs.harvard.edu/abs/2016MNRAS.463.1780L} {463, 1780}

\bibitem[\protect\citeauthoryear{{Lillo-Box}, {Barrado}  \& {Bouy}}{{Lillo-Box}
  et~al.}{2012}]{lillo-box12}
{Lillo-Box} J.,  {Barrado} D.,   {Bouy} H.,  2012, \mn@doi [\aap]
  {10.1051/0004-6361/201219631}, \href
  {http://ads.nao.ac.jp/abs/2012A%26A...546A..10L} {546, A10}

\bibitem[\protect\citeauthoryear{{Lillo-Box} et~al.,}{{Lillo-Box}
  et~al.}{2014a}]{2014A&A...562A.109L}
{Lillo-Box} J.,  et~al., 2014a, \mn@doi [\aap] {10.1051/0004-6361/201322001},
  \href {https://ui.adsabs.harvard.edu/abs/2014A&A...562A.109L} {562, A109}

\bibitem[\protect\citeauthoryear{{Lillo-Box}, {Barrado}  \& {Bouy}}{{Lillo-Box}
  et~al.}{2014b}]{lillo-box14b}
{Lillo-Box} J.,  {Barrado} D.,   {Bouy} H.,  2014b, \mn@doi [\aap]
  {10.1051/0004-6361/201423497}, \href
  {http://adsabs.harvard.edu/abs/2014A%26A...566A.103L} {566, A103}

\bibitem[\protect\citeauthoryear{{Lillo-Box} et~al.,}{{Lillo-Box}
  et~al.}{2014c}]{2014A&A...568L...1L}
{Lillo-Box} J.,  et~al., 2014c, \mn@doi [\aap] {10.1051/0004-6361/201424587},
  \href {https://ui.adsabs.harvard.edu/abs/2014A&A...568L...1L} {568, L1}

\bibitem[\protect\citeauthoryear{{Lissauer} et~al.,}{{Lissauer}
  et~al.}{2011}]{2011ApJS..197....8L}
{Lissauer} J.~J.,  et~al., 2011, \mn@doi [\apjs] {10.1088/0067-0049/197/1/8},
  \href {https://ui.adsabs.harvard.edu/abs/2011ApJS..197....8L} {197, 8}

\bibitem[\protect\citeauthoryear{{Lithwick}, {Xie}  \& {Wu}}{{Lithwick}
  et~al.}{2012}]{2012ApJ...761..122L}
{Lithwick} Y.,  {Xie} J.,   {Wu} Y.,  2012, \mn@doi [\apj]
  {10.1088/0004-637X/761/2/122}, \href
  {https://ui.adsabs.harvard.edu/abs/2012ApJ...761..122L} {761, 122}

\bibitem[\protect\citeauthoryear{{Livingston} et~al.,}{{Livingston}
  et~al.}{2018a}]{2018AJ....156...78L}
{Livingston} J.~H.,  et~al., 2018a, \mn@doi [\aj] {10.3847/1538-3881/aaccde},
  \href {https://ui.adsabs.harvard.edu/abs/2018AJ....156...78L} {156, 78}

\bibitem[\protect\citeauthoryear{{Livingston} et~al.,}{{Livingston}
  et~al.}{2018b}]{2018AJ....156..277L}
{Livingston} J.~H.,  et~al., 2018b, \mn@doi [\aj] {10.3847/1538-3881/aae778},
  \href {https://ui.adsabs.harvard.edu/abs/2018AJ....156..277L} {156, 277}

\bibitem[\protect\citeauthoryear{{Lopez} \& {Fortney}}{{Lopez} \&
  {Fortney}}{2014}]{2014ApJ...792....1L}
{Lopez} E.~D.,  {Fortney} J.~J.,  2014, \mn@doi [\apj]
  {10.1088/0004-637X/792/1/1}, \href
  {https://ui.adsabs.harvard.edu/abs/2014ApJ...792....1L} {792, 1}

\bibitem[\protect\citeauthoryear{{Luger}, {Agol}, {Kruse}, {Barnes}, {Becker},
  {Foreman-Mackey}  \& {Deming}}{{Luger} et~al.}{2016}]{2016AJ....152..100L}
{Luger} R.,  {Agol} E.,  {Kruse} E.,  {Barnes} R.,  {Becker} A.,
  {Foreman-Mackey} D.,   {Deming} D.,  2016, \mn@doi [\aj]
  {10.3847/0004-6256/152/4/100}, \href
  {https://ui.adsabs.harvard.edu/abs/2016AJ....152..100L} {152, 100}

\bibitem[\protect\citeauthoryear{{Luger}, {Kruse}, {Foreman-Mackey}, {Agol}  \&
  {Saunders}}{{Luger} et~al.}{2018}]{2018AJ....156...99L}
{Luger} R.,  {Kruse} E.,  {Foreman-Mackey} D.,  {Agol} E.,   {Saunders} N.,
  2018, \mn@doi [\aj] {10.3847/1538-3881/aad230}, \href
  {https://ui.adsabs.harvard.edu/abs/2018AJ....156...99L} {156, 99}

\bibitem[\protect\citeauthoryear{{Luri} et~al.,}{{Luri}
  et~al.}{2018}]{2018A&A...616A...9L}
{Luri} X.,  et~al., 2018, \mn@doi [\aap] {10.1051/0004-6361/201832964}, \href
  {https://ui.adsabs.harvard.edu/abs/2018A&A...616A...9L} {616, A9}

\bibitem[\protect\citeauthoryear{{Mandel} \& {Agol}}{{Mandel} \&
  {Agol}}{2002}]{2002ApJ...580L.171M}
{Mandel} K.,  {Agol} E.,  2002, \mn@doi [\apjl] {10.1086/345520}, \href
  {https://ui.adsabs.harvard.edu/abs/2002ApJ...580L.171M} {580, L171}

\bibitem[\protect\citeauthoryear{{Mann} et~al.,}{{Mann}
  et~al.}{2017}]{2017AJ....153...64M}
{Mann} A.~W.,  et~al., 2017, \mn@doi [\aj] {10.1088/1361-6528/aa5276}, \href
  {https://ui.adsabs.harvard.edu/abs/2017AJ....153...64M} {153, 64}

\bibitem[\protect\citeauthoryear{{Marigo}, {Girardi}, {Bressan}, {Groenewegen},
  {Silva}  \& {Granato}}{{Marigo} et~al.}{2008}]{2008A&A...482..883M}
{Marigo} P.,  {Girardi} L.,  {Bressan} A.,  {Groenewegen} M.~A.~T.,  {Silva}
  L.,   {Granato} G.~L.,  2008, \mn@doi [\aap] {10.1051/0004-6361:20078467},
  \href {https://ui.adsabs.harvard.edu/abs/2008A&A...482..883M} {482, 883}

\bibitem[\protect\citeauthoryear{{Mayo} et~al.,}{{Mayo}
  et~al.}{2018}]{2018AJ....155..136M}
{Mayo} A.~W.,  et~al., 2018, \mn@doi [\aj] {10.3847/1538-3881/aaadff}, \href
  {https://ui.adsabs.harvard.edu/abs/2018AJ....155..136M} {155, 136}

\bibitem[\protect\citeauthoryear{{Montet} et~al.,}{{Montet}
  et~al.}{2015}]{2015ApJ...809...25M}
{Montet} B.~T.,  et~al., 2015, \mn@doi [\apj] {10.1088/0004-637X/809/1/25},
  \href {https://ui.adsabs.harvard.edu/abs/2015ApJ...809...25M} {809, 25}

\bibitem[\protect\citeauthoryear{{Morton}}{{Morton}}{2012}]{2012ApJ...761....6M}
{Morton} T.~D.,  2012, \mn@doi [\apj] {10.1088/0004-637X/761/1/6}, \href
  {http://adsabs.harvard.edu/abs/2012ApJ...761....6M} {761, 6}

\bibitem[\protect\citeauthoryear{{Morton}}{{Morton}}{2015a}]{2015ascl.soft03010M}
{Morton} T.~D.,  2015a, {isochrones: Stellar model grid package}, Astrophysics
  Source Code Library (\mn@eprint {ascl} {1503.010})

\bibitem[\protect\citeauthoryear{{Morton}}{{Morton}}{2015b}]{2015ascl.soft03011M}
{Morton} T.~D.,  2015b, {VESPA: False positive probabilities calculator},
  Astrophysics Source Code Library (\mn@eprint {ascl} {1503.011})

\bibitem[\protect\citeauthoryear{{Morton}, {Bryson}, {Coughlin}, {Rowe},
  {Ravichandran}, {Petigura}, {Haas}  \& {Batalha}}{{Morton}
  et~al.}{2016}]{2016ApJ...822...86M}
{Morton} T.~D.,  {Bryson} S.~T.,  {Coughlin} J.~L.,  {Rowe} J.~F.,
  {Ravichandran} G.,  {Petigura} E.~A.,  {Haas} M.~R.,   {Batalha} N.~M.,
  2016, \mn@doi [\apj] {10.3847/0004-637X/822/2/86}, \href
  {https://ui.adsabs.harvard.edu/abs/2016ApJ...822...86M} {822, 86}

\bibitem[\protect\citeauthoryear{{Owen} \& {Wu}}{{Owen} \&
  {Wu}}{2013}]{2013ApJ...775..105O}
{Owen} J.~E.,  {Wu} Y.,  2013, \mn@doi [\apj] {10.1088/0004-637X/775/2/105},
  \href {https://ui.adsabs.harvard.edu/abs/2013ApJ...775..105O} {775, 105}

\bibitem[\protect\citeauthoryear{Parviainen}{Parviainen}{2015}]{Parviainen2015}
Parviainen H.,  2015, \mn@doi [MNRAS] {10.1093/mnras/stv894}, 450, 3233

\bibitem[\protect\citeauthoryear{{Parviainen} \& {Korth}}{{Parviainen} \&
  {Korth}}{2020}]{2020MNRAS.499.3356P}
{Parviainen} H.,  {Korth} J.,  2020, \mn@doi [\mnras] {10.1093/mnras/staa2953},
  \href {https://ui.adsabs.harvard.edu/abs/2020MNRAS.499.3356P} {499, 3356}

\bibitem[\protect\citeauthoryear{{Paxton} et~al.,}{{Paxton}
  et~al.}{2015}]{2015ApJS..220...15P}
{Paxton} B.,  et~al., 2015, \mn@doi [\apjs] {10.1088/0067-0049/220/1/15}, \href
  {https://ui.adsabs.harvard.edu/abs/2015ApJS..220...15P} {220, 15}

\bibitem[\protect\citeauthoryear{{Petigura}, {Howard}  \& {Marcy}}{{Petigura}
  et~al.}{2013}]{2013PNAS..11019273P}
{Petigura} E.~A.,  {Howard} A.~W.,   {Marcy} G.~W.,  2013, \mn@doi [Proceedings
  of the National Academy of Science] {10.1073/pnas.1319909110}, \href
  {https://ui.adsabs.harvard.edu/abs/2013PNAS..11019273P} {110, 19273}

\bibitem[\protect\citeauthoryear{{Petigura} et~al.,}{{Petigura}
  et~al.}{2018}]{2018AJ....155...21P}
{Petigura} E.~A.,  et~al., 2018, \mn@doi [\aj] {10.3847/1538-3881/aa9b83},
  \href {https://ui.adsabs.harvard.edu/abs/2018AJ....155...21P} {155, 21}

\bibitem[\protect\citeauthoryear{{Pope}, {Parviainen}  \& {Aigrain}}{{Pope}
  et~al.}{2016}]{2016MNRAS.461.3399P}
{Pope} B. J.~S.,  {Parviainen} H.,   {Aigrain} S.,  2016, \mn@doi [\mnras]
  {10.1093/mnras/stw1373}, \href
  {https://ui.adsabs.harvard.edu/abs/2016MNRAS.461.3399P} {461, 3399}

\bibitem[\protect\citeauthoryear{{Ragozzine} \& {Holman}}{{Ragozzine} \&
  {Holman}}{2010}]{2010arXiv1006.3727R}
{Ragozzine} D.,  {Holman} M.~J.,  2010, arXiv e-prints, \href
  {https://ui.adsabs.harvard.edu/abs/2010arXiv1006.3727R} {p. arXiv:1006.3727}

\bibitem[\protect\citeauthoryear{{Rein} \& {Liu}}{{Rein} \&
  {Liu}}{2012}]{2012A&A...537A.128R}
{Rein} H.,  {Liu} S.~F.,  2012, \mn@doi [\aap] {10.1051/0004-6361/201118085},
  \href {https://ui.adsabs.harvard.edu/abs/2012A&A...537A.128R} {537, A128}

\bibitem[\protect\citeauthoryear{{Rizzuto}, {Mann}, {Vanderburg}, {Kraus}  \&
  {Covey}}{{Rizzuto} et~al.}{2017}]{2017AJ....154..224R}
{Rizzuto} A.~C.,  {Mann} A.~W.,  {Vanderburg} A.,  {Kraus} A.~L.,   {Covey}
  K.~R.,  2017, \mn@doi [\aj] {10.3847/1538-3881/aa9070}, \href
  {https://ui.adsabs.harvard.edu/abs/2017AJ....154..224R} {154, 224}

\bibitem[\protect\citeauthoryear{{Rowe} et~al.,}{{Rowe}
  et~al.}{2014}]{2014ApJ...784...45R}
{Rowe} J.~F.,  et~al., 2014, \mn@doi [\apj] {10.1088/0004-637X/784/1/45}, \href
  {https://ui.adsabs.harvard.edu/abs/2014ApJ...784...45R} {784, 45}

\bibitem[\protect\citeauthoryear{{Schmitt} et~al.,}{{Schmitt}
  et~al.}{2014}]{2014ApJ...795..167S}
{Schmitt} J.~R.,  et~al., 2014, \mn@doi [\apj] {10.1088/0004-637X/795/2/167},
  \href {https://ui.adsabs.harvard.edu/abs/2014ApJ...795..167S} {795, 167}

\bibitem[\protect\citeauthoryear{{Schmitt}, {Hartman}  \& {Kipping}}{{Schmitt}
  et~al.}{2019}]{2019arXiv191008034S}
{Schmitt} A.~R.,  {Hartman} J.~D.,   {Kipping} D.~M.,  2019, arXiv e-prints,
  \href {https://ui.adsabs.harvard.edu/abs/2019arXiv191008034S} {p.
  arXiv:1910.08034}

\bibitem[\protect\citeauthoryear{{Schwamb} et~al.,}{{Schwamb}
  et~al.}{2012}]{2012ApJ...754..129S}
{Schwamb} M.~E.,  et~al., 2012, \mn@doi [\apj] {10.1088/0004-637X/754/2/129},
  \href {https://ui.adsabs.harvard.edu/abs/2012ApJ...754..129S} {754, 129}

\bibitem[\protect\citeauthoryear{{Shporer} et~al.,}{{Shporer}
  et~al.}{2017a}]{2017AJ....154..188S}
{Shporer} A.,  et~al., 2017a, \mn@doi [\aj] {10.3847/1538-3881/aa8bb9}, \href
  {https://ui.adsabs.harvard.edu/abs/2017AJ....154..188S} {154, 188}

\bibitem[\protect\citeauthoryear{{Shporer} et~al.,}{{Shporer}
  et~al.}{2017b}]{2017ApJ...847L..18S}
{Shporer} A.,  et~al., 2017b, \mn@doi [\apjl] {10.3847/2041-8213/aa8bff}, \href
  {https://ui.adsabs.harvard.edu/abs/2017ApJ...847L..18S} {847, L18}

\bibitem[\protect\citeauthoryear{{Skrutskie} et~al.,}{{Skrutskie}
  et~al.}{2006}]{2006AJ....131.1163S}
{Skrutskie} M.~F.,  et~al., 2006, \mn@doi [\aj] {10.1086/498708}, \href
  {https://ui.adsabs.harvard.edu/abs/2006AJ....131.1163S} {131, 1163}

\bibitem[\protect\citeauthoryear{{Stassun} \& {Torres}}{{Stassun} \&
  {Torres}}{2018}]{2018ApJ...862...61S}
{Stassun} K.~G.,  {Torres} G.,  2018, \mn@doi [\apj]
  {10.3847/1538-4357/aacafc}, \href
  {https://ui.adsabs.harvard.edu/abs/2018ApJ...862...61S} {862, 61}

\bibitem[\protect\citeauthoryear{{Strehl}}{{Strehl}}{1902}]{strehl1902}
{Strehl} K.,  1902, \mn@doi [Astronomische Nachrichten]
  {10.1002/asna.19021580604}, \href
  {http://adsabs.harvard.edu/abs/1902AN....158...89S} {158, 89}

\bibitem[\protect\citeauthoryear{{Torres} et~al.,}{{Torres}
  et~al.}{2011}]{2011ApJ...727...24T}
{Torres} G.,  et~al., 2011, \mn@doi [\apj] {10.1088/0004-637X/727/1/24}, \href
  {https://ui.adsabs.harvard.edu/abs/2011ApJ...727...24T} {727, 24}

\bibitem[\protect\citeauthoryear{{Vanderburg} \& {Johnson}}{{Vanderburg} \&
  {Johnson}}{2014}]{2014PASP..126..948V}
{Vanderburg} A.,  {Johnson} J.~A.,  2014, \mn@doi [\pasp] {10.1086/678764},
  \href {https://ui.adsabs.harvard.edu/abs/2014PASP..126..948V} {126, 948}

\bibitem[\protect\citeauthoryear{{Vanderburg} et~al.,}{{Vanderburg}
  et~al.}{2016}]{2016ApJS..222...14V}
{Vanderburg} A.,  et~al., 2016, \mn@doi [\apjs] {10.3847/0067-0049/222/1/14},
  \href {https://ui.adsabs.harvard.edu/abs/2016ApJS..222...14V} {222, 14}

\bibitem[\protect\citeauthoryear{{Yu} et~al.,}{{Yu}
  et~al.}{2018}]{2018AJ....156...22Y}
{Yu} L.,  et~al., 2018, \mn@doi [\aj] {10.3847/1538-3881/aac6e6}, \href
  {https://ui.adsabs.harvard.edu/abs/2018AJ....156...22Y} {156, 22}

\bibitem[\protect\citeauthoryear{{Zaninetti}}{{Zaninetti}}{2008}]{2008SerAJ.177...73Z}
{Zaninetti} L.,  2008, \mn@doi [Serbian Astronomical Journal]
  {10.2298/SAJ0877073Z}, \href
  {https://ui.adsabs.harvard.edu/abs/2008SerAJ.177...73Z} {177, 73}

\bibitem[\protect\citeauthoryear{{Zechmeister} \& {K{\"u}rster}}{{Zechmeister}
  \& {K{\"u}rster}}{2009}]{2009A&A...496..577Z}
{Zechmeister} M.,  {K{\"u}rster} M.,  2009, \mn@doi [\aap]
  {10.1051/0004-6361:200811296}, \href
  {https://ui.adsabs.harvard.edu/abs/2009A&A...496..577Z} {496, 577}

\bibitem[\protect\citeauthoryear{Zeng et~al.,}{Zeng et~al.}{2019}]{Zeng9723}
Zeng L.,  et~al., 2019, \mn@doi [Proceedings of the National Academy of
  Sciences] {10.1073/pnas.1812905116}, 116, 9723

\bibitem[\protect\citeauthoryear{{Zink} et~al.,}{{Zink}
  et~al.}{2019}]{2019RNAAS...3...43Z}
{Zink} J.~K.,  et~al., 2019, \mn@doi [Research Notes of the American
  Astronomical Society] {10.3847/2515-5172/ab0a02}, \href
  {https://ui.adsabs.harvard.edu/abs/2019RNAAS...3...43Z} {3, 43}

\bibitem[\protect\citeauthoryear{{Zsom}, {Seager}, {de Wit}  \&
  {Stamenkovi{\'c}}}{{Zsom} et~al.}{2013}]{2013ApJ...778..109Z}
{Zsom} A.,  {Seager} S.,  {de Wit} J.,   {Stamenkovi{\'c}} V.,  2013, \mn@doi
  [\apj] {10.1088/0004-637X/778/2/109}, \href
  {https://ui.adsabs.harvard.edu/abs/2013ApJ...778..109Z} {778, 109}

\bibitem[\protect\citeauthoryear{{de Leon} et~al.,}{{de Leon}
  et~al.}{2021}]{2021MNRAS.508..195D}
{de Leon} J.~P.,  et~al., 2021, \mn@doi [\mnras] {10.1093/mnras/stab2305},
  \href {https://ui.adsabs.harvard.edu/abs/2021MNRAS.508..195D} {508, 195}

\makeatother
\end{thebibliography}







\bsp	
\label{lastpage}
\end{document}